\UseRawInputEncoding
\documentclass[onecolumn,secnumarabic,amssymb, nobibnotes, aps, prx, 10pt, floatfix]{revtex4-2}
\usepackage{amsmath, amssymb}
\usepackage{color, hyperref}
\usepackage[all]{hypcap} 
\usepackage{graphicx}
\usepackage{xcolor}
\usepackage{placeins}
\usepackage[normalem]{ulem}
\usepackage{bm}
\usepackage{tabularx}
\usepackage{multirow}

\usepackage{graphicx}
\usepackage{dcolumn}
\usepackage{bm}
\usepackage{amsmath}
\usepackage{color}
\usepackage{hyperref}
\usepackage{subcaption}   

\usepackage{float}
\hypersetup{
    colorlinks=true,
    linkcolor=blue,
    citecolor=blue,
    filecolor=blue,      
    urlcolor=blue,
}

\usepackage{graphicx}
\usepackage{dcolumn}
\usepackage{bm,amsmath,amssymb,amsbsy}     
\usepackage{braket}
\usepackage{appendix}
\usepackage{array}
\usepackage{booktabs}
\usepackage{soul} 
\setstcolor{red}

\usepackage{mathtools}
\usepackage{amsthm}
\usepackage{amsfonts}
\usepackage{float}
\usepackage{pstricks}           
\usepackage{mathrsfs}
\usepackage{hyperref}           
\usepackage{hypernat}           
\usepackage{soul}
\usepackage{slashed}    
\usepackage{color}
\usepackage{cleveref}
\def\COtwo{\text{CO}_2}
\def\COtwop{\text{CO}_2^+}

\def\CtwoSigmagp{\text{C}^2\Sigma_\text{g}^+}

\begin{document}

\title{Disentangling the Effect of Ionic Coupling and Multiple Interfering Terms in Attosecond Molecular Interferometry}

\author{Ioannis Makos$^{1}$, Jakub Benda$^{2}$, David Busto$^{1,3}$, Benjamin Steiner$^{1}$, Barbara Merzuk$^{1}$, Serguei Patchkovskii$^{4}$, Van-Hung Hoang$^{5}$, Uwe Thumm$^{5}$, Zden\v{e}k Ma\v{s}\'{i}n$^{2}$, Giuseppe Sansone$^{1,6}$}
 
\affiliation{$^{1}$ Institute of Physics, University of Freiburg, Hermann-Herder-Stra{\ss}e 3, 79104 Freiburg, Germany}
\affiliation{$^{2}$ Institute of Theoretical Physics, Faculty of Mathematics and Physics, Charles University, V~Hole\v{s}ovi\v{c}k\'ach 2, 180 00, Prague 8, Czech Republic}
\affiliation{$^{3}$ Department of Physics, Lund University, PO Box 118, SE-221 00 Lund, Sweden}
\affiliation{$^{4}$ Max Born Institute, Max-Born-Str. 2A, D-12489 Berlin, Germany}
\affiliation{$^{5}$ J. R. Macdonald Laboratory, Department of Physics, Kansas State University, Manhattan, Kansas 66506, USA}
\affiliation{$^{6}$ Freiburg Institute for Advanced Studies (FRIAS), University of Freiburg, Albertstrasse 19, 79104 Freiburg, Germany}
\email{corresponding author: giuseppe.sansone@physik.uni-freiburg.de}

\date{\today}
\begin{abstract}
Attosecond interferometry in a two-color field is central to attosecond metrology and spectroscopy. In this technique, a photoelectron wave packet is released when a single photon from an extreme ultraviolet comb is absorbed. The wave packet then either emits or absorbs one or more near-infrared photons, leading to the formation of sidebands of the main photoelectron peaks. This picture applies well to atoms and assumes that the near-infrared laser pulse only acts on the photoelectron leaving the parent ion. The effect of the near-infrared pulse on the electronic structure of the cation is not considered, since the field usually cannot induce transitions between its electronic levels. 
Here, we demonstrate how dynamics induced by the near-infrared field in the cation can significantly impact the amplitude and phases of the sideband signal of the photoelectrons associated with specific dissociative channels of CO$_2$ molecules. This coupling of the near-infrared field with the molecular cation opens a third  quantum pathway contributing to the signal measured in attosecond interferometry. Through angle- and energy-resolved characterization of the sideband oscillations, we observe reduction of interference amplitude over specific energy range upon angle integration. By  comparison with theoretical predictions, we can isolate the contributions of specific interfering pathways to the two-color multi-pathway photoionization process. The scheme investigated in our work is general, and our observations highlight the importance of the additional pathway for accurately interpreting attosecond interferometry experiments involving molecules and more complex quantum systems. 
\end{abstract}
\maketitle
 
\section{Introduction}
In 2001, the Reconstruction of Attosecond Beating by Interference of Two-photon transitions (\hbox{RABBIT}) technique was used to reconstruct the temporal profile of an attosecond pulse train~\cite{paulObservationTrainAttosecond2001}. Oscillations in the photoelectron spectra of the two-color photoionization process  depend on the relative delay $\Delta t$ between the extreme ultraviolet (XUV) comb of odd harmonics of a driving near-infrared (IR) field and the synchronized replica of the IR pulse. The modulations of the sideband yield contain information about the phase difference between consecutive harmonics, thereby providing access to the attosecond chirp~\cite{varju_FrequencyChirpHarmonic2005}. This technique is the workhorse of attosecond science driven by multi-cycle femtosecond pulses and has been used to investigate several electronic processes with unprecedented time resolution~\cite{sansoneElectronCorrelationReal2012, calegariAdvancesAttosecondScience2016, nisoliAttosecondElectronDynamics2017}.
This method can be understood as an interferometric process involving two different pathways connecting the initial state (a neutral atom, typically in its ground state) to the final state (an ion in its ground or in an excited state accompanied by a photoelectron wave packet with energy $E$). In the first pathway, the atom absorbs an XUV photon of the harmonic $(2q-1)$ (with $q$ an integer) releasing a photoelectron wave packet into the continuum, followed by the absorption of a IR photon. In the second pathway, the absorption of a photon of the harmonic $2q+1$ is followed by the emission of a IR photon. This leads to the same final state, which is characterized by an ion in its ground (or excited state) and a photoelectron with energy $E$. In this interpretation, the interaction of the IR field with the ion-photoelectron system solely changes the energy of the photoelectron wave packet.

Attosecond interferometry has been used to investigate the effect of shape resonances in the photoionization of molecules~\cite{huppertAttosecondDelaysMolecular2016, nandiAttosecondTimingElectron2020, gongAsymmetricAttosecondPhotoionization2022a} and was found to be sensitive to the influence of the molecular potential on the outgoing photoelectron wave packet~\cite{vosOrientationdependentStereoWigner2018, ahmadiAttosecondPhotoionisationTime2022}. Similarly, this technique has been implemented for the investigation of photoemission from surfaces~\cite{2017_Kasmi_RABBITT_CuSurface,2019_Ambrosio_RABBITT_CuSurface}, demonstrating the delicate response of interferometric photoelectron spectra and phases to the propagation of excited electrons in and outside metal surfaces~\cite{2019_Ambrosio_RABBITT_CuSurface,2020_Liao_PE_dispersion_RABBITT}. 

The IR field can also directly influence the residual molecular ion by inducing the absorption or emission of photons between cationic states, due to the presence of a complex energy structure.
In molecular systems, the effect of the coupling between cationic electronic states was first predicted~\cite{bendaAnalysisRABITTTime2022} between the $B^2\Sigma^+_u$ and $C^2\Sigma^+_g$ states of the CO$^+_2$ molecule, which present an energy difference of approximately 1.3~eV. This energy matches the photon energy of femtosecond pulses centered around 1030~nm, which are widely used in attosecond interferometry experiments. The IR field can efficiently couple the parallel electric dipole transition between the $B^2\Sigma^+_u$ and $C^2\Sigma^+_g$ states~\cite{2024_Hung_CO2_PRA, rubertiMultichannelDynamicsHigh2018}.
The signature of the coupling between the states was observed experimentally by monitoring the attosecond time delays measured in coincidence with the $B^2\Sigma^+_u$ ionic state~\cite{makosEntanglementPhotoionisationReveals2025}. The observed additional coupling delay stems from the entangled nature of the ion + photoelectron system and the action of the IR field on the bipartite system. In this new picture, the interferometer that produces the sideband yield oscillations results from three possible pathways, leading to multiple interfering terms that contribute to the photoelectron spectrum.
The IR-induced ionic coupling mechanism is expected to be general whenever two cationic states are in resonance with the external field. Its contribution has been predicted also in other molecular systems~\cite{delgadoThreepathInterferencesReconstruction2025}. 
The interaction of CO$_2$ molecules with intense femtosecond laser fields and attosecond pulses exhibits rich dynamics: Hole dynamics in the cation were observed in the process of high-order harmonic generation~\cite{smirnovaHighHarmonicInterferometry2009}. Nonadiabatic dynamics between the $B^2\Sigma^+_u$ and $A^2\Pi_u$ states were observed by exploiting  IR-induced transitions between the $B^2\Sigma^+_u$ and $C^2\Sigma^+_g$ state, which eventually predissociated due to spin-orbit interaction~\cite{timmersCoherentElectronHole2014a}. Finally, the effect of electronic-correlation on the photoelectron emission in CO$_2$ has also been reported~\cite{kamalovElectronCorrelationEffects2020}.

Due to the presence of different final cationic states and fragmentation pathways, the coincidence measurement of the photoelectron with the corresponding photoion provides additional experimental information for disentangling different photoionization pathways contributing to the photoelectron spectra~\cite{moshammer4pRecoilionElectron1996,dornerColdTargetRecoil2000a}. Furthermore, coincidence measurements allow one to measure photoionization time delays of different molecular species~\cite{gongAttosecondDelaysDissociative2023, ertelInfluenceNuclearDynamics2023} and cluster sizes~\cite{gongAttosecondSpectroscopySizeresolved2022} under the same experimental conditions, as well as angle-resolved photoionization delays in molecular systems~\cite{ertelAnisotropyParametersTwoColor2024} and, more recently, in chiral molecules~\cite{hanAttosecondControlMeasurement2025}. 

In this work, we combine attosecond interferometry and attosecond coincidence spectroscopy~\cite{sabbarCombiningAttosecondXUV2014,srinivasHighrepetitionRateAttosecond2022,ertelUltrastableHighrepetitionrateAttosecond2023} to demonstrate the effect of ionic coupling on the photoionization process leading to the fragmentation of CO$_2$ molecules.
We present experimental evidences for the existence of a third path connecting the initial and final states in attosecond interferometry, characterized by the exchange (absorption) of an IR photon in the cation formed by single-XUV-photon ionization. The presence of this additional path induces a large jump in the photoionization time delays in excellent agreement with the predictions of state-of-the-art quantum mechanical calculations.
Energy- and angle-resolved characterization of the photoelectron spectra for a specific ionic channel provides additional information about the pathways contributing to the two-color signal, offering the possibility of disentangling the different interfering terms in the complex photoelectron spectra. Throughout  this work we use Hartree atomic units (a.u.; e=$\hbar$=m$_e$=1), unless specified otherwise.

\section{Dissociation spectroscopy of CO$_2$}\label{Spectroscopy}
\begin{figure}[t!]
\centering 
\includegraphics[width=0.9\textwidth]{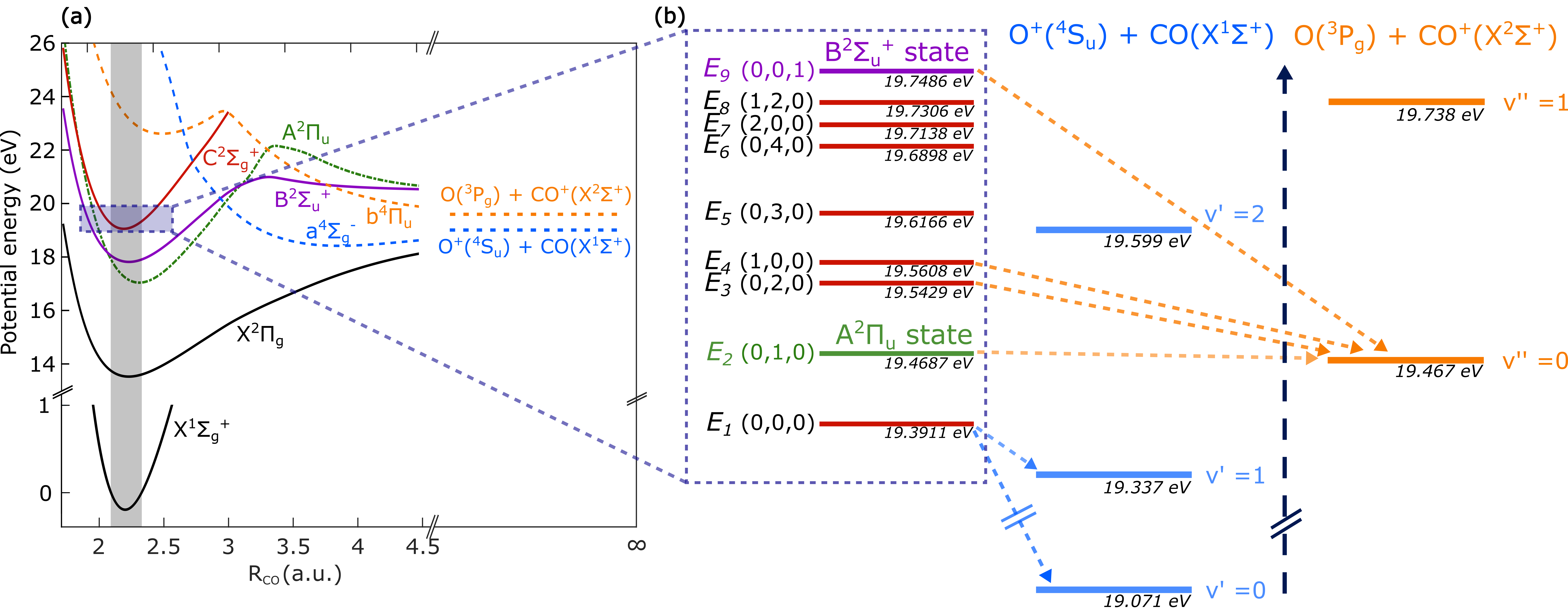}
\caption{a) Cut of the PESs for the electronic ground state of the neutral CO$_2$ molecule
and the ground and lowest excited electronic states of the CO$_2^+$ molecular ion as a function of one of the C-O distances. The other C-O distance is kept at the equilibrium bond length of the CO$_2$ ground state (2.24 a.u.) and the bond angle at 180$^\circ$ (linear configuration). 
b) Ground and excited vibrational levels of the cationic $C^2\Sigma^+_g$ state relative to the  lowest dissociation limits for the production of CO (blue) and CO$^+$ (orange). The state labels $E_n$ correlate to specific combinations of symmetric stretch ($v_s$), bending ($v_b$) and antisymmetric stretch ($v_a$) vibrational quantum numbers. The excited states $E_2$ and $E_9$ borrow intensity from the cationic states $A^2\Pi_u$ and $B^2\Sigma^+_u$, respectively.
The vibrational quantum numbers for the CO and CO$^+$ fragments are denoted as $v^{\prime}$ and $v^{\prime\prime}$, respectively.%
}
\label{Fig1}
\end{figure}

Several states of the cation can be populated by the absorption of a single photon from the neutral ground state of CO$_2$ molecules using the XUV spectrum employed in the experiment (see Fig.~\ref{Fig1} and Fig.~\ref{Fig1SM}).
The ground state $X ^2\Pi_g$ and the first excited states $A^2\Pi_u$ and $B^2\Sigma^+_u$ do not lead to dissociation (over a microsecond timescale), when considering a vertical transition from the neutral molecule~\cite{elandPredissociationTriatomicIons1972}. 
Cations in the $C^2\Sigma^+_g$ state completely predissociate through spin-orbit coupling via the dissociative states $a^4\Sigma_g^{-}$ and $b^4\Pi_u$, leading to the formation of O$^+$ and CO$^+$ ions (see Fig.~\ref{Fig1}(a), (b))~\cite{2024_Hung_CO2_PRA, praetUnimolecularReactionPaths1982,2003_Liu_CO2_JCP,mengTheoreticalStudyPredissociation2009,yang1+1TwophotonDissociation2008, lebergSynchrotronRadiationStudy1994}. 

According to energy conservation, the absorbed photon energy $\hbar\omega_{\mathrm{XUV}}$ is redistributed into the kinetic energy of the photoelectron leaving the system with (asymptotic) energy  $E$ and ionization energy $E_i$, which is given by the energy difference between the ground state of the initial neutral molecule, CO$_2$[$X^1\Sigma^+_g$(0,0,0)], and  the residual CO$_2^+$ molecular ion in a specific electronic and ro-vibrational state. 
In Fig.~\ref{Fig1} and Table~\ref{table1} we refer to the residual cationic molecular states by the energies $E_i$.
Dissociation of the residual molecular cation is energetically possible if $E_i$ exceeds the dissociation limit (relative to the energy of the initial CO$_2$[$X^1\Sigma^+_g$(0,0,0)] state), DL. This energy difference is eventually shared between the kinetic energy release (KER) of the fragments and the internal ro-vibrational energy ($E_{\mathrm{int}}$) of the neutral or charged diatomic fragment,
\begin{equation}\label{energy} 
\hbar\omega_{\mathrm{XUV}} = E_i+E = \mathrm{DL}+\mathrm{KER}+E_{\mathrm{int}} +E.
\end{equation}

For the XUV photon-energy range we consider, ionization from the electronic and vibrational ground state of the neutral molecule, $X^1\Sigma^+_g$(0,0,0), can efficiently populate the vibrational ground state in the third excited  electronic state of the molecular ion, $C^2\Sigma^+_g$(0,0,0)~\cite{bombachBranchingRatiosPartition1983,2003_Liu_CO2_JCP}. 
The reportedly large Franck-Condon factor of 0.93 (our calculation, cf., Table~\ref{table1}) and 0.82~\cite{elandFormationPredissociationCO+21977} is due to the almost matched equilibrium bond lengths of the $X^1\Sigma^+_g$ and $C^2\Sigma^+_g$ potential energy surfaces (PESs).
Experimental photoelectron-photoion coincidence studies have shown that the ground and excited vibrational levels of the electronic $C^2\Sigma^+_g$ state completely dissociate according to two pathways, 
\begin{eqnarray}
  \mathrm{CO}^{+}_2 &\rightarrow& \mathrm{O}^+(^4S_u)+\mathrm{CO}(X^1\Sigma^+;v^{\prime}=0,1,..) \label{Eq1_1}\\
  \mathrm{CO}^{+}_2  &\rightarrow& \mathrm{O}(^3P_g)+\mathrm{CO}^+(X^2\Sigma^+;v^{\prime\prime}=0,1,..)\label{Eq1_2},
\end{eqnarray}
where $v^{\prime}$ and $v^{\prime\prime}$ indicate the quantum numbers of the vibrational state of the CO and CO$^+$ fragments, respectively.
The dissociation thresholds for these pathways leading to the formation of O$^+$ and CO$^+$ fragments are DL(CO;$v^{\prime}$ =0) = 19.071~eV and DL(CO$^+$,$v^{\prime\prime}$ =0) = 19.467~eV, respectively~\cite{bombachBranchingRatiosPartition1983,elandFormationPredissociationCO+21977}. 


The vibrational ground state in CO$_2^+[C^2\Sigma^+_g$] ~\cite{elandFormationPredissociationCO+21977} lies energetically above the two lowest vibrational states ($v^\prime=0, 1$) and below the energy required for dissociation to the second excited ($v^\prime=2$) vibrational state of the CO fragment in the O$^+$+ CO dissociation pathway (see Eq.~\eqref{Eq1_1} and Fig.~\ref{Fig1}(b)). It lies below the threshold for the O + CO$^+$ dissociation. Consequently, the $C^2\Sigma^+_g$(0,0,0) state dissociates following the pathways given in Eq.~\eqref{Eq1_1}, resulting in neutral CO fragments in the ground ($v^{\prime}=0$) and first excited ($v^{\prime}=1$) vibrational states. For low-lying rotational states in CO$_2^+[C^2\Sigma^+_g$](0,0,0), these are the two only open fragmentation channels. 
For high-lying rotational excitations in the $C^2\Sigma^+_g$ (0,0,0) state to energies above the
DL(CO$^+$,0) dissociation threshold, CO$^+$ formation is possible.
In this case, the expected fragment KER is extremely small, since the ground vibrational state of the CO$^+$ ion lies 76~meV above the $C^2\Sigma^+_g$ ro-vibrational ground state $E_1$~\cite{bombachBranchingRatiosPartition1983}.

\begin{table*}[h!]    
\centering
\caption{Energies and Franck-Condon factors for selected vibrational levels ($v_s,v_b,v_a$) in the adiabatic $\CtwoSigmagp$ state of $\COtwop$ 
relative to the vibrational and electronic ground state of $\COtwo$.
The Franck-Condon coefficients are calculated as the squared magnitude of the specified vibrational state's wave function overlap with the vibrational ground state wave function in the $X^1\Sigma^+_g$ state of CO$_2$ (see supplementary material for additional information).
The quantum numbers $v_s$, $v_b$, and $v_a$ designate symmetric stretch, bending, and antisymmetric stretch modes, respectively. 
Branching ratios $w^\text{X}_{v_s,v_b,v_a,v}$ to the asymptotic dissociation channels 
$X = \text{O}^+(^4\text{S}_\text{u}) + \text{CO}(\text{X}^1\Sigma^+,v^\prime)$ 
and
$X = \text{O}(^3\text{P}_\text{g}) + \text{CO}^+(\text{X}^2\Sigma^+,v^{\prime\prime})$ 
from specific $\COtwop\left[\CtwoSigmagp(v_s,v_b,v_a)\right]$ levels are according to Ref.~\citenum{2003_Liu_CO2_JCP}. The branching ratios from the ($v_s,v_b,v_a$)= $(0,3,0)$ and $(0,0,1)$ states are interpolations and listed in brackets. 
}\vspace{0.2cm}
\begin{tabular}{c|ccc|c|c|c|c|cc}
\hline 
\hline
Label &$v_s$ & $v_b$ & $v_a$ & Energy & Franck-Condon  & \; $v^\prime$, $v^{\prime\prime}$ \; & \multicolumn{2}{c}{$w^\text{X}_{v_s,v_b,v_a,v^{\prime}-v^{\prime\prime}}$}   \\ 
      &      &       &       &[eV]   &  factor & &       \;     $\text{O}^+ +\text{CO}(v^\prime)$ \;   
                                            & \; $\text{O} + \text{CO}^+(v^{\prime\prime})$  \;  \\      
\hline
$E_1$ &0&0&0    & 19.3911     & 0.9348   &0& $0.80\pm0.05 $       & $0.0$ \\
& &  &    &             &          &1& $0.20\pm0.05 $        &   $0.0$    \\
$E_2$ &0&1&0    & 19.4687     & 0.0000   &0& $0.57\pm0.02 $        & $0.12\pm0.01$ \\
& &  &    &             &          &1& $0.31\pm0.02 $        &  $0.0$    \\
$E_3$ &0&2&0    & 19.5429     & 0.0121   &0& $0.23\pm0.02 $       & $0.65\pm 0.01 $ \\
 & &  &   &             &          &1& $0.12\pm0.02 $            & $0.0$\\
$E_4$ &1&0&0    & 19.5608      & 0.0486   &0& $0.15\pm0.02 $       & $0.78\pm 0.02 $ \\
 & &  &   &             &          &1& $0.07\pm0.01 $ &$0.0$       \\
$E_5$ &0&3&0    & 19.6166     & 0.0000   &0& ($0.15\pm0.02 $)        & ($0.77\pm0.02$) \\
 && &     &             &          &1& ($0.08\pm0.02$)         & ($0.00\pm0.02$)  \\
$E_6$ &0&4&0    & 19.6898    & 1.194$ \times 10^{-4}$  &0&          --         &        --         \\
 & &  &   &             &          &1&       -- & --                \\
%
%
$E_7$ &1&2&0    & 19.7138    & 5.783 $\times 10^{-4}$  &0&          --         &           --      \\
 & & &    &             &          &1&       -- & --                \\
$E_8$ &2&0&0    & 19.7306    & 5.930$\times 10^{-4}$    &0&        --           &        --         \\
 & &  &   &             &          &1&           -- & --            \\
$E_9$ &0&0&1    & 19.7486          &0&     0 & ($0.13\pm0.02$)               &        ($0.67\pm0.02$)         \\
 & &  &   &             &          &1&                      ($0.06\pm0.02$) & ($0.14\pm0.02$) \\
%
%
\hline
\hline
\end{tabular}
\label{table1}
\end{table*}

For ionization into the excited states $E_2$ and $E_3$, our calculated energies in Table~\ref{table1} agree with the ionization energies in Liu {\it et al.}~\cite{2003_Liu_CO2_JCP} in all displayed six digits. For $E_4$, $E_5$, and $E_9$, Liu {\it et al.} find slightly larger values of 19.5634, 19.6228, and 19.7627~eV, respectively. 
Higher-lying vibrational states in the $C^2\Sigma^+_g$ PES with energies above the 
DL(CO$^+$,0) = 19.467~eV dissociation threshold
predominantly decay along the O+CO$^+$ dissociation pathway (Eq.~\eqref{Eq1_2} and dashed orange arrows in Fig.~\ref{Fig1}(b))~\cite{bombachBranchingRatiosPartition1983, elandFormationPredissociationCO+21977}. 
More specifically, the bending excitation 
$E_2$ borrows intensity from the cationic $A^2\Pi_u$ state~\cite{rathboneModespecificPhotoelectronScattering2004a} and dissociates mainly into the O$^{+}$ dissociation channel. While the mechanism behind this intensity sharing is not fully understood to the best of our knowledge, based on vibrationally-resolved measured branching ratios for the population by photoionization of excited CO$_2^+$ states, the (0,1,0) bending excitation was interpreted as due to vibronic symmetry breaking enabled by the zero-point motion of the vibronic ground state (see Ref.~\citenum{rathboneModespecificPhotoelectronScattering2004a} and Refs. therein). 
The corresponding branching ratios for CO ($v^{\prime}$=0,1) 
and CO$^+$ ($v^{\prime\prime}$=0) production are given in Table~\ref{table1}~\cite{2003_Liu_CO2_JCP}.
Similarly, the antisymmetric stretch mode $E_9$ receives intensity from the cationic $B^2\Sigma^+_u$ state~\cite{rathboneModespecificPhotoelectronScattering2004a} and dissociates from the cationic $C^2\Sigma^+_g$ state into, preferentially, the O + CO$^+$ channel (Table~\ref{table1}). 
Within the Born-Oppenheimer and Franck-Condon approximations, mainly even quanta $v_a$ of the antisymmetric stretch vibrational mode are expected to be excited. However, vibronic mixing with the $B^2\Sigma^+_u$ state increases the population of the vibrational level $(0,0,1)$~\cite{kovacHePhotoelectronSpectra1983, 2003_Liu_CO2_JCP}.
The bending mode $E_3$ and symmetric stretch mode $E_4$ have small Franck-Condon overlap factors and are thus weakly populated in comparison with the vibrational ground state $E_1$. Located energetically above the threshold for CO$^+$ production, these vibrational modes primarily dissociate into the CO$^+$($v^{\prime\prime}$=0) channel (Table~\ref{table1}). 
The remaining vibrational states in the $C^2\Sigma^+_g$ PES listed in Table~\ref{table1} have extremely small ($E_6,E_7,E_8$) or vanishing ($E_5$) Franck-Condon factors, and
their contribution to the experimental data presented here could not be clearly identified.  Likewise, additional energetically allowed dissociation pathways that are not included in Fig.~\ref{Fig1}(b), cannot be ruled out, but their contribution to our experimental data could not be unequivocally assigned. We note that the (0,3,0) bending mode ($E_5$) and an additional mode, $C^2\Sigma^+_g$ (1,1,0), were observed in photoelectron spectra for a photon energy of 42~eV~\cite{rathboneModespecificPhotoelectronScattering2004a} that contributes very little to the XUV spectrum in the present study (cf., Fig.~\ref{Fig1SM}).

With regard to the interpretation of our experimental data below, we note that even for photon energies up to 35~eV, i.e., well above the DL(CO$^+$,0) dissociation threshold, the Franck-Condon factors of excited CO$_2^+$ [$C^2\Sigma^+_g$] states are about one order of magnitude smaller than the Franck-Condon factor for vibrational ground state $E_1$ in CO$_2^+$ [$C^2\Sigma^+_g$]~\cite{royPhotoionizationShapeResonance1984}.
   
\section{Experimental measurements}\label{Setup}
\subsection{XUV-only photoelectron spectra}
\begin{figure}[h]
    \centering
    \includegraphics[width=0.9\linewidth]{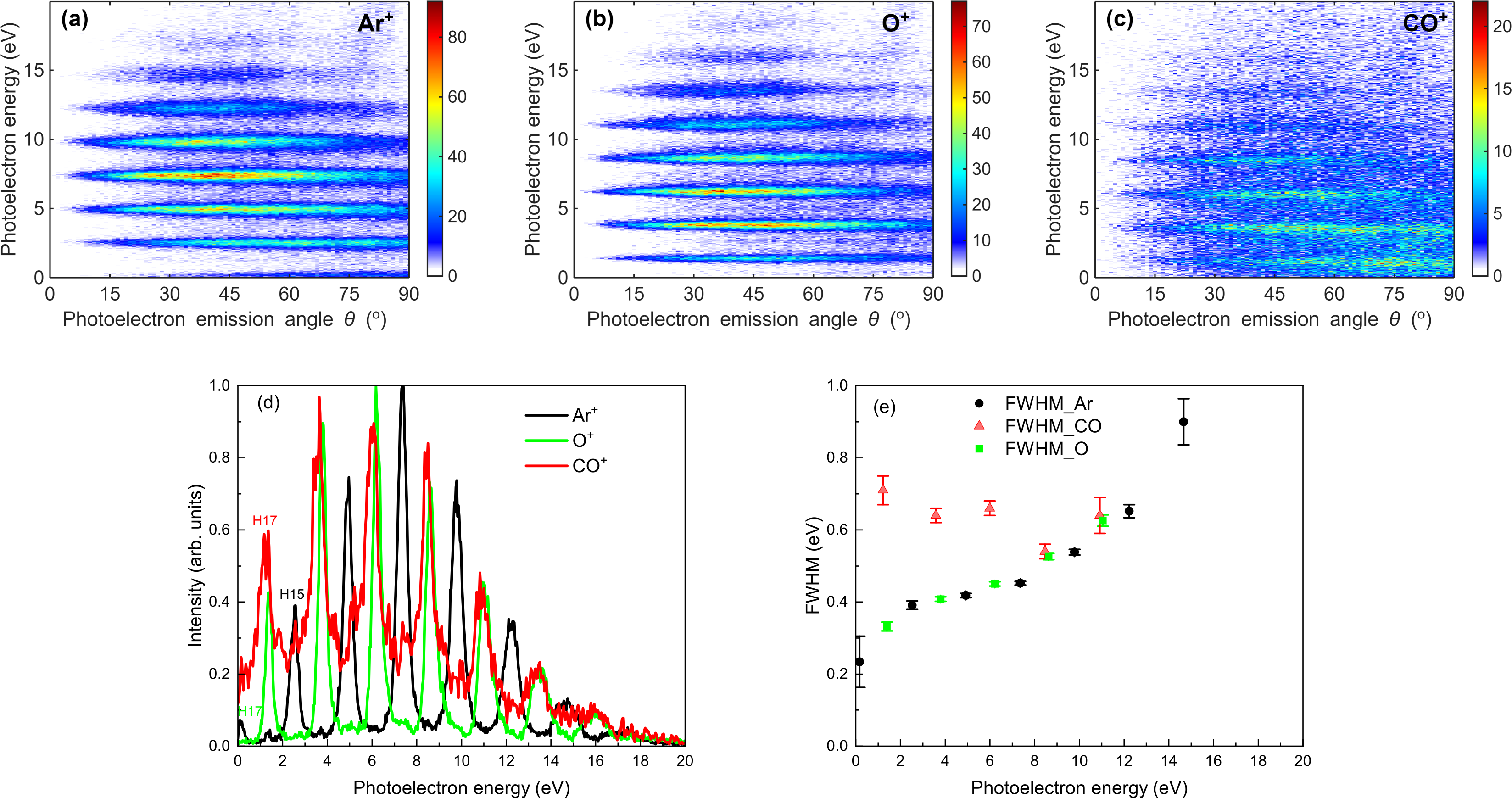}
    
\caption{Angle-resolved XUV-only spectra measured in coincidence with the ions Ar$^+$ (a), O$^+$ (b), and CO$^+$ (c). (d) Photoelectron spectra integrated over the emission angle $0^{\circ}\leq\theta\leq30^{\circ}$ for photoelectrons measured in coincidence with Ar$^+$ (black line), O$^+$  (green line), and CO$^+$ (red line). (e) The FWHM of the single harmonic peaks is derived from a multiple Gaussian fit of the spectra presented in panel (d). The error bars were derived from the fitting procedure.}
\label{Fig2}
\end{figure}

In our first experiment, CO$_2$ molecules were exposed to a comb of XUV harmonics (see~Fig. \ref{Fig1SM}) driven by a laser field centered at 1018~nm. The experimental setup has been described in detail in Refs.~\cite{ahmadiCollinearSetupDelay2020, ertelUltrastableHighrepetitionrateAttosecond2023}. The photoelectrons and photoions released by the absorption of a single XUV photon were measured using a coincidence spectrometer~\cite{moshammer4pRecoilionElectron1996, dornerColdTargetRecoil2000a}, determining their three-dimensional momenta. Energy- and angle-resolved photoelectron spectra, as well as KER-resolved photoelectron spectra, were reconstructed from the momentum distributions. 
The experimental data were acquired in a mixture of CO$_2$ molecules and Ar atoms (see time-of-flight spectra in Fig.~\ref{Fig2SM}).
Angular-resolved photoelectron distributions ($\theta$ indicates the photoemission angle with respect to the common polarization direction of the XUV and IR pulses) were measured in coincidence with the fragments Ar$^+$, O$^+$, and CO$^+$ using only the XUV attosecond pulse trains and are presented in Figs.~\ref{Fig2}(a), \ref{Fig2}(b), and \ref{Fig2}(c), respectively.

Figure~\ref{Fig2}d shows the normalized photoelectron spectra integrated within the range~$0^{\circ}\leq\theta\leq30^{\circ}$. The photoelectron spectra were fitted with a set of Gaussians to determine the central energy and the width of each photoelectron peak (see supplementary material and Table~\ref{TableS1}).
A linear fit of the photoelectron peak energies measured in coincidence with the three different ions yields ionization energies of $I_p=15.76\pm0.06$~eV, $E=19.32\pm0.06$~eV, and $E=19.55\pm0.23$~eV for Ar$^+$, O$^+$, and CO$^+$, respectively. 
The ionization energy for the photoelectrons measured in coincidence with the O$^+$ ions is very close to the value of 19.3911~eV of the ground vibrational state $E_1\equiv(0,0,0)$ in the $C^2\Sigma^+_g$ state, supporting the conclusion that this state predissociates, leading to the formation of O$^+$ ions. The ionization energy for the photoelectrons measured in coincidence with the CO$^+$ ions is intermediate between the energies of the two excited vibrational states $E_3\equiv (0,2,0)$ and $E_4\equiv(1,0,0)$ (see Table~\ref{table1}). This value is consistent with the observation that both states lead to predissociation, resulting in the formation of the CO$^+$ ions.

The analysis of the full-width at half maximum (FWHM) of the photoelectron peaks reveals that the first three peaks, measured in coincidence with Ar$^+$ and O$^+$, have comparable widths. This is due to the intrinsic width of the harmonics, estimated at about 200~meV and the spectral resolution of the spectrometer in this energy range. These comparable peak widths support the conclusion that photoionization leading to the formation of O$^+$ is due solely to (or at least is dominated by) photoionization to the ground vibrational state of the $C^2\Sigma^+_g$ state. Conversely, the widths of the first photoelectron peaks measured in coincidence with the CO$^+$ ions are significantly larger. We attribute this difference to the fact that, while dissociation of the vibrational ground state of the $C^2\Sigma^+_g$ PES dominates the O$^+$ ion yield, the generation of the CO$^+$ final fragments results from several excited states of the same PES, as well as states presenting vibronic coupling with other PESs (see Fig.~\ref{Fig1}(b))~\cite{bombachBranchingRatiosPartition1983}. 

\begin{figure}[h!]
    \centering
    \includegraphics[width=0.7\textwidth]{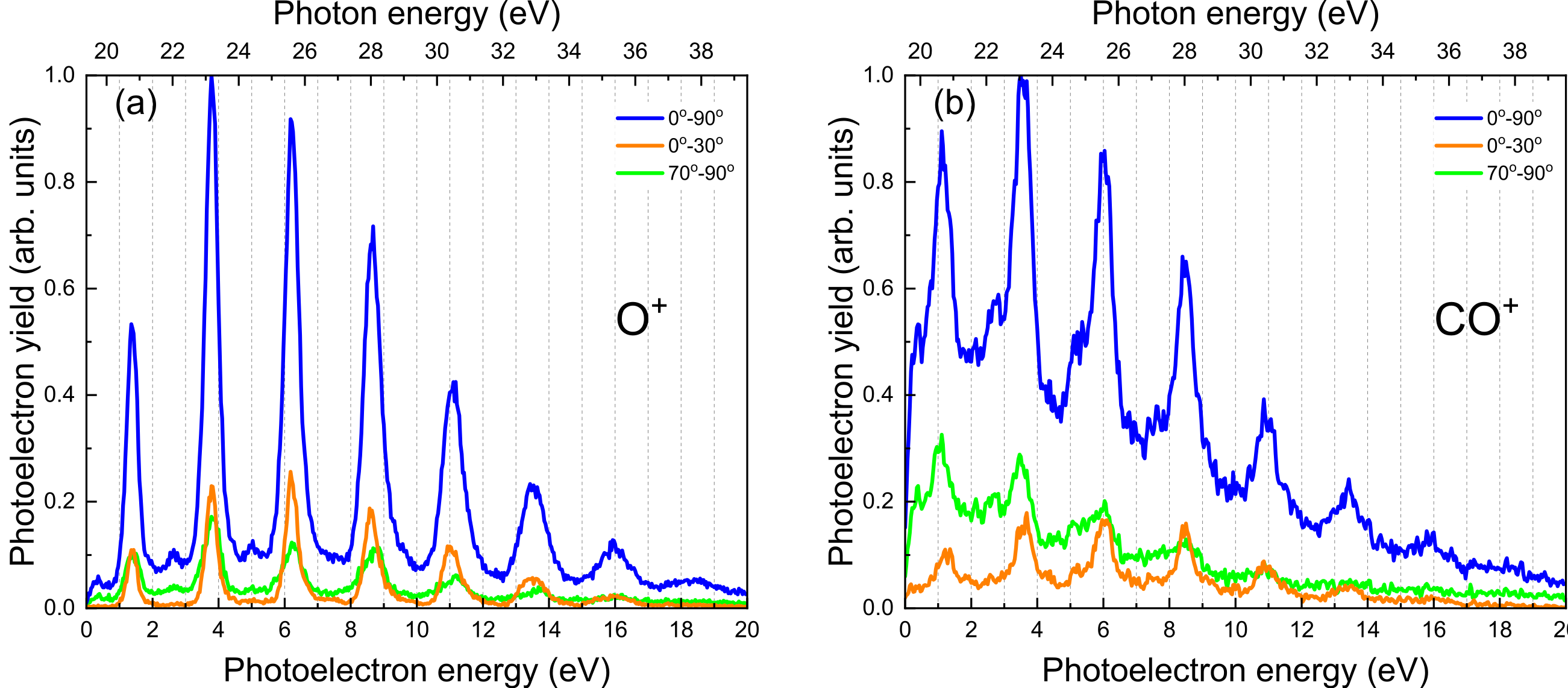}
\caption{a) XUV-only spectra measured in coincidence with O$^+$ (a) and CO$^+$ ions (b) and integrated over the angle intervals $[0^{\circ}-90^{\circ}]$ (blue line), $[0^{\circ}-30^{\circ}]$ (orange line; parallel case), and $[70^{\circ}-90^{\circ}]$ (green line; perpendicular case).} 
\label{Fig3}
\end{figure}

The angle-resolved photoelectron spectra measured in coincidence with the O$^+$ and CO$^+$ ions are presented in Figs.~\ref{Fig3}(a) and~\ref{Fig3}(b), respectively. The shapes of the photoelectron spectra for emission parallel and perpendicular with respect to the XUV polarization direction do not differ significantly in the case of measurements in coincidence with the O$^+$ ion. The asymmetry parameters $\beta$ extracted from the photoelectron angular distribution are in very good agreement with those of the $C^2\Sigma^+_g$ state reported in the literature~\cite{siggelShapeResonanceEnhanced1993}, as shown in Fig.~\ref{Fig3SM}(a). 
Conversely, the shape changes significantly for parallel and perpendicular emission of the photoelectrons measured in coincidence with the CO$^+$ ion (Fig.~\ref{Fig3}(b)). In particular, for the lowest harmonic orders, we observe a significant increase of the signal strength for the perpendicular direction. Furthermore, the photoelectron signal drops significantly after approximately 6~eV for the perpendicular direction, while the drop is much less pronounced for the parallel direction.

These observations suggest that the photoelectrons emitted perpendicular to the XUV polarization direction originate from the state with energy $E_9=19.7486$~eV, which is associated with a strong vibronic interaction with the $B^2\Sigma^+_u$ PES~\cite{2003_Liu_CO2_JCP}. Indeed, for this state, the photoelectron emission exhibits a negative $\beta$ (Fig.~\ref{Fig3SM}(a))~\cite{grimmAngleresolvedPhotoelectronSpectroscopy1981}, and the total cross section is characterized by a significant drop in this energy range (see Fig.~\ref{Fig3SM}b)~\cite{siggelShapeResonanceEnhanced1993}.
In the supplementary material we provide further experimental evidences that support this conclusion.
 
The experimentally measured ratio of the CO$^+$ and O$^+$ ion yield is approximately 0.31. This is  significantly larger than the calculated value of approximately $0.048$ obtained from the Franck-Condon factors for vertical photoionization from the ground state of the neutral molecule and the observation that the ground state dissociates along the pathways described by Eqs.~\eqref{Eq1_1} and ~\eqref{Eq1_2} according to the branching ratios reported in Table~\ref{table1}. Several factors may contribute to this discrepancy. First, the contribution of vibrational states with strong vibronic coupling to either the $A^2\Pi^+_u$~\cite{2003_Liu_CO2_JCP, rathboneResonantlyAmplifiedVibronic2001} or $B^2\Sigma^+_u$~\cite{2003_Liu_CO2_JCP, rathboneModespecificPhotoelectronScattering2004a} states is not considered in the ratio of the Franck-Condon factor. Furthermore, as observed in Ref.~\citenum{bombachBranchingRatiosPartition1983}, the ground vibrational state of the $C^2\Sigma^+_g$ curve can also contribute to the predissociation in the CO$^+$ channel.

\subsection{XUV-only KER-resolved photoelectron spectra}
\begin{figure}[h!]
    \centering
    \includegraphics[width=0.9\linewidth]{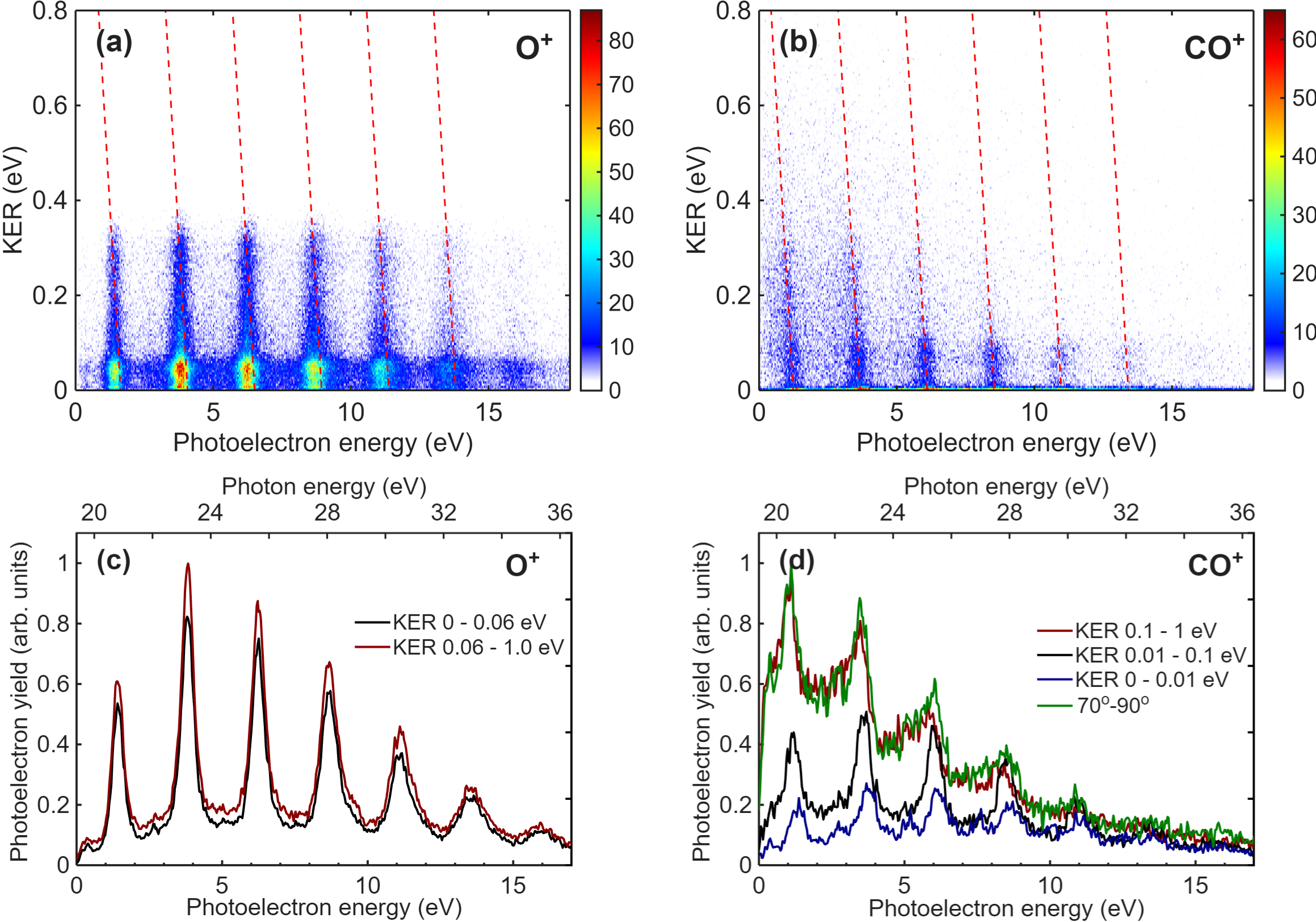}
\caption{KER-resolved photoelectron spectra measured in the XUV-only case in coincidence with O$^+$ (a) and CO$^+$ (b). The dashed red lines indicate  energy conservation as expressed by Eq.~\eqref{energy}. Photoelectron spectra obtained by integrating over the KER in coincidence with O$^+$ (panel c; black line KER between 0 and 0.06~eV; red line KER between 0.06 and 1~eV) and CO$^+$ (panel d; blue line KER between 0 and 0.01 eV; black line KER between 0.01 and 0.1~eV; red line KER between 0.1 and 1~eV) ions. The normalized photoelectron distribution measured in coincidence with CO$^+$ ions for perpendicular electron emission ($70^{^\circ}\leq\theta\leq90^{\circ}$) is also plotted in panel d.}
\label{Fig4}
\end{figure}

Photoelectron spectra for different KERs for the O$^+$ and CO$^+$ photodissociation channels for the XUV-only case are reported in Fig.~\ref{Fig4}.  The red lines represent the graphical representation of Eq.~(\ref{energy}) of the energy conservation due to the sharing of the photon energies of the XUV harmonics ($\hbar\omega_{\mathrm{XUV}}$) between the photoelectron energy ($E$) and the KER. The internal energy was set to zero ($E_{\mathrm{int}}=0$) and the corresponding dissociation limits (DLs) presented in Fig.~\ref{Fig1} were used.

For the O$^+$ channel, the correlated photoelectron-KER energy map in Fig.~\ref{Fig4}(a) is characterized by lines parallel to the KER axis, which correspond to different harmonic orders. The slope of these features deviate from that of the red lines, indicating that a single vibrational energy level (ground level of the $C^2\Sigma^+_g$ state) leads to fragmentation, resulting in the formation of O$^+$ ions. This conclusion further confirms the results obtained from the analysis of the FWHM of the photoelectron peaks presented in Fig.~\ref{Fig2}(e).

The KER distributions (see Fig.~\ref{Fig4}(a)) exhibit a main contribution extending up to approximately 0.06~eV and a secondary contribution extending up to approximately 0.3~eV.
The photoelectron spectra obtained by integrating over these two KER regions present very similar distributions as shown in Fig.~\ref{Fig4}(c).
These experimental observations are consistent with the finding that predissociation populates the ground ($v^\prime=0$) and first excited ($v^\prime=1$) states of the CO fragment, which have threshold energies of 19.071~eV and 19.337~eV, respectively (Fig.~\ref{Fig1})~\cite{bombachBranchingRatiosPartition1983}. Considering that the energy of the ground vibrational state of the $C^2\Sigma^+_g$ potential energy surface is 19.3911~eV, the maximum energy available as KER in the dissociation process (neglecting the rotational energy of the diatomic fragment) is 0.320~eV and 0.054~eV, respectively, in good agreement with the experimental observations. Similar distributions have been reported previously in Ref.~\citenum{chengKineticEnergyRelease2018}. Furthermore, in Ref.~\citenum{bombachBranchingRatiosPartition1983} a bimodal distribution of the KER in O$^+$ ions was observed.
Integrating over the two KER regions yields populations of the $v'=0$ and $v'=1$ vibrational levels of 0.55 and 0.45, respectively, in good agreement with the previously reported branching ratios in Ref.~\citenum{bombachBranchingRatiosPartition1983} ($0.56\pm0.1$ and $0.38\pm0.1$, respectively). Note, however, that ratios of 0.23 and 0.77, and 0.3 and 0.7 for the $v'=0$ and $v'=1$ vibrational levels have been reported in Refs.~\citenum{elandPredissociationTriatomicIons1972} and \citenum{freyPhotoionizationResonanceSpectra1977}, respectively. 

As shown in Fig.~\ref{Fig4}(b), the KER as a function of the photoelectron energy measured in combination with the CO$^+$ channel presents a more complex KER - photoelectron energy correlation. In general, the signal qualitatively follows the tilted lines expressing energy conservation. Furthermore, the photoelectron spectra features strongly depend on the KER, as indicated by the photoelectron energy distributions for the low and high KER regions presented in Fig.~\ref{Fig4}(d). These observations suggest that several vibrational levels of the cationic $C^2\Sigma^+_g$ state get populated during the photoionization, leading to the formation of CO$^+$ ions. This conclusion is consistent with the observation that the FWHM of the photoelectron peaks measured in coincidence with CO$^+$ ions is larger than the widths measured in coincidence with the Ar$^+$ reference atomic ion  [Table~\ref{TableS1} and Fig.~\ref{Fig2}(d),(e)]. 

To identify the different contributions to the KER maps of CO$^+$, we refer to the notation introduced in Fig.~\ref{Fig1} and Table~\ref{table1}. The first level encountered, starting from the lowest energy is the $E_2\equiv(0,1,0)$ vibrational state, owes its intensity from vibronic mixing with the $A^2\Pi_u$ state~\cite{kovacHePhotoelectronSpectra1983,reineckHighresolutionUVPhotoelectron1983,2003_Liu_CO2_JCP}. The energy of this level is close to the dissociation threshold for the formation of CO$^+$ ions~\cite{elandFormationPredissociationCO+21977} and may significantly contribute to the KER region between 0~eV and 0.01~eV.

The next states encountered as the energy increases are the vibrational states $E_3$ and $E_4$ of the $C^2\Sigma^+_g$ PES (see Fig.~\ref{Fig1}). These excited vibrational states present the largest Franck-Condon factor for excitation of the $C^2\Sigma^+_g$ state from the ground state of the neutral molecule, after the ground vibrational state of the $C^2\Sigma^+_g$ state. These states decay preferentially to the vibrational ground state of the CO$^+$ fragment, releasing a KER of up to approximately $0.1$~eV. For this reason, we conclude that the excited vibrational states of the  $C^2\Sigma^+_g$ PES mostly contribute to the KER up to 0.1~eV.
As other experimental investigations~\cite{2003_Liu_CO2_JCP}, we could not clearly identify the contribution of the next higher-lying states of the $C^2\Sigma^+_g$ curve with energies $E_5$, $E_6$, $E_7$, and $E_8$ due to their small Franck-Condon factors (see Table~\ref{table1}).

The last relevant state is the $(0,0,1)$ vibrational level, which has an energy of $E_9=19.7486$~eV (Fig.~\ref{Fig1}). This state owes its intensity from vibronic mixing with the $B^2\Sigma^+_u$. The $(0,0,1)$ vibrational level can dissociate in two ways: to the ground state of the CO$^+$ ion, releasing a KER of about 300~meV; or to the first excited state, releasing a small KER of about 20~meV. For this reason, this state should dominate the region with a KER greater than about 0.1 eV. This is confirmed by comparing the photoelectron spectra observed in coincidence with high KER ($>0.1$~eV) and those measured for all KERs but in the perpendicular direction ($70^{^\circ}\leq\theta\leq90^{\circ}$), as shown in Fig.~\ref{Fig4}(d) (green curve) (see also Fig.~\ref{Fig3}(b)). The good match between the two spectra further supports the conclusion that the energy level $E_9$ mostly contributes to generating high-KER fragments and emitting low-energy photoelectrons (with energies between 0 and 5~eV) in the perpendicular direction. The weak signal observed in the correlation maps for KER larger than 300~meV might be due to the excitation of higher-lying vibrational states of the $C^2\Sigma^+_g$ PES characterized by a significant coupling with the $B^2\Sigma^+_u$ state. 

\clearpage
\newpage

\subsection{XUV-IR photoelectron spectra}
\begin{figure}[h!]
    \centering
    \includegraphics[width=0.9\linewidth]{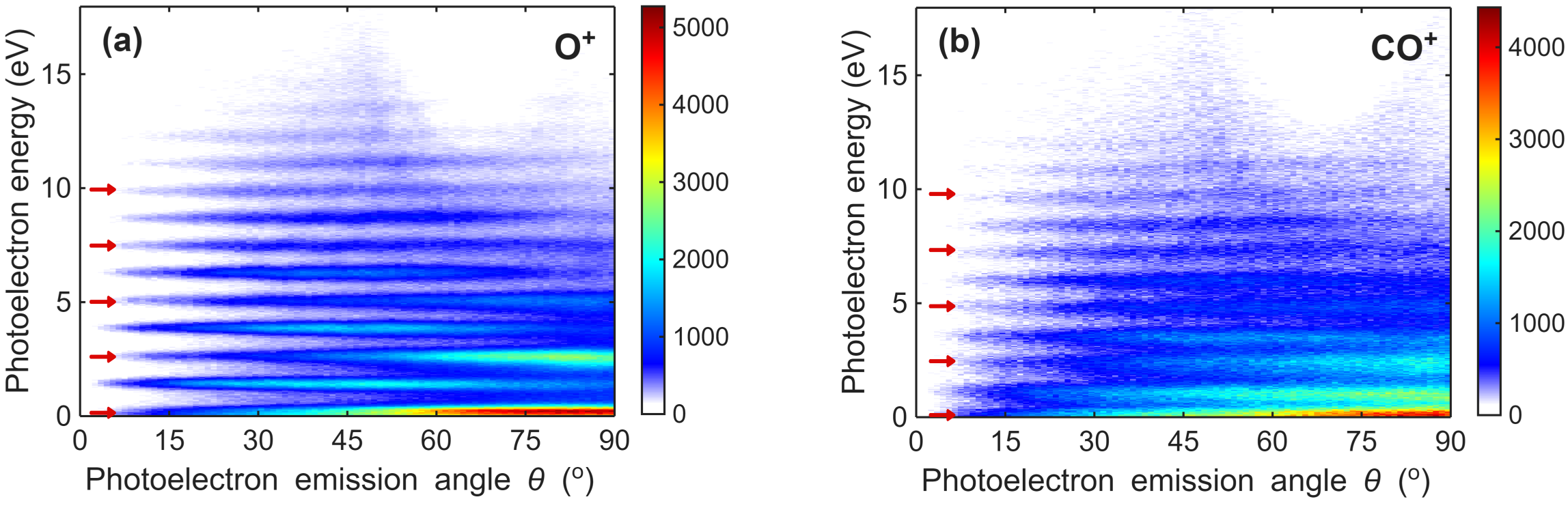}
\caption{
Delay-averaged photoelectron angular distributions measured in coincidence with O$^+$ (a) and CO$^+$ (b) in the XUV-IR case. The red arrows on the left side highlight the photoelectron peaks associated with the sideband signal.
}\label{Fig5}
\end{figure}

\begin{figure}[h!]
   \centering
    \includegraphics[width=0.9\linewidth]{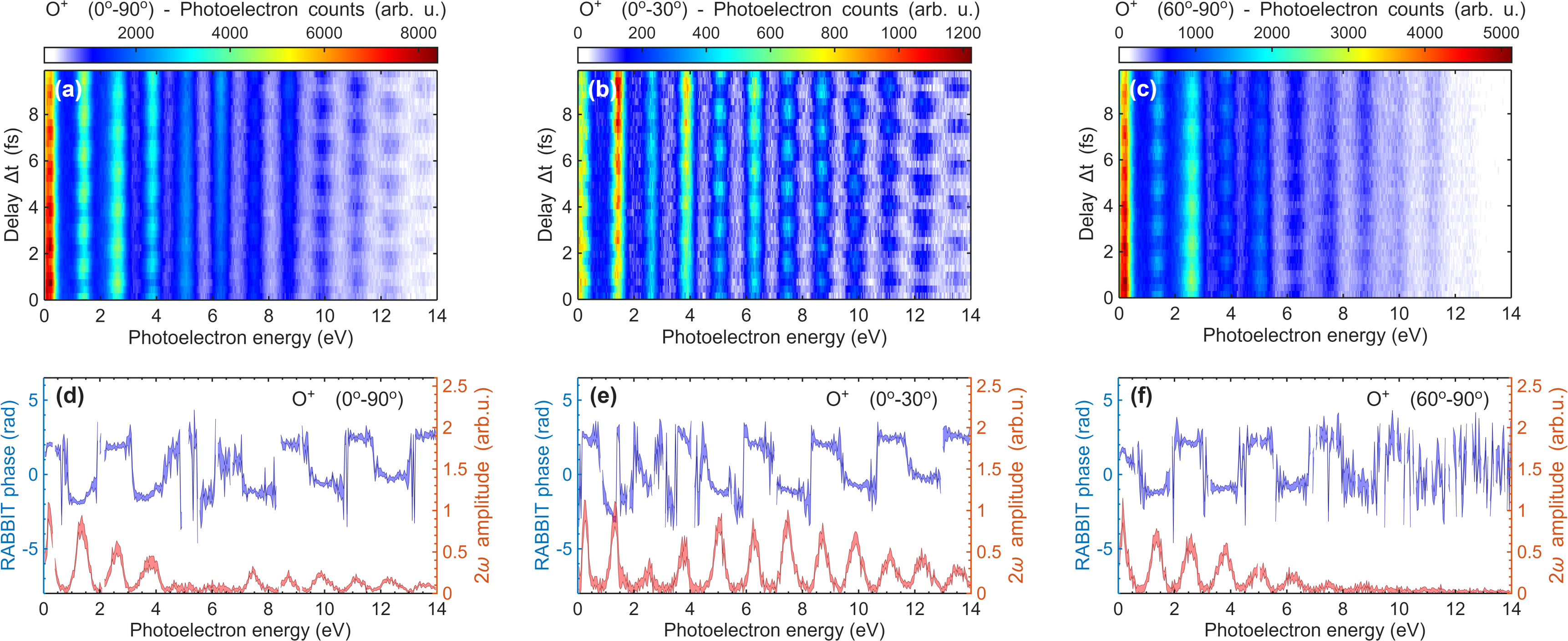}
\caption{RABBIT trace measured in coincidence with the O$^+$ ions, obtained by integrating the photoelectron angular distribution in the intervals $[0^{\circ}-90^{\circ}]$ (a), $[0^{\circ}-30^{\circ}]$ (b), and $[60^{\circ}-90^{\circ}]$ (c). (d-f) Amplitude $A_2$ (red lines) and phase $\Phi$ (blue lines) from Eq.~\eqref{Eq_sideband} as a function of the photoelectron energies derived by an energy resolved fit of the RABBIT traces presented in panels a-c, respectively. The shaded areas represent the error bars estimated a single standard deviation obtained from the fit. Energy bins with phase uncertainties greater than $\pi/2$ are omitted from the plots.}
\label{Fig6}
\end{figure}

RABBIT measurements were acquired by changing the relative delay between the XUV and IR pulses in steps of 300~as in an interval of about 12~fs. The time-integrated, angle-resolved photoelectron spectra associated with the two ionic fragments are shown in Figs.~\ref{Fig5}(a) and~\ref{Fig5}(b) for O$^+$ and CO$^+$, respectively. For photoelectron energies below 7~eV, the sidebands, centered at energies of approximately 0.2~eV, 2.6~eV and 5.0~eV, are emitted preferentially perpendicular to the laser polarization direction ($\theta=90^{\circ}$). For increasing photoelectron energies, the angular distributions of the sidebands approach those of the corresponding harmonic peaks.
The angle-resolved RABBIT traces measured in coincidence with the ion O$^+$ are shown in Fig.~\ref{Fig6}, for three different integration angles. The oscillations of the sideband signal can be modeled according to the relation
\begin{equation}\label{Eq_sideband}
  \mathrm{SB}(\theta,E,\Delta t)=A_0(\theta,E)+A_2(\theta,E)\cos\left[2\omega_{\mathrm{IR}}\Delta t-\Phi(\theta,E)\right],
\end{equation}
where $A_0$, $A_2$, and $\Phi$ indicate the constant offset, the amplitude of the oscillation, and the photoionization phase, respectively. In general, these three quantities depend on the photoemission angle $\theta$ and on the energy $E$ of the emitted photoelectron.
By applying a fitting procedure to each angle $\theta$, we can extract the corresponding angle-resolved parameters $A_0$, $A_2$, and $\Phi$ as a function of the photoelectron energy $E$ and integrate their values over the desired interval. 

The evolution of these quantities as a function of the photoelectron energy $E$ for the three angle intervals presented in Figs.~\ref{Fig6}(a), \ref{Fig6}(b), and \ref{Fig6}(c) are shown in Fig.~\ref{Fig6}(d), \ref{Fig6}(e), and \ref{Fig6}(f), respectively. As can be seen in Fig.~\ref{Fig6}(d), the modulation depth presents a minimum around 4-7~eV for the signal integrated between $0^{\circ}$ and $90^{\circ}$ degrees. Additionally, the phase of the sideband photoelectron peak oscillations exhibit a $\pi$ phase shift when crossing the energy region of minimum modulation.
Additional information can be  gained by analyzing the angle-resolved sideband oscillations, as shown Figs.~\ref{Fig6}(b) and \ref{Fig6}(c). For integration angles parallel to the laser polarization ($0^{\circ}\leq\theta\leq30^{\circ}$; Fig.~\ref{Fig6}(b)), the spectrum extends to higher kinetic energies and the modulation amplitude $A_2$ presents a minimum around 2-3~eV, thus being shifted with respect to the total integration angle. Conversely, for integration angles around the perpendicular direction ($60^{\circ}\leq\theta\leq90^{\circ}$; Fig.~\ref{Fig6}(c)), the spectrum is concentrated in the low kinetic energy region, and the modulation amplitude monotonically decreases with increasing photon energy. 

The RABBIT traces and the energy-resolved fits of the photoelectron spectra measured in coincidence with the CO$^+$ are presented in the supplementary material (see Fig.~\ref{Fig4SM}).

\clearpage
\newpage

\section{Theoretical Model}
The combined effect of the XUV and IR fields was theoretically simulated at the level of second order of perturbation theory, 
by calculating the two-photon amplitudes 
\begin{equation}
    T_{f,\text{XUV}\pm\text{IR}}^{(2)}
    = \langle \Psi_{f\bm{k}}^{(-)} | V_\text{IR} G^{(+)} V_\text{XUV} | \Psi_i \rangle
\end{equation}
in the framework of the multiphoton above-threshold R-matrix method~\cite{bendaMultiphotonThresholdIonization2021}, as implemented in the UKRmol+ program suite~\cite{masinUKRmolSuiteModelling2020}. Here \(\Psi_i\) is the initial bound state of the neutral molecule, and \(\Psi_{f\bm{k}}^{(-)}\) is the final continuum state with the photoionization boundary condition, assuming the residual cation is in the state \(f\) and the photoelectron has an asymptotic momentum \(\bm{k}\). The perturbations \(V_\alpha = \bm{E}_\alpha\cdot\bm{D}\) represent interactions of the many-electron system with the respective field in the dipole approximation, and \(G^{(+)}\) is the retarded Green's function. 
The total sideband signal is given by:
\begin{eqnarray}\label{Eqsb}
    \mathrm{SB}(\theta,E,\Delta t)&=&\langle |T_{\text{XUV}+\text{IR}}^{(2)}+T_{\text{XUV}-\text{IR}}^{(2)}|^2 \rangle=\langle |\underbrace{T_{\text{XUV}+\text{IR}}^{(2)\text{cc}}}_{\text{Path 1}}+\underbrace{T_{\text{XUV}-\text{IR}}^{(2)}}_{\text{Path 2}}+\underbrace{T_{\text{XUV}+\text{IR}}^{(2)\text{ii}}}_{\text{Path 3}}|^2 \rangle \,.
\end{eqnarray}
The brackets denote averaging over molecular orientations, which was performed analytically using angular momentum algebra. The superscripts “ii” and “cc” denote two-photon amplitudes in which, following the initial absorption of an XUV photon, the IR photon is absorbed by the residual ion or by the photoelectron, respectively. 
The paths 1-3 corresponding to the different terms of the second-order perturbation theory are presented in Fig.~\ref{Fig7} and will be introduced in the next section.
In general, the different terms depend on the photoelectron energy $E$ and the photoemission angle $\theta$.

The two-photon amplitudes were used to predict the parameters \(A_0\), \(A_2\) and \(\Phi\) of the RABBIT signal as
\begin{eqnarray}
    A_0 &=& \langle |T_{\text{XUV}+\text{IR}}^{(2)}|^2 \rangle
         + \langle |T_{\text{XUV}-\text{IR}}^{(2)}|^2 \rangle \,,\label{A0} \\
    A_2 &=& 2 |\langle T_{\text{XUV}+\text{IR}}^{(2)*} T_{\text{XUV}-\text{IR}}^{(2)} \rangle| \,\,\mathrm{and}\label{A2} \\
    \Phi &=& \arg\langle T_{\text{XUV}+\text{IR}}^{(2)cc*} T_{\text{XUV}-\text{IR}}^{(2)}
    + T_{\text{XUV}+\text{IR}}^{(2)ii*} T_{\text{XUV}-\text{IR}}^{(2)} \rangle
    = \arg ( Q_{cc} + Q_{ii} ) \,. \label{eq:QccQii}
\end{eqnarray}
The contributions to the RABBIT phase can be conveniently grouped in two terms as
\begin{align}
    Q_{cc} &= \langle T_{\text{XUV}+\text{IR}}^{(2)cc*} T_{\text{XUV}-\text{IR}}^{(2)} \rangle \,\, \mathrm{and} \label{eq:Qcc} \\
    Q_{ii} &= \langle T_{\text{XUV}+\text{IR}}^{(2)ii*} T_{\text{XUV}-\text{IR}}^{(2)} \rangle \,, \label{eq:Qii}
\end{align}
which clearly separate two contributing signals to be discussed below.

The molecular model was based on optimized molecular orbitals obtained in Molpro~\cite{wernerMolpro2020} using the cc-pVTZ basis set and complete active space self-consistent field method, with the 1--2\(\sigma_g\) and \(1\sigma_u\) core orbitals frozen and orbitals 3--5\(\sigma_g\), 2--3\(\sigma_u\), 1\(\pi_g\), and 1--2\(\pi_u\) active. Of these orbitals, antisymmetrized spin-adapted electronic configurations of the residual ion were built using a complete active space (CAS) approach. Three hundred states obtained by diagonalization of the residual ion Hamiltonian in this configuration basis, coupled to continuum orbitals, were then used in the close-coupling scattering model, together with square-integrable configurations of the neutral system based on the same CAS. The continuum was represented using a central B-spline basis, including a partial wave expansion up to \(\ell = 7\). The molecule was modeled in its neutral equilibrium geometry, and only electronic vertical transitions were considered. 

\clearpage
\newpage
\section{Discussion}
\begin{figure*}[hb] 
\centering 
\includegraphics[width=0.8\textwidth]{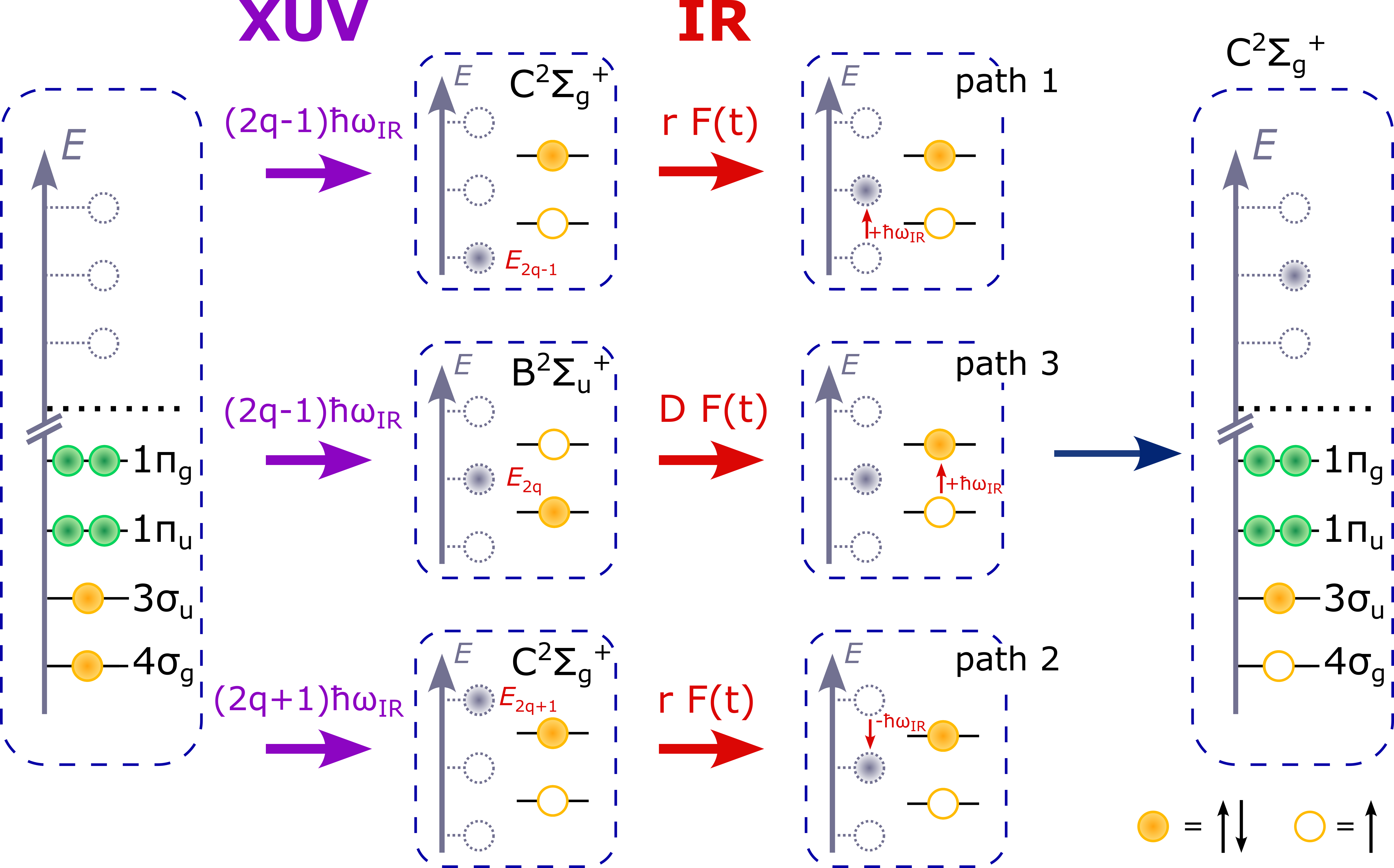}
\caption{Schematic representation of the three-path RABBIT mechanism with the illustration of the path 1 (upper one), path 2 (bottom one), and path 3 (central one).}
\label{Fig7}
\end{figure*}

The energy difference between the $B^2\Sigma^+_u$ and $C^2\Sigma^+_g$ states is close to one IR photon energy: $I_C-I_B\approx\hbar\omega_{\mathrm{IR}}$, where $I_C$ and $I_B$ indicate the ionization energy of the $C^2\Sigma^+_g$ and $B^2\Sigma^+_u$ state, respectively~\cite{pranjalResonantPhotoionizationCO22024}. This condition introduces the possibility of the IR field acting on the cations through a dipole coupling between the two cationic states~\cite{bendaAnalysisRABITTTime2022, bendaDipolelaserCouplingDelay2024}. We will refer to this coupling as $B-C$ coupling hereafter . This mechanism introduces an additional path that contributes to the sideband signal for the emission of photoelectrons leaving the residual cation in the $C^2\Sigma^+_g$ state, as shown in Fig.~\ref{Fig7}. Traditional RABBIT paths (indicated as path 1 and  path 2) require the absorption of an XUV photon with an energy of either $\hbar\omega_{\mathrm{XUV}}=(2q-1)\omega_{\mathrm{IR}}$ or $\hbar\omega_{\mathrm{XUV}}=(2q+1)\omega_{\mathrm{IR}}$, followed by the absorption or emission of an additional IR photon by the outgoing photoelectron wave packet, respectively. These paths are indicated in Eq.~\eqref{Eqsb} as path~2 and path~1, respectively. The final state of the system is a photoelectron with energy $E_{2q}=2q\omega_{\mathrm{IR}}-I_C$ and a cation in the $C^2\Sigma^+_g$ state, which will dissociate along the pathway described in Eqs.~\eqref{Eq1_1} and~\eqref{Eq1_2}.
Due to the $B-C$ coupling an additional path (path 3) can contribute to the final state. Along this path the neutral molecule first absorbs an XUV photon with energy $\hbar\omega_{\mathrm{XUV}}=(2q-1)\omega_{\mathrm{IR}}$ releasing a photoelectron with energy $E_{2q}=(2q-1)\omega_{\mathrm{IR}}-I_B\approx2q\omega_{\mathrm{IR}}-I_C$ and populating the cationic state $B^2\Sigma^+_u$. The cation can then absorb an additional IR photon, which determines a transition between the $B^2\Sigma^+_u$ and $C^2\Sigma^+_g$ states. Since the final state is the same as that of the previous paths, an additional coherent path contributes to the sideband population.

\begin{figure*}[h]   
\centering 
   \includegraphics[width=0.9\linewidth]{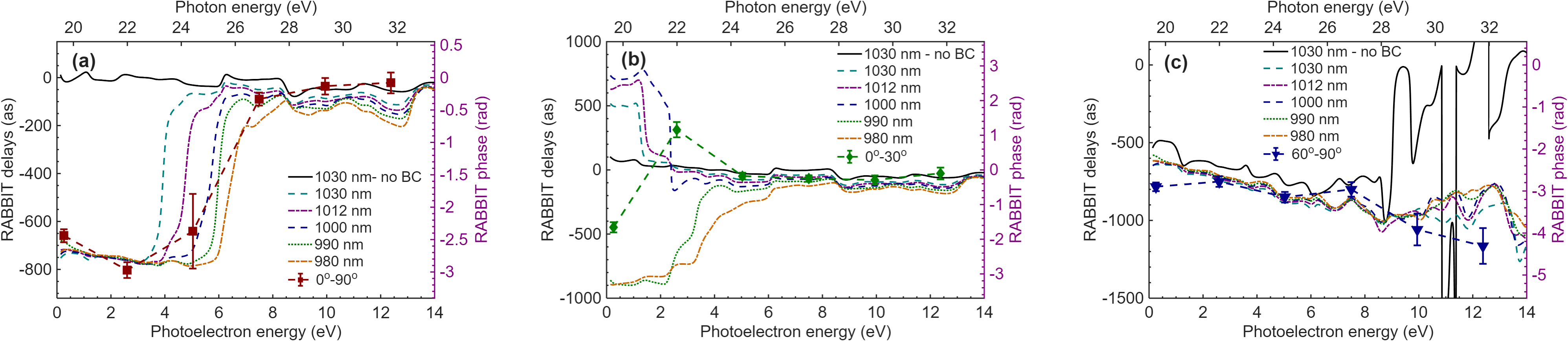}
\caption{
Simulated (lines of different styles, as indicated in the legend) and measured (symbols with error bars connected by dashed lines) photoionization delays obtained integrating the photoelectron angular distributions measured in coincidence with O$^+$ ions over the intervals $0^{\circ}-90^{\circ}$ (red experimental points) (a), $0^{\circ}-30^{\circ}$ (green points) (b), and $60^{\circ}-90^{\circ}$ (blue points) (c). The delays were corrected for the photoionization time delays measured in coincidence with argon to eliminate the contribution of the attochirp.}
\label{Fig8}
\end{figure*}

 \begin{figure*} [h] 
\centering 
\includegraphics[width=0.7\textwidth]{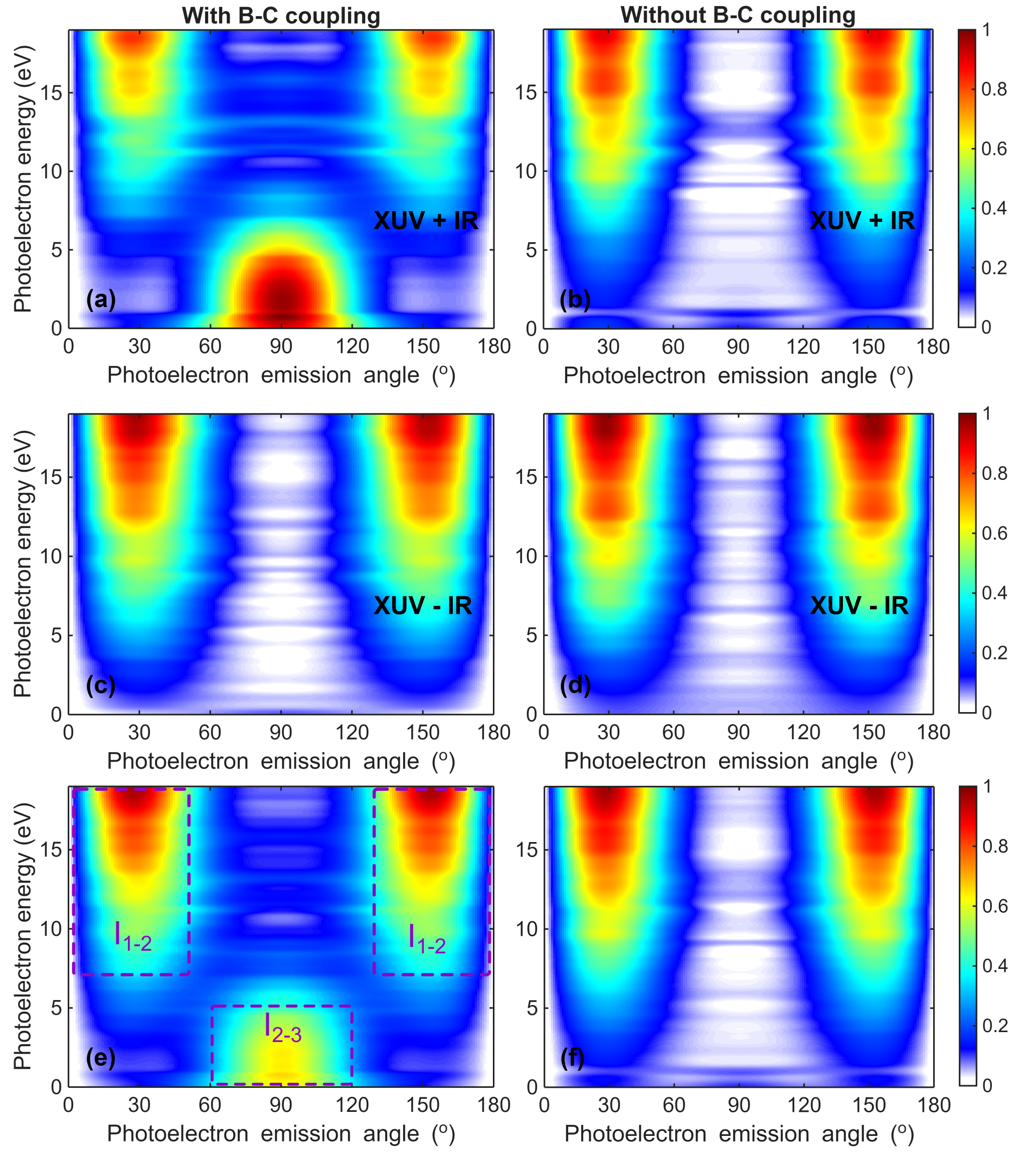}
\caption{Contribution of the paths characterized by the absorption of an XUV photon and an additional IR photon (a,b) or the emission of an IR photon (c,d) with (a,c) and without (b,d) the contribution of the $B-C$ coupling. Photoelectron spectra averaged over the relative delay between XUV and IR fields corresponding to a full oscillation of the sideband signal including both absorption and emission of an IR photon with (e) and without (f) $B-C$ coupling. 
The lowest two panels represent the amplitude $A_0(\theta,E)$ introduced in Eqs.~\eqref{Eq_sideband} and \eqref{A0}.}
\label{Fig9}
\end{figure*}

The effect of the ionic coupling on the attosecond time delays measured in coincidence with the $B^2\Sigma^+_u$ state was analyzed in Ref.~\citenum{makosEntanglementPhotoionisationReveals2025}. The presence of the additional path introduces an additional contribution to the delay, leading to an angle-dependent shift in the attosecond time delays.

The effect of ionic coupling on the attosecond time delay in photoionization associated with the $C^2\Sigma^+_g$ state is presented in Fig.~\ref{Fig8} for different wavelengths of the IR field.
In particular, for the photoelectron angular distribution integrated over the entire angle (Fig.~\ref{Fig8}(a)), the evolution of the delay is characterized by a significant jump in the photoelectron energy range 4-7~eV photon energy (XUV photon energy range of approximately 24-27~eV, considering the energy of $E_1$=19.3911 eV of the ground vibrational state of the $C^2\Sigma^+_g$) and depends on the wavelength of the driving field. The best matching is achieved for a wavelength of approximately $\lambda=1000$~nm. The overall variation of the RABBITT delay is approximately 780~as. Remarkably, this variation is a direct manifestation of the $B-C$ coupling, as demonstrated by the solid black line obtained by switching off the $B-C$ coupling, which does not exhibit any jump in the attosecond time delays. 
The presence of a strong variation in the attosecond time delays is also visible in the photoelectron spectra integrated between $0^{\circ}$ and $30^{\circ}$ degrees (Fig.~\ref{Fig8}(b)). Also in this case, the evolution of the drastic phase variation strongly depends on the driving IR wavelength, and the jump disappears by switching off the $B-C$ coupling.
Finally, the evolution of the attosecond time delays for photoelectrons emitted in the perpendicular direction ($60^{\circ}\leq\theta\leq90^{\circ}$; Fig.~\ref{Fig8}(c)) does not exhibit variation and does not present a significant dependence on the $B-C$ coupling in the low photoelectron energy region, while differences are observed above approximately 9~eV photoelectron energies.

The discrepancy between the wavelength that provides the best agreement with the theoretical predictions ($\lambda = 1000$~nm) and the wavelength driving the harmonic generation ($\lambda = 1018$~nm) may arise either from a slight mismatch between the actual energy separation of the $B^2\Sigma^+_u$ and $C^2\Sigma^+_g$ states and the value implemented in the theoretical model, or from a small difference in the central wavelength of the pulse that drives the HHG process and that of the portion of the beam responsible for the generation of the sideband signal. 

The summary of the evolution evolution of the attosecond time delays extracted from the experimental data measured in coincidence with the ionic fragments O$^+$ and CO$^+$ is presented in Fig.~\ref{Fig5SM}(a) and \ref{Fig5SM}(b), respectively. The evolution of the attosecond time delays measured in coincidence with the ion CO$^+$ do not present a similar phase jump over all integration intervals.

To investigate the origin of the jump in the photoionization time delay observed in the experimental and theoretical simulations, and its connection to the $B-C$ coupling,
it is convenient to start by considering the contributions of the individual terms contributing to the photoelectron signal.
Figure~\ref{Fig9} presents the intensity of the XUV+IR (Fig.~\ref{Fig9}(a)) and XUV-IR (Fig.~\ref{Fig9}(c)) terms, as well as their coherent superposition (Fig.~\ref{Fig9}(e)) with the $B-C$ coupling, averaged over one full delay-dependent oscillation of the RABBIT signal. The latter one corresponds to the term $A_0$ introduced in Eq.~\eqref{A0}. These two-dimensional maps can be compared with those obtained by switching off the $B-C$ coupling as shown in Figs.~\ref{Fig9}(b), \ref{Fig9}(d), and \ref{Fig9}(f).
Simulations with the \(B\)-\(C\) coupling suppressed were performed by resetting the calculated transition dipole between the state $B^2\Sigma^+_u$ and $C^2\Sigma^+_g$ of the residual ion (approximately one~\(ea_0\), where $e$ indicates the electron charge and $a_0$ the Bohr radius) to zero.
The comparison between Fig.~\ref{Fig9}(a) and \ref{Fig9}(c) shows that photoelectrons with low kinetic energies emitted in the perpendicular direction are present only in the XUV+IR path and not in the XUV-IR path. Furthermore, this contribution is absent when the $B-C$ coupling is switched off (between Fig.~\ref{Fig9}(a) and \ref{Fig9}(b)). 

\begin{figure*}   
\centering 
\includegraphics[width=.75\textwidth]{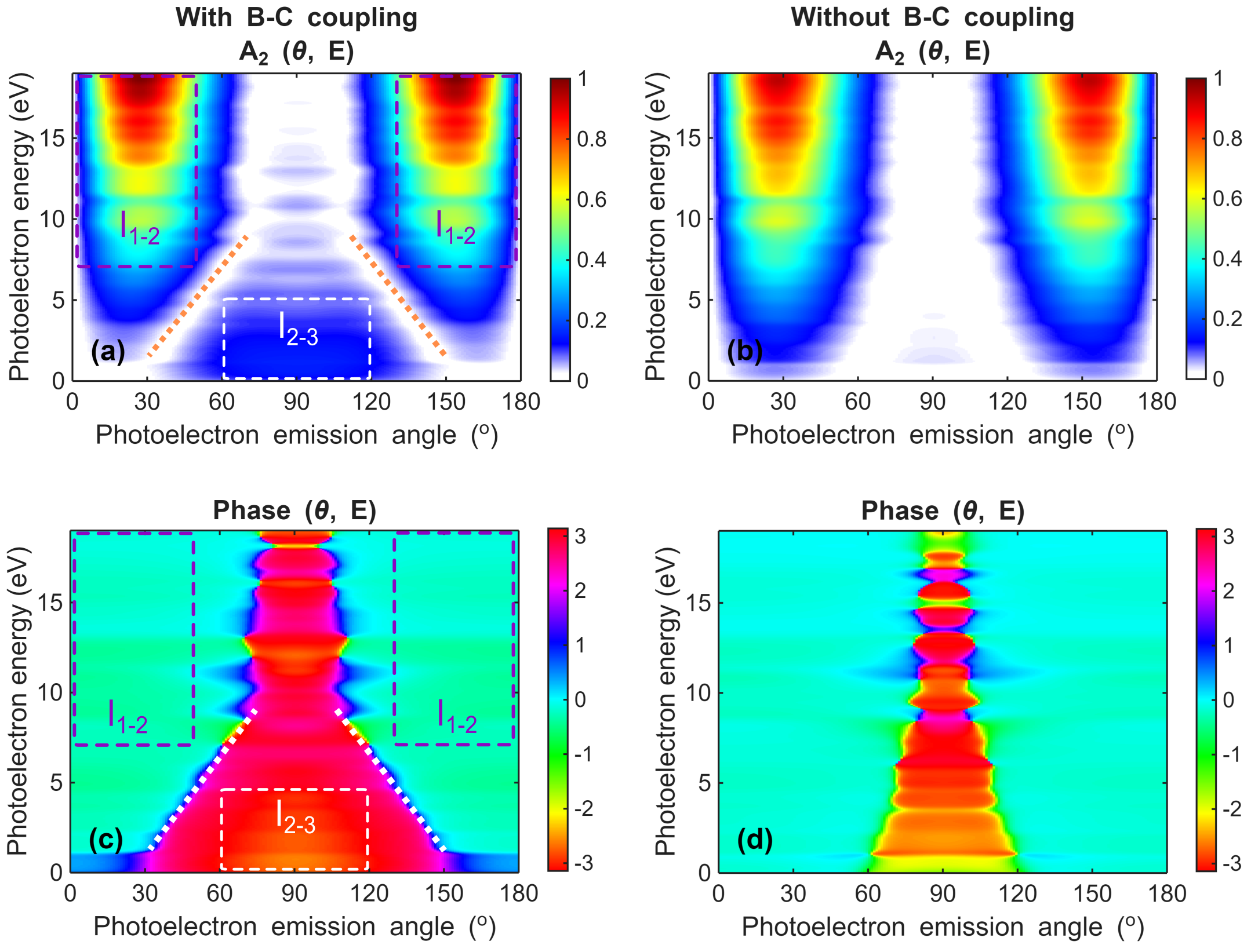}
\caption{ Simulated two-dimensional maps of $A_2(\theta,E)$ (a-c) and $\Phi(\theta,E)$ (b-d), obtained with (a,b) and without (c,d) the contribution of the $B-C$ coupling (a,b), according to Eqs.~\eqref{A2} and \eqref{eq:QccQii}. The red dotted lines in panel (a) mark the points in the $\theta-E$ plane at which the sideband modulation amplitude $A_2$ becomes zero. The white dotted lines in panel (c) mark the points in the $\theta-E$ plane at which a sudden change in the phase $\Phi$ is observed. The effect of the chirp of the attosecond pulses was not included in the simulations.}
\label{Fig10}
\end{figure*}

The comparison between the two lowest panels clearly indicates that the $B-C$ coupling (path 3) significantly contributes to the photoelectron signal at low energies ($0\leq E\leq5$~eV) and angles around the perpendicular direction ($60^{\circ}\leq\theta\leq120^{\circ}$). We will show in Fig.~\ref{Fig10} that the sideband oscillations in this region are dominated by the interference between the paths 2 and 3. We therefore indicate this region as I$_{2-3}$ (see Fig.~\ref{Fig9}(e)). Conversely, the contributions of the two other paths (paths 1 and 2) dominate the RABBIT delay-integrated signal for high photoelectron energies ($7\leq E\leq20$~eV) and angles around the parallel direction ($0^{\circ}\leq\theta\leq50^{\circ}$) (region I$_{1-2}$). 

The contribution of path 3 is also clearly visible in the amplitude and phase of the oscillations of the photoelectron signal expressed by the functions $A_2(\theta,E)$ and $\Phi(\theta,E)$ introduced in Eq.~(\ref{Eq_sideband}), which are presented in Fig.~\ref{Fig10} for the cases with (Fig.~\ref{Fig10}(a) and \ref{Fig10}(c)) and without $B-C$ coupling (Fig.~\ref{Fig10}(b) and \ref{Fig10}(d)), respectively. 
In particular, for low photoelectron kinetic energies and for emission along the direction perpendicular to the polarization axis, the oscillatory behavior is dominated by the interference between the direct XUV-IR ionization (XUV absorption and IR emission) pathway and the pathway associated with the $B-C$ ionic coupling (paths 3 and 2 in Fig.~\ref{Fig7}). The combined contributions of all three pathways defines areas in which the modulation amplitude vanishes (tilted orange dashed lines between 2 and 9~eV and $30^{\circ}$ and $70^{\circ}$ degrees, between 2 and 9~eV and $110^{\circ}$ and $150^{\circ}$ degrees), as presented in Fig.~\ref{Fig10}(a)). Along these tilted lines, correspondingly sharp variations in the phase of the oscillations are observed in Fig.~\ref{Fig10}(c) (white dotted lines).
We can conclude that the low photoelectron kinetic energy range is dominated in the perpendicular direction by the interference between the paths 2 and 3. This region is indicated as I$_{2-3}$ in Fig.~\ref{Fig10}(a) and \ref{Fig10}(c). In contrast, the high photoelectron kinetic energy range parallel to the emission direction is mostly due to the interference between paths 1 and 2, corresponding to region I$_{1-2}$ (Fig.~\ref{Fig10}(a) and \ref{Fig10}(c)).
The pronounced phase variations observed in the simulations as a function of emission angle at fixed photoelectron energy (and vice versa) marks the locations in the $\theta\!-\!E$ plane where the two contributions have similar amplitudes.
The interpretation of the evolution of the amplitude $A_2$ and phase $\Phi$ is in complete agreement with the conclusions previously obtained for the term $A_0$ from Fig.~\ref{Fig9}(e).

\begin{figure} [h] 
    \centering
    \includegraphics[width=0.5\textwidth]{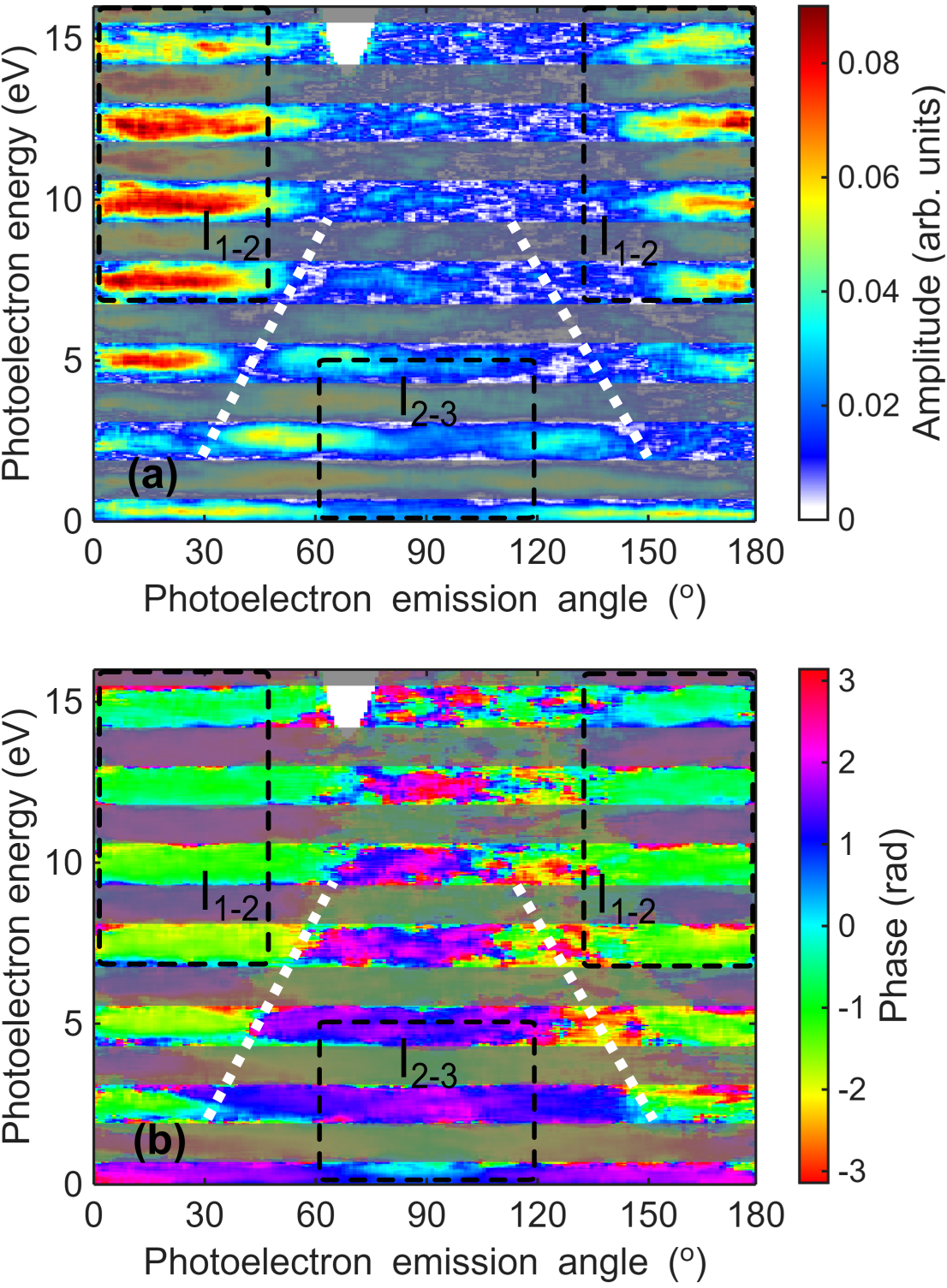}
\caption{Experimental angle- and energy-resolved analysis of the experimental RABBIT traces showing the $A_2(\theta,E)$ (a) and $\Phi(\theta,E)$ (b) measured in coincidence with O$^+$ ions. The energy intervals corresponding to the harmonics have been shadowed in gray for visual clarity. The white dotted lines in panels (a) and (b) mark the position in the $\theta-E$ plane of the orange and white dotted lines reported in Fig.~\ref{Fig10}(a) and~\ref{Fig10}(c), respectively. The effect of the chirp of the attosecond pulses was not subtracted in these experimental data.}
\label{Fig11}
\end{figure}

The same qualitative evolution can be observed in the experimental data for the amplitude $A_2$ and phase $\Phi$ of the sideband oscillations reported in Fig.~\ref{Fig11}. For the sake of clarity, the regions corresponding to the main lines due to the absorption of a single XUV photon, have been obscured by gray overlays. Furthermore, in the experimental data we can identify inclined lines of minimal oscillations, along which strong phase variations take place (white dotted lines), appearing at the same locations as those found in the simulations (see Fig.~\ref{Fig10}(a) and \ref{Fig10}(c)).
We note that the quality of the experimental data at large emission angles ($120^{\circ}\leq\theta\leq170^{\circ}$) is affected by the presence of a magnetic node, which influences the reconstruction of the photoelectron three-momentum distribution. Consequently, a partial asymmetry between the angular regions below and above $\theta=90^{\circ}$ is observed in the experimental data.

The two-dimensional representation, together with the identification of the two regions $I_{1-2}$ and $I_{2-3}$ in the $\theta-$–$E$ maps, makes it straightforward to interpret how the energy position of the phase jump observed in Fig.~\ref{Fig6} depends on the integration angle. When the integration is restricted along the parallel direction (Figs.~\ref{Fig6}(d) and \ref{Fig8}(b)), the minimum of the amplitude $A_2$ -which is associated with the abrupt variation of the phase $\Phi$- appears at photoelectron energies of about 2~eV, in agreement with the theoretical simulations in Figs.~\ref{Fig10}(a) and \ref{Fig10}(c) and the experimental data shown in Figs.~\ref{Fig11}(a) and \ref{Fig10}(b).  By enlarging the integration angle (Figs.~\ref{Fig6}(d) and \ref{Fig8}(a)), we increase the contribution of the region $I_{2-3}$, which imposes a shift of the position of the phase jump towards higher photoelectron energies. 
Finally, for integration over the perpendicular emission angle, the photoelectron signal extends up to roughly 8~eV (see Fig.~\ref{Fig6}(c)), and the interference between paths 2 and 3 governs the total signal across the entire photoelectron energy range (particularly in the region labeled $I_{2-3}$) with minimal contribution of the other interfering term due to the paths 1 and 2. As a result, no pronounced phase jump is visible, neither in the experimental measurements (Fig.~\ref{Fig8}(c)) nor in the corresponding theoretical simulations (Fig.~\ref{Fig10}(c)).

\clearpage

\subsection*{Partial-wave analysis}
\begin{figure}[htbp]
    \centering
    \includegraphics[width=\textwidth]{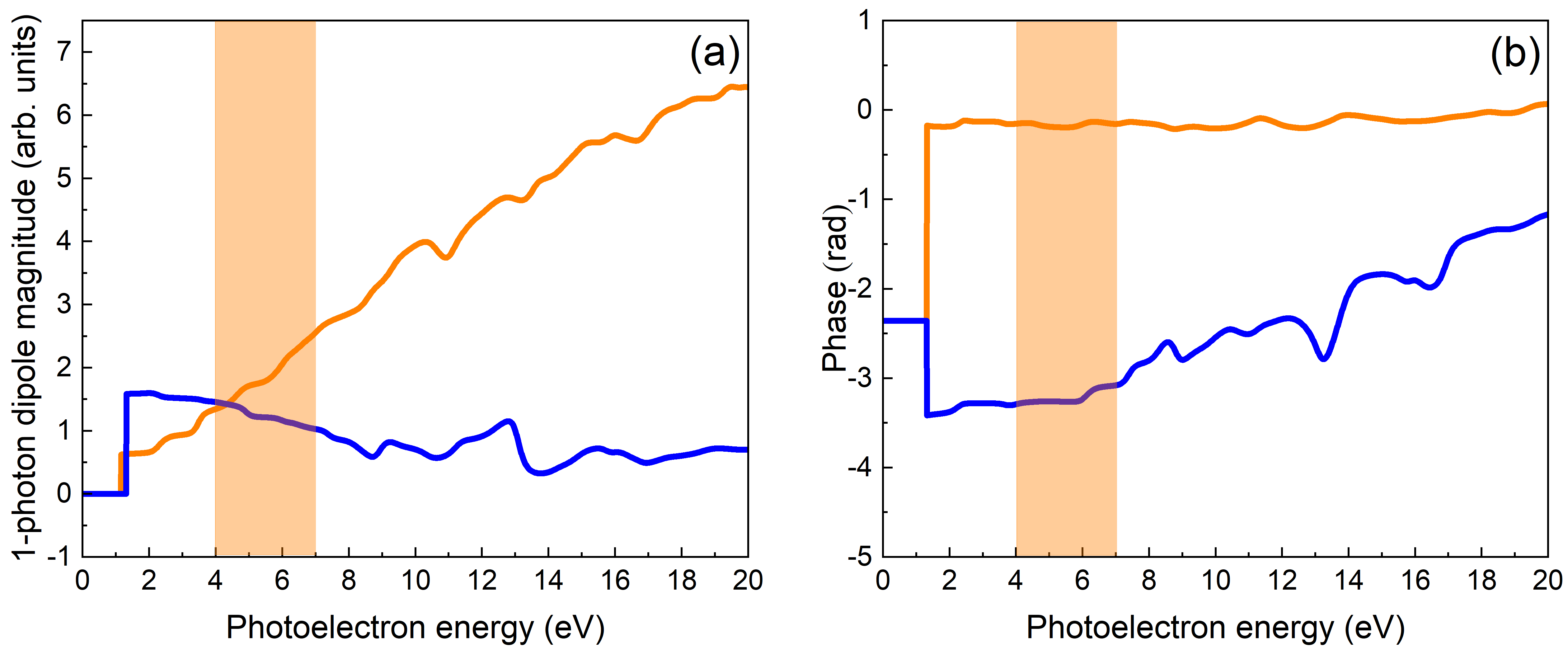}
    \caption{Magnitudes (a) and phases (b) of the orientation-averaged and emission-integrated free-free and ion-ion transition
    contributions to the complete RABBIT interference terms $Q_{cc}$ (orange lines) and $Q_{ii}$ (blue lines), as introduced in Eqs.~\eqref{eq:Qcc}, \eqref{eq:Qii}. The shaded areas indicate the photoelectron energy regions corresponding to the sharp variation in photoionization time delays.}
    \label{Fig12}
\end{figure}

The theoretical model can be used to further investigate the origin of the energy dependence of the photoionization time delays. 
The terms $Q_{cc}$ and $Q_{ii}$ in Eqs.~\eqref{eq:QccQii}--\eqref{eq:Qii} correspond to the interference terms contributed by the paths 1-2 and 2-3, respectively.
The amplitude and phase of these two terms is presented in Fig.~\ref{Fig12}(a) and \ref{Fig12}(b), respectively. As already discussed, the magnitude of these two terms are comparable in the region where the amplitude of the oscillations is minimal. Furthermore, the oscillations are approximately out of phase in these regions. To understand the first property, we observe that the terms differ only in the factors \(T_\text{XUV+IR}\) and we invoke the standard asymptotic approximation for these two terms~\cite{bendaAngularMomentumDependence2025},
\begin{align}
    Q_{cc} \sim T_\text{XUV+IR}^{(2)cc*} \sim A_{cc+}^* d_C^{(1)*} \,, \qquad
    Q_{ii} \sim T_\text{XUV+IR}^{(2)ii*} \sim A_{ii+}^* d_B^{(1)*} \,.
\end{align}
Here \(A_{cc+}\) and \(A_{ii+}\) stand for a dipole integral and an overlap integral, respectively, between the intermediate and final photoelectron wavepackets (i.e., before and after IR field interaction; for a detailed expression see the supplementary material and Figs.~\ref{fig:Akk}, \ref{fig:pwasy}, \ref{fig:asy}, and \ref{fig:simplified}).
While $Q_{cc}$ is proportional to the single-photon dipole matrix element \(d_C^{(1)}\) between the ground state of the molecule and the final state where photoelectron couples to the $C^2\Sigma^+_g$ state, $Q_{ii}$ is proportional to the analog quantity \(d_B^{(1)}\) between the ground state of the molecule and the final state, where photoelectron couples to the $B^2\Sigma^+_u$ state. While the cross section of the $B^2\Sigma^+_u$ state starts at large values and decreases at higher energies, the cross section into the $C^2\Sigma^+_g$ state is small close to the threshold and gets progressively larger
with increasing energy, up to photon energies of about 36~eV, as shown in Fig.~\ref{Fig3SM}(b). The energy at which the two curves cross is approximately 4.45~eV, in good agreement with the experimental observation for the minimum of the amplitude of the sideband oscillations in the range 4-7~eV (Fig.~\ref{Fig6}(d)). For lower energies, the RABBIT traces integrated over the entire solid angle and averaged over the molecular orientation are dominated by the $Q_{ii}$ term, which corresponds to the interference between the paths 2 and 3. For high photoelectron energies, the signal is dominated by the $Q_{cc}$ term, which corresponds to the interference between the paths 1 and 2.

The $\pi$ phase difference between the two terms causes a delay jump, when crossing the point in which the two contributions are equal in magnitude, given by:
\begin{equation}
    \Delta t \approx \pi/(2\omega_{\mathrm{IR}})=T_{\mathrm{IR}}/4=830\,\mathrm{as},
\end{equation}
in good agreement with the experimental value observed in Fig.~\ref{Fig8}(a) (approximately 750 as).
The $\pi$-phase difference between \(Q_{cc}\) and \(Q_{ii}\) can be understood by observing that, while the phase of the term $Q_{cc}$ is close to zero (Fig.~\ref{Fig12}(b)), corresponding to a usual small RABBIT sideband delay, the term $Q_{ii}$ exhibits a phase of almost $\pi$. This exotic phase is the result of two phase terms, whose contributions can be understood by considering a more complete expression for \(Q_{ii}\), where we also expand the XUV--IR amplitude factor (path 2),
\begin{equation}
    Q_{ii} \sim T_\text{XUV+IR}^{(2)ii*} T_\text{XUV--IR}^{(2)} \sim A_{ii+}^* d_B^{(1)*} A_{cc-} d_C^{(1)} .
    \label{eq:Qiisimp1}
\end{equation}

\begin{figure}[htbp]
    \centering
    \includegraphics[width=\textwidth]{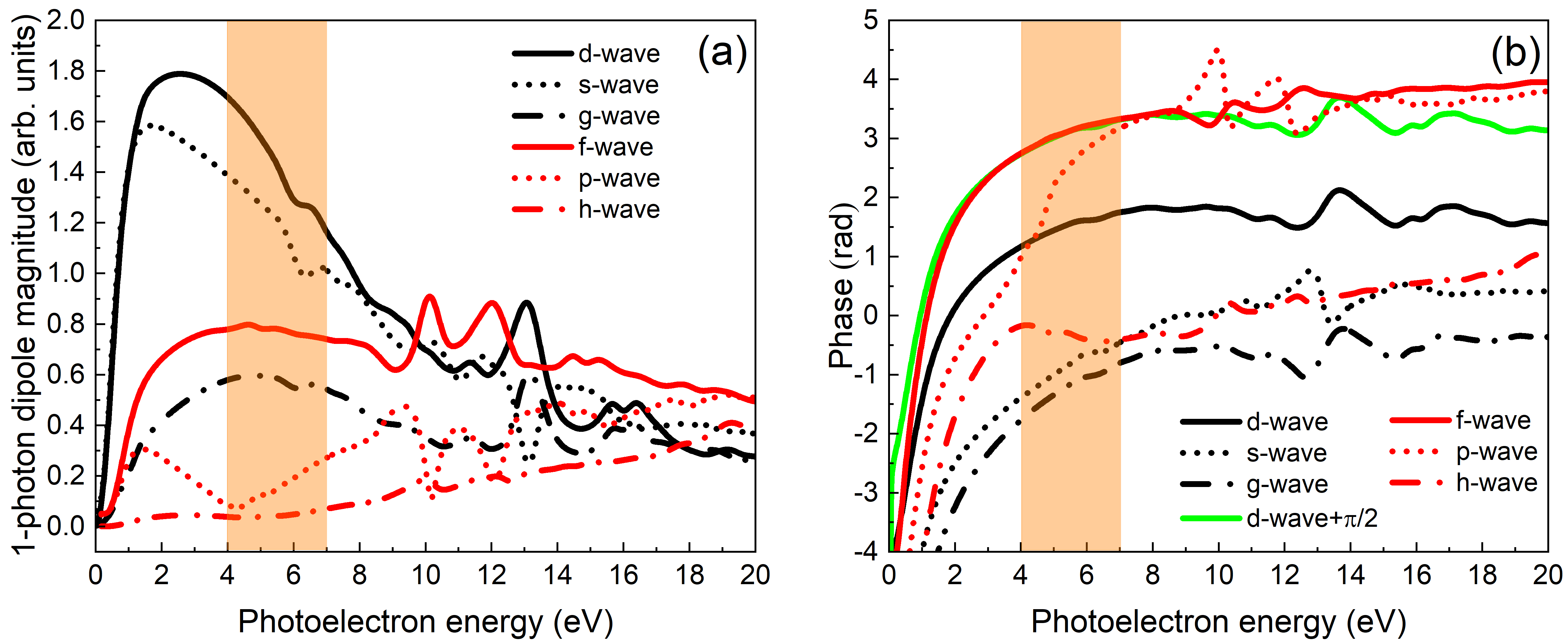}
    \caption{Magnitudes (a) and phases (b) of the partial-wave resolved one-photon ionization dipoles for the $B^2\Sigma^+_u$ (black lines) and $C^2\Sigma^+_g$ (red lines) channel. Only partial waves for
\(m = 0\) are shown. (Magnitudes for dipoles with \(m \ne 0\) are smaller than those for \(m = 0\) and are neglected in this
analysis.) The shifted  green curve in the panel (b) illustrates that the phase of d-wave component in $B^2\Sigma^+_u$ channel is behind
in phase roughly by $\pi$/2 compared to the f -wave component in $C^2\Sigma^+_g$ channel. The shaded areas indicate the photoelectron energy regions corresponding to the sharp variation in photoionization time delays.}
    \label{Fig13}
\end{figure}

In Eq.~\eqref{eq:Qiisimp1} the integral \(A_{ii+}\) is real, but the complex integral $A_{cc-}$ has a phase close to $\pi/2$  and is related to Fourier transform of the dipole operator (see also Fig.~\ref{fig:Akk}).
The second contributions to the phase can be understood by noticing that the low energies present a very clear partial wave picture: The angular momentum \(l = 2\) corresponds
to the dominant partial wave contribution in the $B^2\Sigma^+_u$ channel, while the angular momentum \(\lambda = 3\) corresponds to the
dominant partial wave contribution in the $C^2\Sigma^+_g$ channel, as shown in Fig.~\ref{Fig13}(a). The dominant partial waves correspond to zero axial angular momentum projections, \(m = \mu = 0\) (see also Figs.~\ref{fig:pws} and \ref{fig:dominant-dipoles}). 
Additionally, in the emission-integrated and orientation-averaged \(\langle Q_{ii} \rangle\) any direction-dependent factors are integrated out, leaving only the raw partial-wave components of the one-photon dipole amplitudes (see also Eq.~\eqref{eq:Qiisimp}).

Thus, at low energies, the only relevant contribution to \(Q_{ii}\)
comes from the combination of a \(d\)-wave (black solid line) in the $B^2\Sigma^+_u$ channel and the \(f\)-wave (red solid line) in the $C^2\Sigma^+_g$ channel (see Fig.~\ref{Fig13}(a)). All other terms exhibit a smaller contribution. These two
partial waves have almost constant phase difference of \(+\pi/2\), as shown in Fig.~\ref{Fig13}(b), which reports the phase of the partial waves associated to photoionization of the $B^2\Sigma^+_u$ (black lines) and $C^2\Sigma^+_g$ (red lines) levels. The phases of the dominant contributions are represented by solid lines, together with the phase of the dominant $C^2\Sigma^+_g$-terms shifted by $\pi/2$ (green line). This analysis enables us to ultimately relate the jump in the attosecond time delay to the evolution of the cross sections of the $B^2\Sigma^+_u$ and $C^2\Sigma^+_g$ states, as well as to the contribution of a specific partial wave in the interference term that includes the ionic coupling.

The reduced amplitude of interference in the 4-6~eV range can also be interpreted in terms of entanglement between different degrees of freedom of the photoelectron. In particular, the dominant contributions from partial waves of different parity in the $B^2\Sigma^+_u$ and $C^2\Sigma^+_g$ channels suggest an entanglement between the radial and angular parts of the photoelectron wavefunction~\cite{bustoProbingElectronicDecoherence2022}. In this energy region, where the two pathways have comparable strength, integrating over emission angles ($0^\circ$--$90^\circ$) effectively traces out the angular degrees of freedom, leading to a suppression of the interference amplitude.


\section{Conclusions} \label{Conclusions}
We have presented a detailed theoretical and experimental investigation of attosecond time delays measured in coincidence with the dissociation pathways of CO$_2$ molecules exposed to the combination of an XUV attosecond pulse train and a synchronized IR field. Analyzing the XUV-only data allows us to identify the reaction pathways leading to the emission of the O$^+$ and CO$^+$ ionic fragments. By measuring the KER of the ionic fragments, we can also infer information about the vibrational state of the neutral (CO) or ionic (CO$^+$) molecule generated in the dissociation.
The analysis of the RABBIT trace reveals a third path contributing to the sideband signal, due to the absorption of an IR photon in the molecular ion created by the interaction with the XUV harmonics. 
Our results clearly demonstrate the importance of considering and describing the effect of the IR field not only on the photoelectron wave packet released by the target system, but also on the cation for a correct interpretation of attosecond interferometry of molecules.
The angle-, energy- and ion-channel-resolved measurements allows one to disentangle the contributions of the different interfering terms building up the RABBIT signal, offering a simple and intuitive explanation of the strong variation of the experimental attosecond time delays. The theoretical analysis and its comparison with the experimental data allow us to attribute the variation of the photoionization time delays to the specific contributions of only two partial waves populated in the photoionization process.
Further investigations should be devoted to identifying this mechanism in other molecular systems to confirm the generality of the ionic coupling contribution to the attosecond time delays we observed in CO$_2$. 

\vspace{-3.0truemm}

\section*{Acknowledgments} 
We thank L. Poletto and F. Frassetto for the implementation of the XUV photon spectrometer, D. Ertel for the support in the data acquisition, and C.D. Schr\"oter and R. Moshammer for the development of the Reaction Microscope. 
G.S. acknowledges financial support by FRIAS, by the Deutsche Forschungsgemeinschaft project Research Training Group DynCAM (RTG 2717), project INST 39/1079 (High-Repetition-Rate Attosecond Source for Coincidence Spectroscopy), and grant  546852490 (Project SA3470/13-1). S.P. acknowledges financial support by Deutsche Forschungsgemeinschaft project PA 2691/3-1. I.M. and G.S. acknowledge financial support by Georg H. Endress Foundation. D.B. acknowledges support from the Swedish Research Council grant 2020-06384 and 2025-03729 and from the Knut and Alice Wallenberg Foundation through the Wallenberg Center for Quantum Technology. G.S. and I.M. acknowledges financial support from the European Union’s Horizon Europe research and innovation programme under the Marie Skłodowska-Curie grant agreement No 101168628 (project Qu-ATTO).
This work has been supported by the Charles University Research Centre program No. UNCE/24/SCI/016 and by the Ministry of Education, Youth and Sports of the Czech Republic through the e-INFRA CZ (ID:90254). J.B. and Z.M. acknowledge the support of the Czech Science Foundation (25-24428L).
V.H. and U.T were supported by the Chemical Sciences, Geosciences, and Biosciences Division, Office of Basic Energy Sciences, Office of Science, US Department of Energy under Award DEFG02-86ER13491. U.T. acknowledges partial support by NSF grant PHY 2409183. We acknowledge access to the BEOCAT Research Computing Cluster at Kansas State University.

\bibliographystyle{apsrev4-2}

\bibliography{library.bib}

@article{2019_Ambrosio_RABBITT_CuSurface,
  title = {Spatiotemporal analysis of a final-state shape resonance in interferometric photoemission from Cu(111) surfaces},
  author = {Ambrosio, M. J. and Thumm, U.},
  journal = {Phys. Rev. A},
  volume = {100},
  issue = {4},
  pages = {043412},
  numpages = {15},
  year = {2019},
  month = {Oct},
  publisher = {American Physical Society},
  doi = {10.1103/PhysRevA.100.043412},
  url = {https://link.aps.org/doi/10.1103/PhysRevA.100.043412}
}

@article{2017_Kasmi_RABBITT_CuSurface,
author = {Lamia Kasmi and Matteo Lucchini and Luca Castiglioni and Pavel Kliuiev and J\"{u}rg Osterwalder and Matthias Hengsberger and Lukas Gallmann and Peter Kr\"{u}ger and Ursula Keller},
journal = {Optica},
number = {12},
pages = {1492--1497},
publisher = {Optica Publishing Group},
title = {Effective mass effect in attosecond electron transport},
volume = {4},
month = {Dec},
year = {2017},
url = {https://opg.optica.org/optica/abstract.cfm?URI=optica-4-12-1492},
doi = {10.1364/OPTICA.4.001492},
}

@article{2020_Liao_PE_dispersion_RABBITT,
  title = {Distinction of Electron Dispersion in Time-Resolved Photoemission Spectroscopy},
  author = {Liao, Qing and Cao, Wei and Zhang, Qingbin and Liu, Kai and Wang, Feng and Lu, Peixiang and Thumm, Uwe},
  journal = {Phys. Rev. Lett.},
  volume = {125},
  issue = {4},
  pages = {043201},
  numpages = {6},
  year = {2020},
  month = {Jul},
  publisher = {American Physical Society},
  doi = {10.1103/PhysRevLett.125.043201},
  url = {https://link.aps.org/doi/10.1103/PhysRevLett.125.043201}
}

@article{2020_Barca_GAMESS_JCP,
author = "{G. M. J. Barca  {\em et al.}}",
title = "{Recent developments in the general atomic and molecular electronic structure system}",
journal = {J. Chem. Phys.},
volume = {152},
number = {15},
pages = {154102},
year = {2020},
doi = {10.1063/5.0005188},
url = { https://doi.org/10.1063/5.0005188},
}

@article{ahmadiAttosecondPhotoionisationTime2022,
  title = {Attosecond Photoionisation Time Delays Reveal the Anisotropy of the Molecular Potential in the Recoil Frame},
  author = {Ahmadi, H. and Pl{\'e}siat, E. and Moioli, M. and Frassetto, F. and Poletto, L. and Decleva, P. and Schr{\"o}ter, C. D. and Pfeifer, T. and Moshammer, R. and Palacios, A. and Martin, F. and Sansone, G.},
  year = {2022},
  month = mar,
  journal = {Nature Communications},
  volume = {13},
  number = {1},
  pages = {1242},
  issn = {2041-1723},
  doi = {10.1038/s41467-022-28783-x}
}

@article{ahmadiCollinearSetupDelay2020,
  title = {Collinear Setup for Delay Control in Two-Color Attosecond Measurements},
  author = {Ahmadi, H and Kellerer, S and Ertel, D and Moioli, M and Reduzzi, M and Maroju, P K and J{\"a}ger, A and Shah, R N and Lutz, J and Frassetto, F and Poletto, L and Bragheri, F and Osellame, R and Pfeifer, T and Schr{\"o}ter, C D and Moshammer, R and Sansone, G},
  year = {2020},
  month = apr,
  journal = {Journal of Physics: Photonics},
  volume = {2},
  number = {2},
  pages = {024006},
  publisher = {IOP Publishing},
  issn = {2515-7647},
  doi = {10.1088/2515-7647/ab823f}
}

@article{bendaAnalysisRABITTTime2022,
  title = {Analysis of {{RABITT}} Time Delays Using the Stationary Multiphoton Molecular \${{R}}\$-Matrix Approach},
  author = {Benda, J. and Ma{\v s}{\'i}n, Z. and Gorfinkiel, J. D.},
  year = {2022},
  month = may,
  journal = {Physical Review A},
  volume = {105},
  number = {5},
  pages = {053101},
  publisher = {American Physical Society},
  doi = {10.1103/PhysRevA.105.053101}
}

@article{bendaDipolelaserCouplingDelay2024,
  title = {Dipole-Laser Coupling Delay in Two-Color {{RABBITT}} Photoionization of Polar Molecules},
  author = {Benda, Jakub and Ma{\v s}{\'i}n, Zden{\v e}k},
  year = {2024},
  month = jan,
  journal = {Physical Review A},
  volume = {109},
  number = {1},
  pages = {013106},
  publisher = {American Physical Society},
  doi = {10.1103/PhysRevA.109.013106}
}

@article{bendaMultiphotonThresholdIonization2021,
  title = {Multi-Photon above Threshold Ionization of Multi-Electron Atoms and Molecules Using the {{R-matrix}} Approach},
  author = {Benda, Jakub and Ma{\v s}{\'i}n, Zden{\v e}k},
  year = {2021},
  month = jun,
  journal = {Scientific Reports},
  volume = {11},
  number = {1},
  pages = {11686},
  issn = {2045-2322},
  doi = {10.1038/s41598-021-89733-z}
}

@article{bombachBranchingRatiosPartition1983,
    author = {Bombach, R. and Dannacher, J. and Stadelmann, J.‐P. and Lorquet, J. C.},
    title = {Branching ratios and partition of the excess energy for the predissociation of CO+2 C̃ 2Σ+g molecular cations},
    journal = {The Journal of Chemical Physics},
    volume = {79},
    number = {9},
    pages = {4214-4220},
    year = {1983},
    month = {11},
    issn = {0021-9606},
    doi = {10.1063/1.446347},
    url = {https://doi.org/10.1063/1.446347},
}

@article{brionPartialOscillatorStrengths1978,
  title = {Partial Oscillator Strengths for the Photoionization of {{N2O}} and {{CO2}} (20--60 {{eV}})},
  author = {Brion, C.E. and Tan, K.H.},
  year = {1978},
  month = nov,
  journal = {Chemical Physics},
  volume = {34},
  number = {2},
  pages = {141--151},
  issn = {0301-0104},
  doi = {10.1016/0301-0104(78)80030-5}
}

@article{calegariAdvancesAttosecondScience2016,
  title = {Advances in Attosecond Science},
  author = {Calegari, Francesca and Sansone, Giuseppe and Stagira, Salvatore and Vozzi, Caterina and Nisoli, Mauro},
  year = {2016},
  month = feb,
  journal = {Journal of Physics B},
  volume = {49},
  number = {6},
  pages = {062001},
  publisher = {IOP Publishing},
  issn = {0953-4075},
  doi = {10.1088/0953-4075/49/6/062001}
}

@article{chengKineticEnergyRelease2018,
  title = {Kinetic Energy Release Distributions from Dissociative Photoionization of Weakly Bound Trimers at 14--27 {{eV}}},
  author = {Cheng, Bing-Ming and Grover, J. R. and Walters, E. A. and Clay, J. T.},
  year = {2018},
  journal = {Physical Chemistry Chemical Physics},
  volume = {20},
  number = {32},
  pages = {21034--21042},
  publisher = {The Royal Society of Chemistry},
  issn = {1463-9076},
  doi = {10.1039/C8CP03013H}
}

@article{delgadoThreepathInterferencesReconstruction2025,
  title = {Three-Path Interferences in the Reconstruction of Attosecond Beatings by Interference of Two-Photon Transitions in Molecules},
  author = {Delgado, Jorge and {Gonz{\'a}lez-Collado}, Celso M. and Decleva, Piero and Palacios, Alicia and Mart{\'i}n, Fernando},
  year = {2025},
  month = jun,
  journal = {Physical Review A},
  volume = {111},
  number = {6},
  pages = {063107},
  publisher = {American Physical Society},
  doi = {10.1103/qzg1-2f7n}
}

@article{dornerColdTargetRecoil2000a,
  title = {Cold {{Target Recoil Ion Momentum Spectroscopy}}: A `Momentum Microscope' to View Atomic Collision Dynamics},
  author = {D{\"o}rner, R. and Mergel, V. and Jagutzki, O. and Spielberger, L. and Ullrich, J. and Moshammer, R. and {Schmidt-B{\"o}cking}, H.},
  year = {2000},
  month = jun,
  journal = {Physics Reports},
  volume = {330},
  number = {2},
  pages = {95--192},
  issn = {0370-1573},
  doi = {10.1016/S0370-1573(99)00109-X}
}

@article{elandFormationPredissociationCO+21977,
  title = {Formation and Predissociation of {{CO}}+2(?\,{{2$\Sigma$}}+g)},
  author = {Eland, John H. D. and Berkowitz, Joseph},
  year = {1977},
  month = sep,
  journal = {The Journal of Chemical Physics},
  volume = {67},
  number = {6},
  pages = {2782--2787},
  issn = {0021-9606},
  doi = {10.1063/1.435194},
  urldate = {2025-08-02}
}

@article{elandPredissociationTriatomicIons1972,
  title = {Predissociation of Triatomic Ions Studied by Photo-Electron-Photoion Coincidence Spectroscopy and Photoion Kinetic Energy Analysis: {{I}}. {{CO2}}+},
  author = {Eland, J.H.D.},
  year = {1972},
  month = sep,
  journal = {International Journal of Mass Spectrometry and Ion Physics},
  volume = {9},
  number = {4},
  pages = {397--406},
  issn = {0020-7381},
  doi = {10.1016/0020-7381(72)80023-8}
}

@article{ertelInfluenceNuclearDynamics2023,
  title = {Influence of Nuclear Dynamics on Molecular Attosecond Photoelectron Interferometry},
  author = {Ertel, Dominik and Busto, David and Makos, Ioannis and Schmoll, Marvin and Benda, Jakub and Ahmadi, Hamed and Moioli, Matteo and Frassetto, Fabio and Poletto, Luca and Schr{\"o}ter, Claus Dieter and Pfeifer, Thomas and Moshammer, Robert and Ma{\v s}{\'i}n, Zden{\v e}k and Patchkovskii, Serguei and Sansone, Giuseppe},
  year = {2023},
  month = sep,
  journal = {Science Advances},
  volume = {9},
  number = {35},
  pages = {eadh7747},
  publisher = {American Association for the Advancement of Science},
  doi = {10.1126/sciadv.adh7747},
  urldate = {2024-02-08}
}

@article{ertelUltrastableHighrepetitionrateAttosecond2023,
  title = {Ultrastable, High-Repetition-Rate Attosecond Beamline for Time-Resolved {{XUV}}--{{IR}} Coincidence Spectroscopy},
  author = {Ertel, D. and Schmoll, M. and Kellerer, S. and J{\"a}ger, A. and Weissenbilder, R. and Moioli, M. and Ahmadi, H. and Busto, D. and Makos, I. and Frassetto, F. and Poletto, L. and Schr{\"o}ter, C. D. and Pfeifer, T. and Moshammer, R. and Sansone, G.},
  year = {2023},
  month = jul,
  journal = {Review of Scientific Instruments},
  volume = {94},
  number = {7},
  pages = {073001},
  issn = {0034-6748},
  doi = {10.1063/5.0139496},
  urldate = {2025-08-18}
}

@article{freyPhotoionizationResonanceSpectra1977,
  title = {Photoionization {{Resonance}} Spectra of {{CO}}+2 and Threshold Electron---Ion Coincidence Measurements of the Fragmentation of {{CO}}+2},
  author = {Frey, R. and Gotchev, B. and Kalman, O.F. and Peatman, W.B. and Pollak, H. and Schlag, E.W.},
  year = {1977},
  month = apr,
  journal = {Chemical Physics},
  volume = {21},
  number = {1},
  pages = {89--100},
  issn = {0301-0104},
  doi = {10.1016/0301-0104(77)85181-1}
}

@article{gongAsymmetricAttosecondPhotoionization2022a,
  title = {Asymmetric {{Attosecond Photoionization}} in {{Molecular Shape Resonance}}},
  author = {Gong, Xiaochun and Jiang, Wenyu and Tong, Jihong and Qiang, Junjie and Lu, Peifen and Ni, Hongcheng and Lucchese, Robert and Ueda, Kiyoshi and Wu, Jian},
  year = {2022},
  month = jan,
  journal = {Physical Review X},
  volume = {12},
  number = {1},
  pages = {011002},
  publisher = {American Physical Society},
  doi = {10.1103/PhysRevX.12.011002}
}

@article{gongAttosecondDelaysDissociative2023,
  title = {Attosecond Delays between Dissociative and Non-Dissociative Ionization of Polyatomic Molecules},
  author = {Gong, Xiaochun and Pl{\'e}siat, {\'E}tienne and Palacios, Alicia and Heck, Saijoscha and Mart{\'i}n, Fernando and W{\"o}rner, Hans Jakob},
  year = {2023},
  month = jul,
  journal = {Nature Communications},
  volume = {14},
  number = {1},
  pages = {4402},
  issn = {2041-1723},
  doi = {10.1038/s41467-023-40120-4}
}

@article{gongAttosecondSpectroscopySizeresolved2022,
  title = {Attosecond Spectroscopy of Size-Resolved Water Clusters},
  author = {Gong, Xiaochun and Heck, Saijoscha and Jelovina, Denis and Perry, Conaill and Zinchenko, Kristina and Lucchese, Robert and W{\"o}rner, Hans Jakob},
  year = {2022},
  month = sep,
  journal = {Nature},
  volume = {609},
  number = {7927},
  pages = {507--511},
  issn = {1476-4687},
  doi = {10.1038/s41586-022-05039-8}
}

@article{grimmAngleresolvedPhotoelectronSpectroscopy1981,
  title = {Angle-resolved Photoelectron Spectroscopy of {{CO2}} with Synchrotron Radiation},
  author = {Grimm, Frederick A. and Allen, Jr., John D. and Carlson, Thomas A. and Krause, Manfred O. and Mehaffy, David and Keller, Paul R. and Taylor, James W.},
  year = {1981},
  month = jul,
  journal = {The Journal of Chemical Physics},
  volume = {75},
  number = {1},
  pages = {92--98},
  issn = {0021-9606},
  doi = {10.1063/1.441859},
  urldate = {2025-06-02}
}

@article{hanAttosecondControlMeasurement2025,
  title = {Attosecond Control and Measurement of Chiral Photoionization Dynamics},
  author = {Han, Meng and Ji, Jia-Bao and Blech, Alexander and Goetz, R. Esteban and Allison, Corbin and Greenman, Loren and Koch, Christiane P. and W{\"o}rner, Hans Jakob},
  year = {2025},
  month = sep,
  journal = {Nature},
  volume = {645},
  number = {8079},
  pages = {95--100},
  issn = {1476-4687},
  doi = {10.1038/s41586-025-09455-4}
}

@article{huppertAttosecondDelaysMolecular2016,
  title = {Attosecond {{Delays}} in {{Molecular Photoionization}}},
  author = {Huppert, Martin and Jordan, Inga and Baykusheva, Denitsa and {von Conta}, Aaron and W{\"o}rner, Hans Jakob},
  year = {2016},
  month = aug,
  journal = {Physical Review Letters},
  volume = {117},
  number = {9},
  pages = {093001},
  publisher = {American Physical Society},
  doi = {10.1103/PhysRevLett.117.093001}
}

@article{kamalovElectronCorrelationEffects2020,
  title = {Electron Correlation Effects in Attosecond Photoionization of {{CO2}}},
  author = {Kamalov, Andrei and Wang, Anna L. and Bucksbaum, Philip H. and Haxton, Daniel J. and Cryan, James P.},
  year = {2020},
  journal = {Physical Review A},
  volume = {102},
  number = {2},
  issn = {1050-2947},
  doi = {10.1103/PhysRevA.102.023118}
}

@article{kovacHePhotoelectronSpectra1983,
  title = {The {{He}}\,i Photoelectron Spectra of {{CO2}}, {{CS2}}, and {{OCS}}: {{Vibronic}} Coupling},
  author = {Kova{\v c}, Branka},
  year = {1983},
  month = feb,
  journal = {The Journal of Chemical Physics},
  volume = {78},
  number = {4},
  pages = {1684--1692},
  issn = {0021-9606},
  doi = {10.1063/1.444967},
  urldate = {2025-06-19}
}

@article{lebergSynchrotronRadiationStudy1994,
  title = {Synchrotron Radiation Study of Photoionization and Photodissociation Processes of {{CO2}} in the 13-21 {{eV}} Region},
  author = {{L E Berg} and {A Karawajczyk} and {C Stromholm}},
  year = {1994},
  month = jul,
  journal = {Journal of Physics B},
  volume = {27},
  number = {14},
  pages = {2971},
  issn = {0953-4075},
  doi = {10.1088/0953-4075/27/14/030}
}

@article{2003_Liu_CO2_JCP,
  title = {Unimolecular Decay Pathways of State-Selected {{CO2}}+ in the Internal Energy Range of 5.2--6.2 {{eV}}: {{An}} Experimental and Theoretical Study},
  author = {Liu, Jianbo and Chen, Wenwu and Hochlaf, M. and Qian, Ximei and Chang, Chao and Ng, C. Y.},
  year = {2003},
  journal = {The Journal of Chemical Physics},
  volume = {118},
  number = {1},
  pages = {149--163},
  issn = {00219606},
  doi = {10.1063/1.1524180}
}

@article{masinElectronCorrelationsPrecollision2018,
  title = {Electron Correlations and Pre-Collision in the Re-Collision Picture of High Harmonic Generation},
  author = {Ma{\v s}{\'i}n, Zden{\v e}k and Harvey, Alex G and Spanner, Michael and Patchkovskii, Serguei and Ivanov, Misha and Smirnova, Olga},
  year = {2018},
  month = jun,
  journal = {Journal of Physics B},
  volume = {51},
  number = {13},
  pages = {134006},
  publisher = {IOP Publishing},
  issn = {0953-4075},
  doi = {10.1088/1361-6455/aac598}
}

@article{masinUKRmolSuiteModelling2020,
  title = {{{UKRmol}}+: {{A}} Suite for Modelling Electronic Processes in Molecules Interacting with Electrons, Positrons and Photons Using the {{R-matrix}} Method},
  author = {Ma{\v s}{\'i}n, Zden{\v e}k and Benda, Jakub and Gorfinkiel, Jimena D. and Harvey, Alex G. and Tennyson, Jonathan},
  year = {2020},
  month = apr,
  journal = {Computer Physics Communications},
  volume = {249},
  pages = {107092},
  issn = {0010-4655},
  doi = {10.1016/j.cpc.2019.107092}
}

@article{mengTheoreticalStudyPredissociation2009,
  title = {Theoretical Study on the Predissociation Mechanism of {{CO}}(2)(+) ({{C}} (2){{Sigma}}(g)(+))},
  author = {Meng, Qingyong and Huang, Ming-Bao and Chang, Hai-Bo},
  year = {2009},
  journal = {The journal of physical chemistry. A},
  volume = {113},
  number = {46},
  pages = {12825--12830},
  doi = {10.1021/jp907351s}
}

@article{moshammer4pRecoilionElectron1996,
  title = {A 4{$\pi$} Recoil-Ion Electron Momentum Analyzer: A High-Resolution ``Microscope'' for the Investigation of the Dynamics of Atomic, Molecular and Nuclear Reactions},
  author = {Moshammer, R. and Unverzagt, M. and Schmitt, W. and Ullrich, J. and {Schmidt-B{\"o}cking}, H.},
  year = {1996},
  month = mar,
  journal = {Nuclear Instruments and Methods in Physics Research Section B: Beam Interactions with Materials and Atoms},
  volume = {108},
  number = {4},
  pages = {425--445},
  issn = {0168-583X},
  doi = {10.1016/0168-583X(95)01259-1}
}

@article{nandiAttosecondTimingElectron2020,
  title = {Attosecond Timing of Electron Emission from a Molecular Shape Resonance},
  author = {Nandi, S. and Pl{\'e}siat, E. and Zhong, S. and Palacios, A. and Busto, D. and Isinger, M. and Neori{\v c}i{\'c}, L. and Arnold, C. L. and Squibb, R. J. and Feifel, R. and Decleva, P. and L'Huillier, A. and Mart{\'i}n, F. and Gisselbrecht, M.},
  year = {2020},
  journal = {Science Advances},
  volume = {6},
  number = {31},
  pages = {eaba7762},
  doi = {10.1126/sciadv.aba7762}
}

@article{nisoliAttosecondElectronDynamics2017,
  title = {Attosecond {{Electron Dynamics}} in {{Molecules}}},
  author = {Nisoli, Mauro and Decleva, Piero and Calegari, Francesca and Palacios, Alicia and Mart{\'i}n, Fernando},
  year = {2017},
  month = aug,
  journal = {Chemical Reviews},
  volume = {117},
  number = {16},
  pages = {10760--10825},
  publisher = {American Chemical Society},
  issn = {0009-2665},
  doi = {10.1021/acs.chemrev.6b00453}
}

@article{paulObservationTrainAttosecond2001,
  title = {Observation of a {{Train}} of {{Attosecond Pulses}} from {{High Harmonic Generation}}},
  author = {Paul, P. M. and Toma, E. S. and Breger, P. and Mullot, G. and Aug{\'e}, F. and Balcou, {\relax Ph}. and Muller, H. G. and Agostini, P.},
  year = {2001},
  month = jun,
  journal = {Science},
  volume = {292},
  number = {5522},
  pages = {1689--1692},
  publisher = {American Association for the Advancement of Science},
  doi = {10.1126/science.1059413},
  urldate = {2024-02-08}
}

@article{2024_Hung_CO2_PRA,
  title = {Dissociative Ionization of {{CO}}{$_2$} in {{XUV}} Pump and {{IR}} or Visible Probe Pulses},
  author = {Hoang, Van-Hung and Thumm, Uwe},
  year = {2024},
  month = mar,
  journal = {Physical Review A},
  volume = {109},
  number = {3},
  pages = {033117},
  publisher = {American Physical Society},
  doi = {10.1103/PhysRevA.109.033117}
}

@article{praetUnimolecularReactionPaths1982,
    author = {Praet, M. Th. and Lorquet, J. C. and Raşeev, G.},
    title = {Unimolecular reaction paths of electronically excited species. IV. The C̃ 2Σ+g state of CO+2},
    journal = {The Journal of Chemical Physics},
    volume = {77},
    number = {9},
    pages = {4611-4618},
    year = {1982},
    month = {11},
    issn = {0021-9606},
    doi = {10.1063/1.444413},
    url = {https://doi.org/10.1063/1.444413},
}

@article{pranjalResonantPhotoionizationCO22024,
  title = {Resonant {{Photoionization}} of {{CO2}} up to the {{Fourth Ionization Threshold}}},
  author = {Pranjal, Prateek and {Gonz{\'a}lez-V{\'a}zquez}, Jes{\'u}s and Bello, Roger Y. and Mart{\'i}n, Fernando},
  year = {2024},
  month = jan,
  journal = {The Journal of Physical Chemistry A},
  volume = {128},
  number = {1},
  pages = {182--190},
  publisher = {American Chemical Society},
  issn = {1089-5639},
  doi = {10.1021/acs.jpca.3c06947}
}

@article{rathboneModespecificPhotoelectronScattering2004a,
  title = {Mode-Specific Photoelectron Scattering Effects on {{CO2}}+({{C}}{$\mkern1mu$}{{2$\Sigma$g}}+) Vibrations},
  author = {Rathbone, G. J. and Poliakoff, E. D. and Bozek, John D. and Lucchese, R. R. and Lin, P.},
  year = {2004},
  month = jan,
  journal = {The Journal of Chemical Physics},
  volume = {120},
  number = {2},
  pages = {612--622},
  issn = {0021-9606},
  doi = {10.1063/1.1630303},
  urldate = {2025-08-19}
}

@article{rathboneResonantlyAmplifiedVibronic2001,
  title = {Resonantly Amplified Vibronic Symmetry Breaking},
  author = {Rathbone, G. J. and Poliakoff, E. D. and Bozek, John D. and Lucchese, R. R.},
  year = {2001},
  month = may,
  journal = {The Journal of Chemical Physics},
  volume = {114},
  number = {19},
  pages = {8240--8243},
  issn = {0021-9606},
  doi = {10.1063/1.1372334},
  urldate = {2025-08-19}
}

@article{reineckHighresolutionUVPhotoelectron1983,
  title = {High-Resolution {{UV}} Photoelectron Spectrum of {{CO2}}},
  author = {Reineck, I. and Nohre, C. and Maripuu, R. and Lodin, P. and {Al-Shamma}, S.H. and Veenhuizen, H. and Karlsson, L. and Siegbahn, K.},
  year = {1983},
  month = aug,
  journal = {Chemical Physics},
  volume = {78},
  number = {3},
  pages = {311--318},
  issn = {0301-0104},
  doi = {10.1016/0301-0104(83)85116-7}
}

@article{royPhotoionizationShapeResonance1984,
  title = {On the Photoionization Shape Resonance Associated to the {{{\~C}2 $\Sigma$2}}+ State of {{CO2}}+},
  author = {Roy, P. and Nenner, I. and Adam, M.Y. and Delwiche, J. and Franskin, M.J.Hubin and Lablanquie, P. and Roy, D.},
  year = {1984},
  month = sep,
  journal = {Chemical Physics Letters},
  volume = {109},
  number = {6},
  pages = {607--614},
  issn = {0009-2614},
  doi = {10.1016/0009-2614(84)85434-2}
}

@article{rubertiMultichannelDynamicsHigh2018,
  title = {Multi-Channel Dynamics in High Harmonic Generation of Aligned {{CO2}}: Ab Initio Analysis with Time-Dependent {{B-spline}} Algebraic Diagrammatic Construction},
  author = {Ruberti, M. and Decleva, P. and Averbukh, V.},
  year = {2018},
  journal = {Physical Chemistry Chemical Physics},
  volume = {20},
  number = {12},
  pages = {8311--8325},
  publisher = {The Royal Society of Chemistry},
  issn = {1463-9076},
  doi = {10.1039/C7CP07849H}
}

@article{sabbarCombiningAttosecondXUV2014,
  title = {Combining Attosecond {{XUV}} Pulses with Coincidence Spectroscopy},
  author = {Sabbar, M. and Heuser, S. and Boge, R. and Lucchini, M. and Gallmann, L. and Cirelli, C. and Keller, U.},
  year = {2014},
  month = oct,
  journal = {Review of Scientific Instruments},
  volume = {85},
  number = {10},
  pages = {103113},
  issn = {0034-6748},
  doi = {10.1063/1.4898017},
  urldate = {2025-09-09}
}

@article{sansoneElectronCorrelationReal2012,
  title = {Electron {{Correlation}} in {{Real Time}}},
  author = {Sansone, Giuseppe and Pfeifer, Thomas and Simeonidis, Konstantinos and Kuleff, Alexander I.},
  year = {2012},
  month = feb,
  journal = {ChemPhysChem},
  volume = {13},
  number = {3},
  pages = {661--680},
  publisher = {John Wiley \& Sons, Ltd},
  issn = {1439-4235},
  doi = {10.1002/cphc.201100528},
  urldate = {2024-02-22}
}

@article{siggelShapeResonanceEnhanced1993,
  title = {Shape--Resonance--Enhanced Continuum--Continuum Coupling in Photoionization of {{CO}} 2},
  author = {Siggel, M. R. F. and West, J. B. and Hayes, M. A. and Parr, A. C. and Dehmer, J. L. and Iga, I.},
  year = {1993},
  journal = {The Journal of chemical physics},
  volume = {99},
  number = {3},
  pages = {1556--1563},
  doi = {10.1063/1.465324}
}

@article{smirnovaHighHarmonicInterferometry2009,
  title = {High Harmonic Interferometry of Multi-Electron Dynamics in Molecules},
  author = {Smirnova, Olga and Mairesse, Yann and Patchkovskii, Serguei and Dudovich, Nirit and Villeneuve, David and Corkum, Paul and Ivanov, Misha Yu.},
  year = {2009},
  month = aug,
  journal = {Nature},
  volume = {460},
  number = {7258},
  pages = {972--977},
  issn = {1476-4687},
  doi = {10.1038/nature08253}
}

@article{srinivasHighrepetitionRateAttosecond2022,
  title = {High-Repetition Rate Attosecond Beamline for Multi-Particle Coincidence Experiments},
  author = {Srinivas, Hemkumar and Shobeiry, Farshad and Bharti, Divya and Pfeifer, Thomas and Moshammer, Robert and Harth, Anne},
  year = {2022},
  month = apr,
  journal = {Optics Express},
  volume = {30},
  number = {8},
  pages = {13630--13646},
  publisher = {Optica Publishing Group},
  doi = {10.1364/OE.454553}
}

@article{timmersCoherentElectronHole2014a,
  title = {Coherent {{Electron Hole Dynamics Near}} a {{Conical Intersection}}},
  author = {Timmers, Henry and Li, Zheng and Shivaram, Niranjan and Santra, Robin and Vendrell, Oriol and Sandhu, Arvinder},
  year = {2014},
  month = sep,
  journal = {Physical Review Letters},
  volume = {113},
  number = {11},
  pages = {113003},
  publisher = {American Physical Society},
  doi = {10.1103/PhysRevLett.113.113003}
}

@article{varju_FrequencyChirpHarmonic2005,
  title = {Frequency Chirp of Harmonic and Attosecond Pulses},
  author = {Varj{\'u}, K. and Mairesse, Y. and Carr{\'e}, B. and Gaarde, M. B. and Johnsson, P. and Kazamias, S. and {L{\'o}pez-Martens}, R. and Mauritsson, J. and Schafer, K. J. and Balcou, {\relax PH}. and L'huillier, A. and Sali{\`e}res, P.},
  year = {2005},
  month = jan,
  journal = {Journal of Modern Optics},
  volume = {52},
  number = {2-3},
  pages = {379--394},
  publisher = {Taylor \& Francis},
  issn = {0950-0340},
  doi = {10.1080/09500340412331301542}
}

@article{vosOrientationdependentStereoWigner2018,
  title = {Orientation-Dependent Stereo {{Wigner}} Time Delay and Electron Localization in a Small Molecule},
  author = {Vos, J. and Cattaneo, L. and Patchkovskii, S. and Zimmermann, T. and Cirelli, C. and Lucchini, M. and Kheifets, A. and Landsman, A. S. and Keller, U.},
  year = {2018},
  month = jun,
  journal = {Science},
  volume = {360},
  number = {6395},
  pages = {1326--1330},
  publisher = {American Association for the Advancement of Science},
  doi = {10.1126/science.aao4731},
  urldate = {2024-02-22}
}

@article{yang1+1TwophotonDissociation2008,
  title = {The [1+1] Two-Photon Dissociation Spectra of {{CO2}}+ via {{{\~A}}}{\textbackslash}{{textgreekPu}},1/22({\textbackslash}textgreeku1{\textbackslash}textgreeku20){\textbackslash}{{leftarrow{\~X}}}{\textbackslash}{{textgreekPg}},1/22(000) Transitions // {{The}} 1+1 Two-Photon Dissociation Spectra of {{CO2}} + via {{A 2Pi}} u,12(Upsilon1upsilon20){\textbackslash}textless--{{X 2Pi}} g,12(000) Transitions},
  author = {Yang, Maoping and Zhang, Limin and Zhuang, Xiujuan and Lai, Likun and Yu, Shuqin},
  year = {2008},
  journal = {The Journal of Chemical Physics},
  volume = {128},
  number = {16},
  pages = {164308},
  issn = {00219606},
  doi = {10.1063/1.2905232}
}

@article{makosEntanglementPhotoionisationReveals2025,
  title = {Entanglement in Photoionisation Reveals the Effect of Ionic Coupling in Attosecond Time Delays},
  author = {Makos, Ioannis and Busto, David and Benda, Jakub and Ertel, Dominik and Merzuk, Barbara and Steiner, Benjamin and Frassetto, Fabio and Poletto, Luca and Schr{\"o}ter, Claus Dieter and Pfeifer, Thomas and Moshammer, Robert and Patchkovskii, Serguei and Ma{\v s}{\'i}n, Zden{\v e}k and Sansone, Giuseppe},
  year = {2025},
  month = sep,
  journal = {Nature Communications},
  volume = {16},
  number = {1},
  pages = {8554},
  issn = {2041-1723},
  doi = {10.1038/s41467-025-64182-8}
}

@article{wernerMolpro2020,
    author = {Werner, Hans-Joachim and Knowles, Peter J. and Manby, Frederick R. and Black, Joshua A. and Doll, Klaus and Heßelmann, Andreas and Kats, Daniel and Köhn, Andreas and Korona, Tatiana and Kreplin, David A. and Ma, Qianli and Miller, Thomas F., III and Mitrushchenkov, Alexander and Peterson, Kirk A. and Polyak, Iakov and Rauhut, Guntram and Sibaev, Marat},
    title = {{The Molpro quantum chemistry package}},
    journal = {The Journal of Chemical Physics},
    volume = {152},
    number = {14},
    pages = {144107},
    year = {2020},
    month = {04},
    doi = {10.1063/5.0005081},
}

@article{bendaAngularMomentumDependence2025,
  title = {Angular Momentum Dependence in Multiphoton Ionization and Attosecond Time Delays},
  author = {Benda, Jakub and Ma{\v s}{\'i}n, Zden{\v e}k and Palakkal, Sreelakshmi and L{\'e}pine, Franck and Nandi, Saikat and Loriot, Vincent},
  year = 2025,
  month = jan,
  journal = {Physical Review A},
  volume = {111},
  number = {1},
  pages = {013110},
  publisher = {American Physical Society},
  doi = {10.1103/PhysRevA.111.013110}
}

@article{ertelAnisotropyParametersTwoColor2024,
  title = {Anisotropy {{Parameters}} for {{Two-Color Photoionization Phases}} in {{Randomly Oriented Molecules}}: {{Theory}} and {{Experiment}} in {{Methane}} and {{Deuteromethane}}},
  author = {Ertel, Dominik and Busto, David and Makos, Ioannis and Schmoll, Marvin and Benda, Jakub and Bragheri, Francesca and Osellame, Roberto and Lindroth, Eva and Patchkovskii, Serguei and Ma{\v s}{\'i}n, Zden{\v e}k and Sansone, Giuseppe},
  year = 2024,
  month = mar,
  journal = {The Journal of Physical Chemistry A},
  volume = {128},
  number = {9},
  pages = {1685--1697},
  publisher = {American Chemical Society},
  issn = {1089-5639},
  doi = {10.1021/acs.jpca.3c06759}
}

@article{bustoProbingElectronicDecoherence2022,
  title = {Probing Electronic Decoherence with High-Resolution Attosecond Photoelectron Interferometry},
  author = {Busto, David and Laurell, Hugo and {Finkelstein-Shapiro}, Daniel and Alexandridi, Christiana and Isinger, Marcus and Nandi, Saikat and Squibb, Richard J. and Turconi, Margherita and Zhong, Shiyang and Arnold, Cord L. and Feifel, Raimund and Gisselbrecht, Mathieu and Sali{\`e}res, Pascal and Pullerits, T{\"o}nu and Mart{\'i}n, Fernando and Argenti, Luca and L'Huillier, Anne},
  year = 2022,
  month = jul,
  journal = {The European Physical Journal D},
  volume = {76},
  number = {7},
  pages = {112},
  issn = {1434-6079},
  doi = {10.1140/epjd/s10053-022-00438-y}
}

\end{document}


\title{Supplementary Material for Disentangling the Effect of Ionic Coupling and Multiple Interfering Terms in Attosecond Molecular Interferometry}

\author{Ioannis Makos$^{1}$, Jakub Benda$^{2}$, David Busto$^{1,3}$, Benjamin Steiner$^{1}$, Barbara Merzuk$^{1}$, Serguei Patchkovskii$^{4}$, Van-Hung Hoang$^{5}$, Uwe Thumm$^{5}$, Zden\v{e}k Ma\v{s}\'{i}n$^{2}$, Giuseppe Sansone$^{1,6}$}
 
\affiliation{$^{1}$ Institute of Physics, University of Freiburg, Hermann-Herder-Stra{\ss}e 3, 79104 Freiburg, Germany}
\affiliation{$^{2}$ Institute of Theoretical Physics, Faculty of Mathematics and Physics, Charles University, V~Hole\v{s}ovi\v{c}k\'ach 2, 180 00, Prague 8, Czech Republic}
\affiliation{$^{3}$ Department of Physics, Lund University, PO Box 118, SE-221 00 Lund, Sweden}
\affiliation{$^{4}$ Max Born Institute, Max-Born-Str. 2A, D-12489 Berlin, Germany}
\affiliation{$^{5}$ J. R. Macdonald Laboratory, Department of Physics, Kansas State University, Manhattan, Kansas 66506, USA}
\affiliation{$^{6}$ Freiburg Institute for Advanced Studies (FRIAS), University of Freiburg, Albertstrasse 19, 79104 Freiburg, Germany}
\email{corresponding author: giuseppe.sansone@physik.uni-freiburg.de}

\date{\today}

\maketitle
\tableofcontents
 
\setcounter{equation}{0}
\setcounter{figure}{0}
\setcounter{table}{0}
\setcounter{enumiv}{1}

\renewcommand{\thefigure}{S\arabic{figure}}
\renewcommand{\theHfigure}{suppfig.\arabic{figure}} 

\renewcommand{\thetable}{S\arabic{table}}
\renewcommand{\theHtable}{supptable.\arabic{table}}

\renewcommand{\theequation}{S\arabic{equation}}
\renewcommand{\theHequation}{suppeq.\arabic{equation}}

\setcounter{page}{1}
\setcounter{section}{1}
\subsection*{Additional experimental information}
\begin{figure} [h] 
    \centering
    \includegraphics[width=0.5\textwidth]{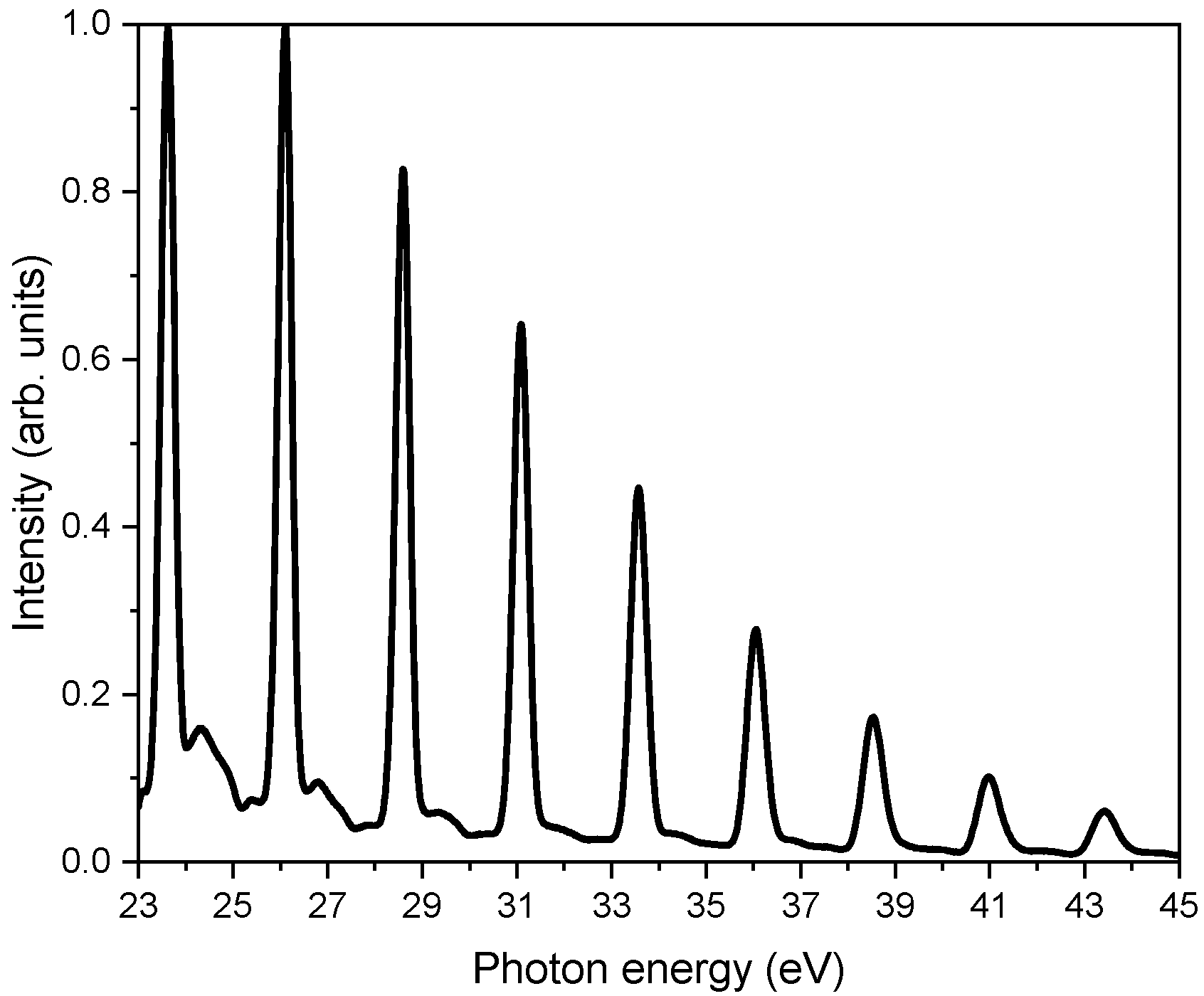}
\caption{Typical XUV spectrum generated in krypton and used for the photoionization of the CO$_2$-Ar mixture.}
\label{Fig1SM}
\end{figure}

\begin{figure*}   
    \centering
    \includegraphics[width=0.8\linewidth]{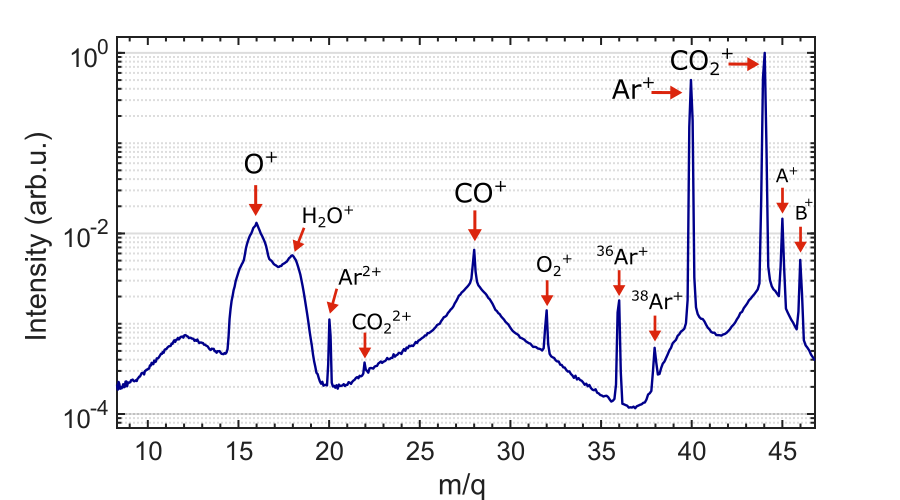}
\caption{Time-of-flight mass over charge (m/q) spectra acquired in the CO$_2$-Ar mixture. $A^+ \equiv ~^{13}\mathrm{CO}_2^+$ ; $B^+ \equiv \mathrm{CO}^{18}\mathrm{O}^+$}
\label{Fig2SM}
\end{figure*}

The XUV spectrum used in the experiment is presented in Fig.~\ref{Fig1SM}.
Figure~\ref{Fig2SM} presents the typical time-of-flight spectrum measured for the XUV-only case in a CO$_2$-Ar gas mixture. 


\clearpage
\subsection*{Analysis of XUV-only spectra}
The XUV-only photoelectron spectra measured in coincidence with the ions Ar$^+$, O$^+$ and CO$^+$ and integrated within the range~$0^{\circ}\leq\theta\leq30^{\circ}$ ($\theta=0^{\circ}$ indicates the polarization direction of the XUV and IR pulses) were fitted using a multiple-gaussian fitting function according to the relation:
\begin{equation}
    S(E)=\Sigma_{i=1}^{N}A_i\exp{\left[-(E-E_{ci})^2/(\Delta E_i)^2\right]}+B_i,
\end{equation}
where $A_i$, $E_{ci}$, and $\Delta E_i$
indicate the amplitude, the central energy and the width of the $i^{th}$-Gaussian function, respectively. The terms $B_i$ are constants that were introduced to take into account small background (on the order of 10$^{-3}$). 
The results of the fitting procedure are reported in Table~\ref{TableS1}.

\begin{table}[h!]
\centering
\caption{Amplitude, central energies and widths of the photoelectron peaks for the XUV-only case measured in coincidence with the ions Ar$^+$, O$^+$ and CO$^+$. I$_p$ indicates the ionization potential of Argon. $E_1$ and $E_3$ indicate the energy of the ground state and excited vibrational state of the $C^2\Sigma^+_g$ state with quantum numbers $(0,0,0)$ and $(0,2,0)$, respectively.}
\medskip

\begin{tabular}{|c|c|c|c|c|}
\hline
& \multicolumn{4}{|c|}{Ar$^+$ ($I_p$=15.79~eV)}  \\
\hline\hline
$i$ &HO ($2q+1$) & $A_i$& $E_{ci}$ (eV)  & $\Delta E_{i}$ (eV)\\
\hline\hline
1 & 13 & $0.058\pm0.019$ & $ 0.143 \pm 0.027$    &  $0.165 \pm 0.043$ \\
2 & 15 & $0.350\pm0.013$&$ 2.522 \pm 0.005$    &  $0.276 \pm 0.007$ \\
3 & 17 & $0.695\pm,0.012$&$ 4.922 \pm 0.002$    &  $0.296 \pm 0.004$ \\	
4 & 19 & $0.964\pm0.012$&$ 7.357 \pm 0.002$    &  $0.320 \pm 0.003 $ \\	
5 & 21 & $0.662\pm0.011$&$ 9.787 \pm 0.003$    &  $0.381 \pm 0.005$ \\	
6 & 23 & $0.321\pm0.009$&$ 12.228 \pm 0.007$   &  $0.461 \pm 0.011$ \\	
7 & 25 & $0.104\pm0.009$&$ 14.662 \pm 0.026$   &  $0.637  \pm 0.039$ \\	
8 & 27 & $0.025\pm0.008$&$ 17.158 \pm 0.131$   &  $0.894  \pm 0.197$ \\	
\hline
\hline
\end{tabular}

\begin{tabular}{|c|c|c|c|c|}
\hline
& \multicolumn{4}{|c|}{O$^+$ ($E_1$=19.3911~eV)} \\
\hline\hline
$i$ &HO ($2q+1$) & $A_i$& $E_{ci}$ (eV) & $\Delta E_{i}$ (eV)\\
\hline\hline
1 & 13 & -- & -- & --    \\
2 & 15 & -- & -- & --    \\
3 & 17 & $0.392 \pm 0.005$ &    $1.392 \pm 0.005$  & $0.199 \pm 0.007$ \\	
4 & 19  & $0.824\pm0.017$&   $3.790 \pm 0.003$     & $0.242 \pm 0.004$ \\	
5 & 21  &  $0.881\pm0.016$&    $6.217 \pm 0.003$  & $0.279 \pm 0.004$ \\	
6 & 23 &  $0.635\pm0.015$&    $8.627 \pm 0.004$   & $0.316 \pm 0.005$ \\	
7 & 25 &  $0.399\pm0.014$&   $11.052 \pm 0.007$  & $0.376 \pm 0.009$ \\	
8 & 27 &  $0.185\pm0.012$&   $13.505 \pm 0.016$  & $0.479 \pm 0.023$ \\	
\hline
\hline
\end{tabular}

\begin{tabular}{|c|c|c|c|c|}
\hline
& \multicolumn{4}{|c|}{CO$^+$ ($E_3$=19.5429~eV)}   \\
\hline\hline
$i$ &HO ($2q+1$) & $A_i$& $E_{ci}$ (eV) & $\Delta E_{i}$ (eV)\\
\hline\hline
1 & 13 & -- & -- & --   \\
2 & 15 & -- & -- & --   \\
3 & 17 & $0.42\pm 0.035$ & $1.22 \pm 0.02$  & $0.43 \pm 0.03$ \\	
4 & 19 & $0.74\pm0.04$  & $3.58 \pm 0.01$  & $0.38 \pm 0.01$ \\	
5 & 21 & $0.75\pm0.04$& $5.99 \pm 0.01$  & $0.39 \pm 0.01$ \\	
6 & 23 & $0.67\pm0.04$& $8.46 \pm 0.01$  & $0.32 \pm 0.01$ \\	
7 & 25 & $0.32\pm0.04$& $10.92 \pm 0.02$ & $0.38 \pm 0.03$ \\	
8 & 27 & $0.12\pm0.04$& $13.39 \pm 0.06$ & $0.39 \pm 0.08$ \\	
\hline
\hline
\end{tabular}\label{TableS1}
\end{table}

\subsection*{Assignment of photoelectron peaks measured in coincidence with O$^+$ and CO$^+$ ions}
\begin{figure} [h] 
    \centering
    \includegraphics[width=0.8\textwidth]{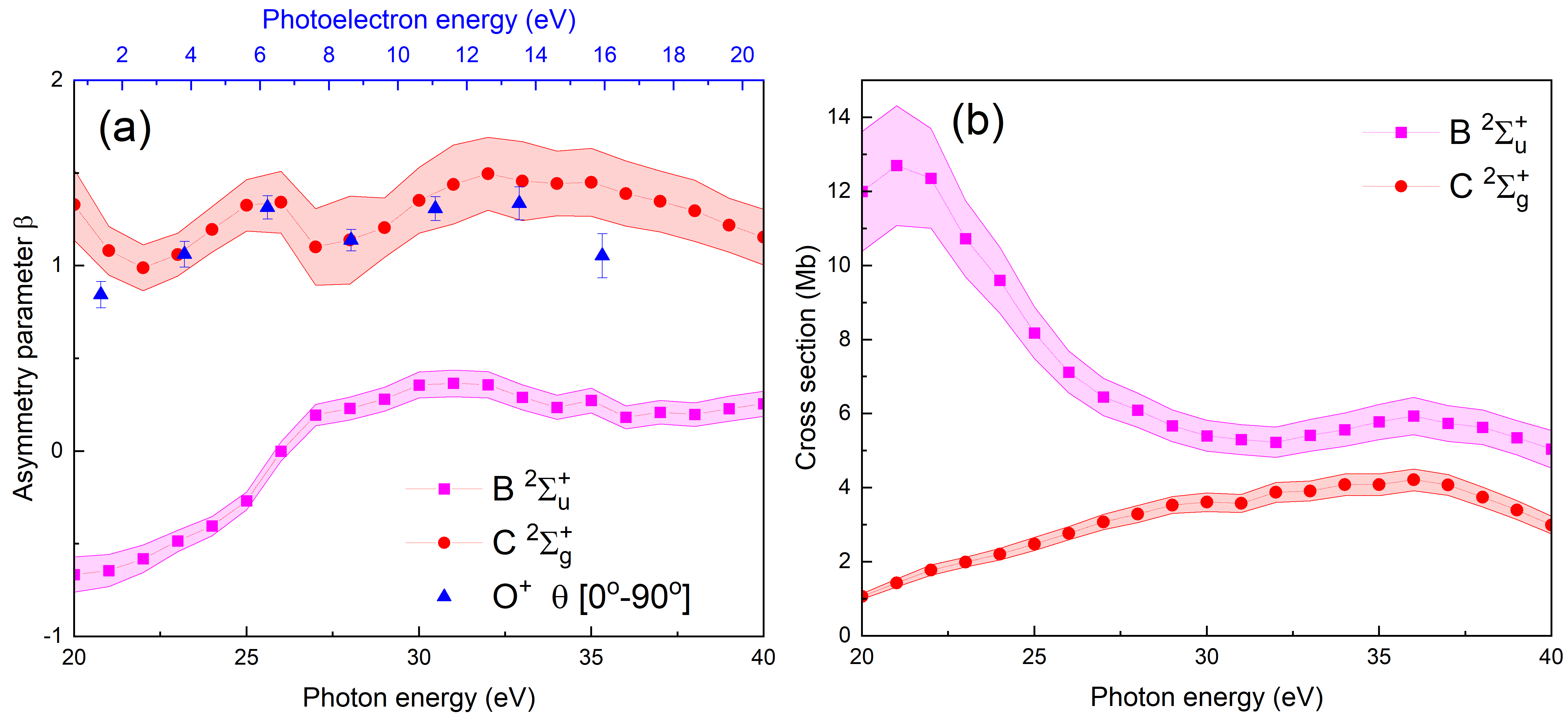}
\caption{Asymmetry parameters $\beta$ (a) and photoionization cross sections (b) for the $B^2\Sigma^+_u$ (magenta squares) and $C^2\Sigma^+_g$ (red circles) states of the neutral CO$_2$ molecule. The shaded areas are the corresponding error bars. Data from Refs.~\citenum{siggelShapeResonanceEnhanced1993, brionPartialOscillatorStrengths1978}. Asymmetry parameter $\beta$ for the photoelectron angular distributions measured in coincidence with the O$^+$ ions (blue triangles).}
\label{Fig3SM}
\end{figure}

The asymmetry parameters $\beta$ and the photoionization cross sections of the neutral CO$_2$ molecules and resulting in the final $B^2\Sigma^+_u$ and $C^2\Sigma^+_g$ ionic state are presented in Fig.~\ref{Fig3SM}(a) and \ref{Fig3SM}(b), respectively. The asymmetry parameters derived from the photoelectron angular distributions measured in coincidence with the O$^+$ (blue triangles) well match those of the $C^2\Sigma^+_g$ state, supporting the conclusion that the photoelectron measured in coincidence with this ion are due to the population of vibrational level(s) (the ground vibrational level) of this electronic state.

\begin{figure} [h] 
    \centering
    \includegraphics[width=0.8\textwidth]{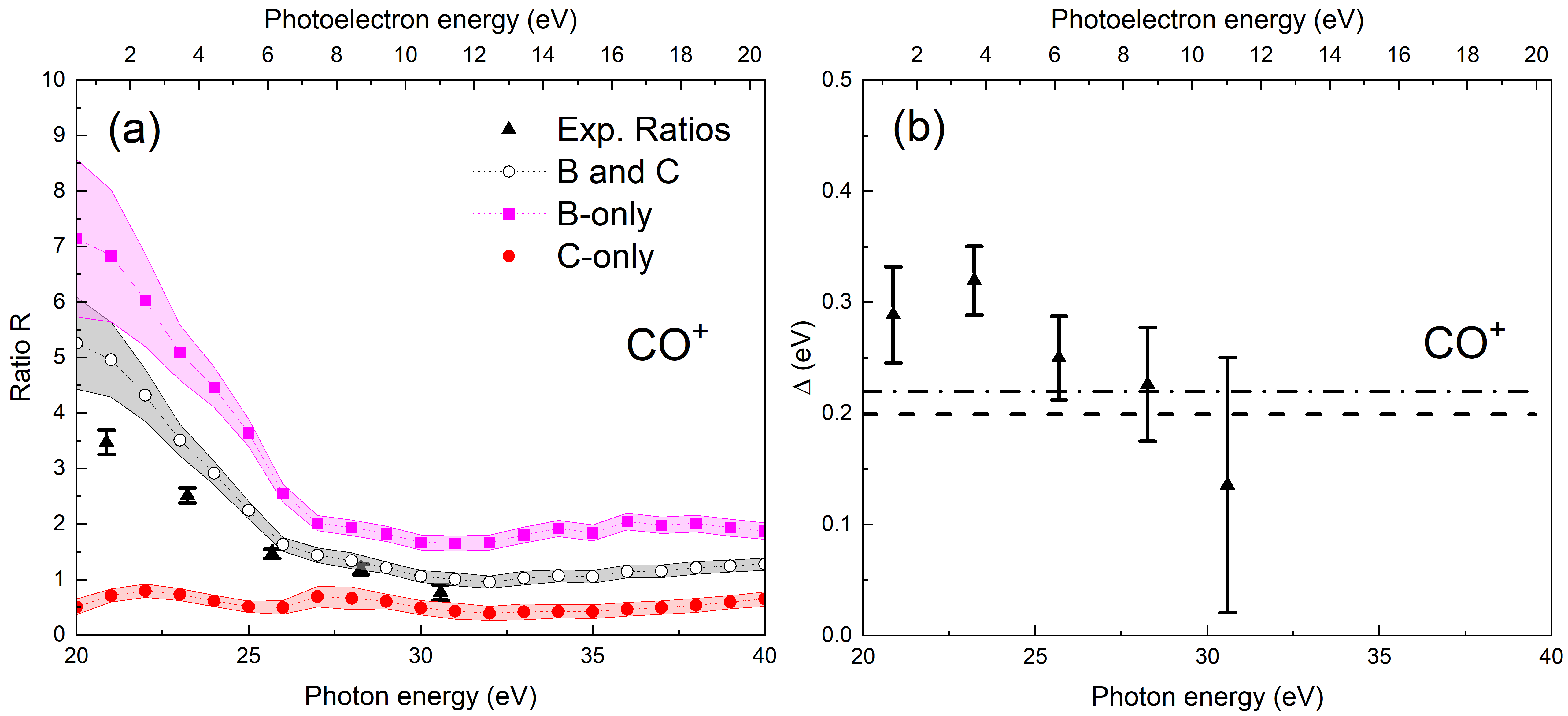}
\caption{(a) Ratio of the areas of the Gaussian fits of the photoelectron peaks measured in coincidence with the CO$^+$ ions emitted perpendicular and parallel to the laser polarization direction (black triangles) according to Eq.~\ref{Eqratio}. Ratio of the angle-integrated cross sections for the $B^2\Sigma^+_u$ and $C^2\Sigma^+_g$ states over the intervals [$70^{\circ}-90^{\circ}$] and [$0^{\circ}-30^{\circ}$] respectively using the experimental data available in Ref. \citenum{siggelShapeResonanceEnhanced1993} for the coefficients $(\alpha_B,\alpha_c)=(1,0)$ (magenta squares), $(0,1)$ (red circles), and $(1,1,)$ (black open circles). The shaded areas represent the error bars derived from the errors in the cross sections and asymmetry parameters available from the same reference. (b) Difference $\Delta$ of the central photoelectron peaks energies obtained from the Gaussian fit for the photoelectron spectra acquired in coincidence with the CO$^+$ ions and measured parallel and perpendicular to the laser polarization direction. The dashed and dash-dotted lines are the differences $E_9-E_3$ and $E_9-E_4$, respectively.}
\label{Fig3SMbis}
\end{figure}

The evolution of the photoelectron spectra as a function of the integration angle, together with their  spectral width (see Fig.~\ref{Fig2}(e)), indicates that multiple electronic states contribute to the dissociation pathway that leads to the emission of CO$^+$ ions. The pronounced enhancement of the signal intensity in the perpendicular direction for the lowest harmonic orders, combined with the marked drop of the photoelectron signal beyond approximately 6 eV in this geometry (see Fig.~\ref{Fig3}(b)), suggests that the level at $E_9 = 19.7486$ eV, which is strongly vibronically coupled to the $B^2\Sigma^+_u$ state, makes a significant contribution to the signal recorded in coincidence with the CO$^+$ ions. To quantify this contribution, we model the total signal measured in the perpendicular ($70^{\circ}\leq\theta\leq90^{\circ}$) and parallel ($0^{\circ}\leq\theta\leq30^{\circ}$) directions as arising from the combined contributions of photoelectrons initially associated with the $B^2\Sigma^+_u$ and $C^2\Sigma^+_g$ states. Following an approach similar to that described in Ref.~\citenum{rathboneModespecificPhotoelectronScattering2004a}, Fig.~\ref{Fig3SMbis}(a) displays the ratio of the areas of the photoelectron peaks obtained from Gaussian fits in the perpendicular and parallel directions (black triangles), along with the integrals of the differential cross sections $d\sigma_B/d\Omega$ and $d\sigma_C/d\Omega$ for the $B^2\Sigma^+_u$ and $C^2\Sigma^+_g$ PESs, respectively, along the perpendicular and parallel directions, according to the expression
\begin{equation}\label{Eqratio}
  R=\frac{\int_{70^{\circ}}^{90^{\circ}}\{\alpha_B\sigma_B\left[1+\beta_BP_2(\cos\theta)\right]+\alpha_C\sigma_C\left[1+\beta_CP_2(\cos\theta)\right]\}\sin\theta\mathrm{d}\theta}{\int_{0^{\circ}}^{30^{\circ}}\{\alpha_B\sigma_B\left[1+\beta_BP_2(\cos\theta)\right]+\alpha_C\sigma_C\left[1+\beta_CP_2(\cos\theta)\right]\}\sin\theta\mathrm{d}\theta}.
\end{equation}
The symbols $\beta_B$ and $\beta_C$ indicate the energy-dependent asymmetry parameters and the coefficients $\alpha_B$ and $\alpha_C$ account for the relative contribution of photoelectrons emitted from the the $B^2\Sigma^+_u$ and $C^2\Sigma^+_g$ states. 
We observe that the experimental data agree neither with the evolution of the ratio predicted when only the $C^2\Sigma^+_g$ state is considered ($\alpha_B=0$; red points and shaded area) nor with that predicted when only the $B^2\Sigma^+_u$ state is taken into account ($\alpha_C=0$; magenta points and shaded area). A satisfactory agreement with the experimental points is obtained only when both contributions are included, i.e., for $\alpha_B=\alpha_C=1$.
The agreement supports the conclusion of strong vibronic coupling with the $B^2\Sigma^+_u$ PES of the state contributing to the photoelectrons emitted in the perpendicular direction.

Furthermore, we observe that the photoelectron peaks in the perpendicular direction appear at lower energies than that of the photoelectrons emitted in the parallel direction. The difference ($\Delta$) obtained from the Gaussian fits of the spectra is presented in Fig.~\ref{Fig3SMbis}(b), showing values of approximately $0.2-0.4$~eV. This is in good agreement with the energy differences between the state at energy $E_9$ and the excited vibrational levels $E_3$ and $E_4$ of the $C^2\Sigma^+_g$ PES (see the dashed and dashed-dotted lines in Fig.~\ref{Fig3SMbis}(b)), which provide a significant contribution to the photoelectrons emitted in the parallel direction.

\clearpage

\subsection*{Energy resolved analysis of RABBIT traces}
RABBIT traces as function of the photoelectron energy $E$ and the relative delay $\Delta t$ between the XUV and IR fields were selected in coincidence with the ionic fragments O$^+$, CO$^+$, and Ar$^+$. The amplitude of the oscillating component at frequency $2\omega_{\mathrm{IR}}$ ($A_2$), and the phase of the oscillation at frequency $2\omega_{\mathrm{IR}}$ ($\Phi$) were extracted by using a Fourier Transform algorithm. This analysis was performed for RABBIT traces obtained by integrating over different intervals of the emission angle $\theta$ of the photoelectron.

\begin{figure} [h] 
    \centering
    \includegraphics[width=0.9\linewidth]{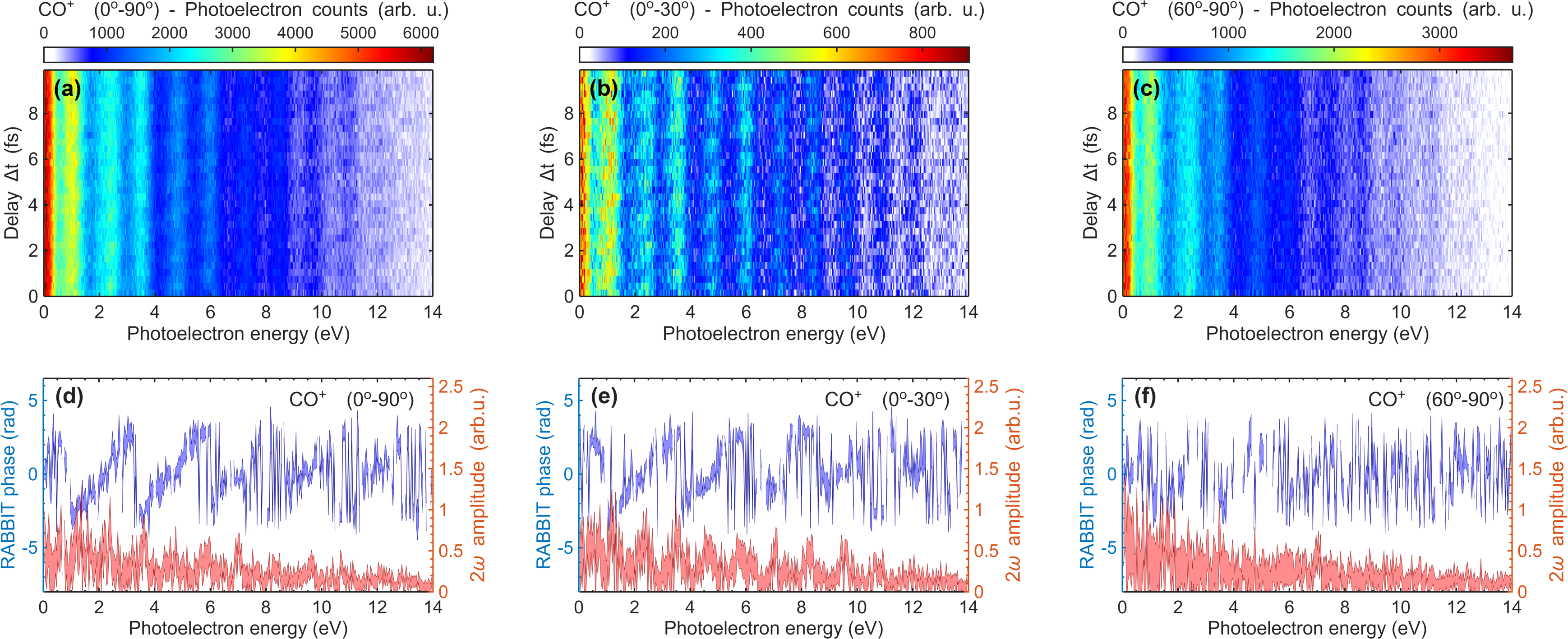}
\caption{RABBIT trace measured in coincidence with the ion CO$^+$ and obtained integrating the photoelectron angular distribution in the intervals $[0^{\circ}-90^{\circ}]$ (a), $[0^{\circ}-30^{\circ}]$ (b), and $[60^{\circ}-90^{\circ}]$ (c). (d-f) Oscillating amplitude $A_2$ (red lines) and phase $\Phi$ (blue lines) from Eq.~\eqref{Eq_sideband} as a function of the photoelectron energies derived by an energy resolved fit of the RABBIT traces presented in panels a-c, respectively. The shaded areas represent the error bars estimated a single standard deviation obtained from the fit. Energy bins with phase uncertainties greater than $\pi/2$ are omitted from the plots.}
\label{Fig4SM}
\end{figure}

In Fig.~\ref{Fig4SM}, we report the RABBIT traces and the Fourier analysis for the photoelectron spectra measured in coincidence with the CO$^+$ ionic fragment for different integration angles. The corresponding data for the photoelectron measured in coincidence with the O$^+$ ion is presented in Fig.~\ref{Fig6} of the main manuscript.

\begin{figure*}   
\centering 
    \includegraphics[width=0.9\linewidth]{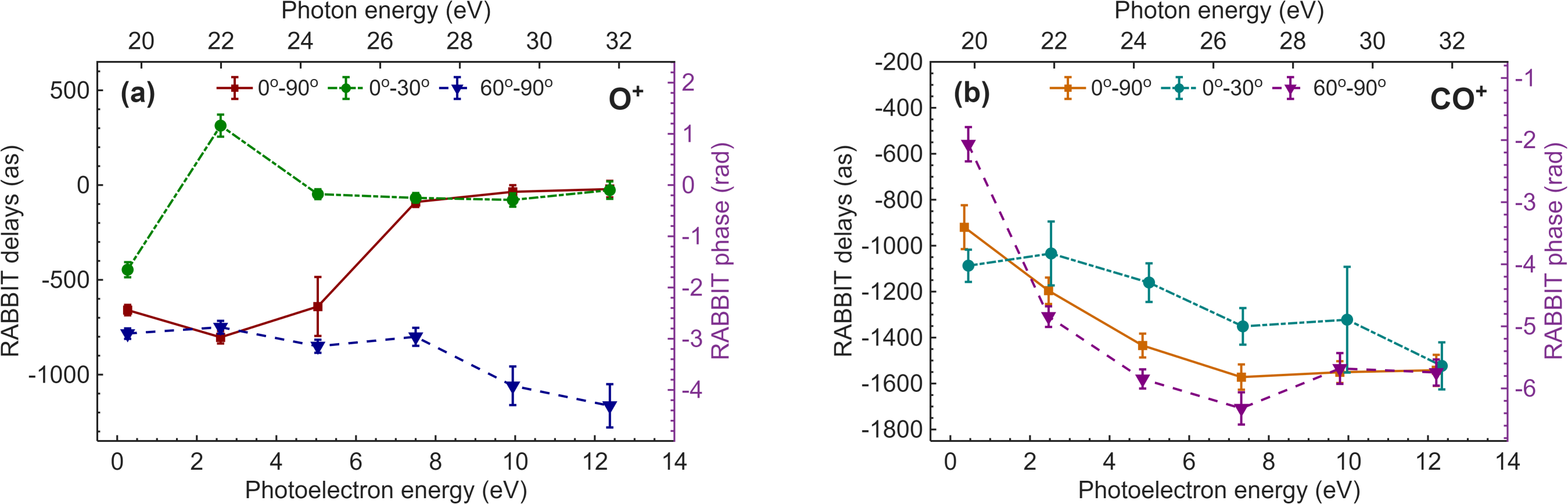}
\caption{
Measured (circles and error bar connected by solid line) photoionization delays obtained integrating the photoelectron angular distributions measured in coincidence with the ions O$^+$ (a) and CO$^+$ (b) over the intervals $0^{\circ}-90^{\circ}$ (red experimental points), $0^{\circ}-30^{\circ}$ (green points), and $60^{\circ}-90^{\circ}$ (blue points). The delays were corrected for the photoionization time delays measured in coincidence with argon to eliminate the contribution of the attochirp. The photoionization time delays measured in coincidence with O$^+$ are equal to those presented in Fig.~\ref{Fig8}.}
\label{Fig5SM}
\end{figure*}

The evolution of the delay for the sideband oscillations measured in coincidence with the CO$^+$ ions differs greatly from that measured in coincidence with the O$^+$ ions, as shown in Fig.~\ref{Fig5SM} (a) and (b). In particular, the the delays measured for the photoelectron associated to the CO$^+$ fragment do not exhibit the significant variation observed for those associated to the O$^+$ ions, when integrating over the entire angle range or in the parallel direction ($0^{\circ}\leq\theta\leq30^{\circ}$). Since the simulations of the RABBIT delay cannot be directly linked to the dissociation mechanism and the resulting ion, we can only speculate about a possible qualitative explanation for this different behavior. 

The evolution of the delays measured in coincidence with CO$^+$ ions for photoelectrons integrated over all emission angles is close to the behavior observed in the perpendicular direction for photoelectrons detected in coincidence with O$^+$ ions. This suggests that the interference between paths 2 and 3 predominantly determines the overall signal recorded in coincidence with CO$^+$ ions.
A possible explanation is that the relative populations of the cationic states leading to CO$^+$ emission via paths 1 and 2, and via path 3, differ from those of the states responsible for O$^+$ emission. This could be due to the fact that the vibrational energy levels $E_3$ and $E_4$, which produce CO$^+$ fragments, are only weakly populated compared to the ground level of the $C^2\Sigma^+_g$ state ($E_1$), which yields O$^+$ fragments. Consequently, for O$^+$ ions the contributions from paths 1 and 2 are expected to be stronger than for CO$^+$. Conversely, the coupling to the $B^2\Sigma^+_u$ state appears to play a more significant role in the formation of CO$^+$ ions than in the case of O$^+$ ions, as already suggested in the XUV-only measurements (see Fig.~\ref{Fig1}, Fig.~\ref{Fig3}b, and Fig.~\ref{Fig4}d and the related discussion of the energy level $E_9$). This may account for why the contribution from path 3 seems to be more pronounced for the RABBIT delays measured in coincidence with CO$^+$ ions.

\clearpage
\subsection*{Calculation of Franck-Condon factors and vibrational energy levels}
The values reported in Table~\ref{table1} are based on adiabatic potential energy surfaces for neutral and singly charged molecules that we calculated by employing the MCSCF quantum-chemistry code GAMESS~\cite{2020_Barca_GAMESS_JCP}. We performed these fixed-nuclei molecular electronic-structure calculations on an equidistant 3D numerical grid with spacings of 0.02~a.u. in the C-O distances and $2^\circ$ increments in the bend angle for frozen K-shell electrons and 15 active electrons in 12 active orbitals. Our calculation thereby includes all atomic valence orbitals in the active space. The employed cc-pVTZ basis set, contracted to the orbitals $\{4s,3p,2d,1f\}$, provides 30 atomic orbitals for each C and O atom and 188760 configuration-state functions. We normalized our energies to the energy of 19.3911~eV for ionization into the vibrational ground state of the cationic adiabatic $C^2\Sigma^+_g$ state given in Liu {\it et al.}.~\cite{2003_Liu_CO2_JCP}. 
This energy slightly exceeds the calculated value of 19.388~eV reported  by 
Bombach {\it et al.}~\cite{bombachBranchingRatiosPartition1983}.

\subsection*{Interpretation of the variation of the phase for the RABBIT traces measured in coincidence with the ion $O^+$}

In the traditional formulation of the asymptotic approximation to the laser-assisted photoionization (RABBIT), the two-photon
amplitude \(T^{(2)}\) corresponding to a free-free transition can be approximately factorized as
\begin{equation}
    T^{(2)}(\bm{k}, \bm{\varepsilon}) \approx A_{\kappa k} [ \hat{\bm{k}} \cdot \hat{\bm{\varepsilon}} ]
    [ \hat{\bm{\varepsilon}} \cdot \bm{d}^{(1)}(\kappa\hat{\bm{k}}) ] \,.
    \label{eq:T2asy}
\end{equation}
In the main text, \(A_{\kappa k}\) was generically labeled as \(A_{cc}\).
Here \(\bm{d}^{(1)}\) is the one-photon ionization dipole, \(\bm{\varepsilon}\) is the common polarizations of the fields,
\(\bm{k}\) is the momentum vector of the detected photoelectron and \(\kappa\) is the intermediate momentum size in between the
XUV and IR interaction steps. At very high energies, one can write
\begin{equation}
    A_{\kappa k} \approx \frac{\mathrm{i}k}{\omega_\text{IR}^2} \,.
    \label{eq:Akasy}
\end{equation}
In partial wave expansion, Eq.~\eqref{eq:T2asy} can be written equivalently as
\begin{equation}
    T^{(2)}(\bm{k}, \bm{\varepsilon}) \approx A_{\kappa k} \sum_{lm\lambda\mu} Y_{lm}(\hat{\bm{k}})
    \langle lm | \hat{\bm{n}} \cdot \bm{\varepsilon} | \lambda\mu \rangle
    [ \hat{\bm{\varepsilon}} \cdot \bm{d}_{\lambda\mu}^{(1)}(\kappa) ] \,.
    \label{eq:T2asy-pw}
\end{equation}
Here \((\lambda,\mu)\) are the angular momentum quantum numbers of a photoelectron before the IR fields interaction, while \((l,m)\) are angular momentum quantum numbers after the interaction, in the final state.

\begin{figure}[htbp]
    \centering
    \includegraphics[width=0.6\textwidth]{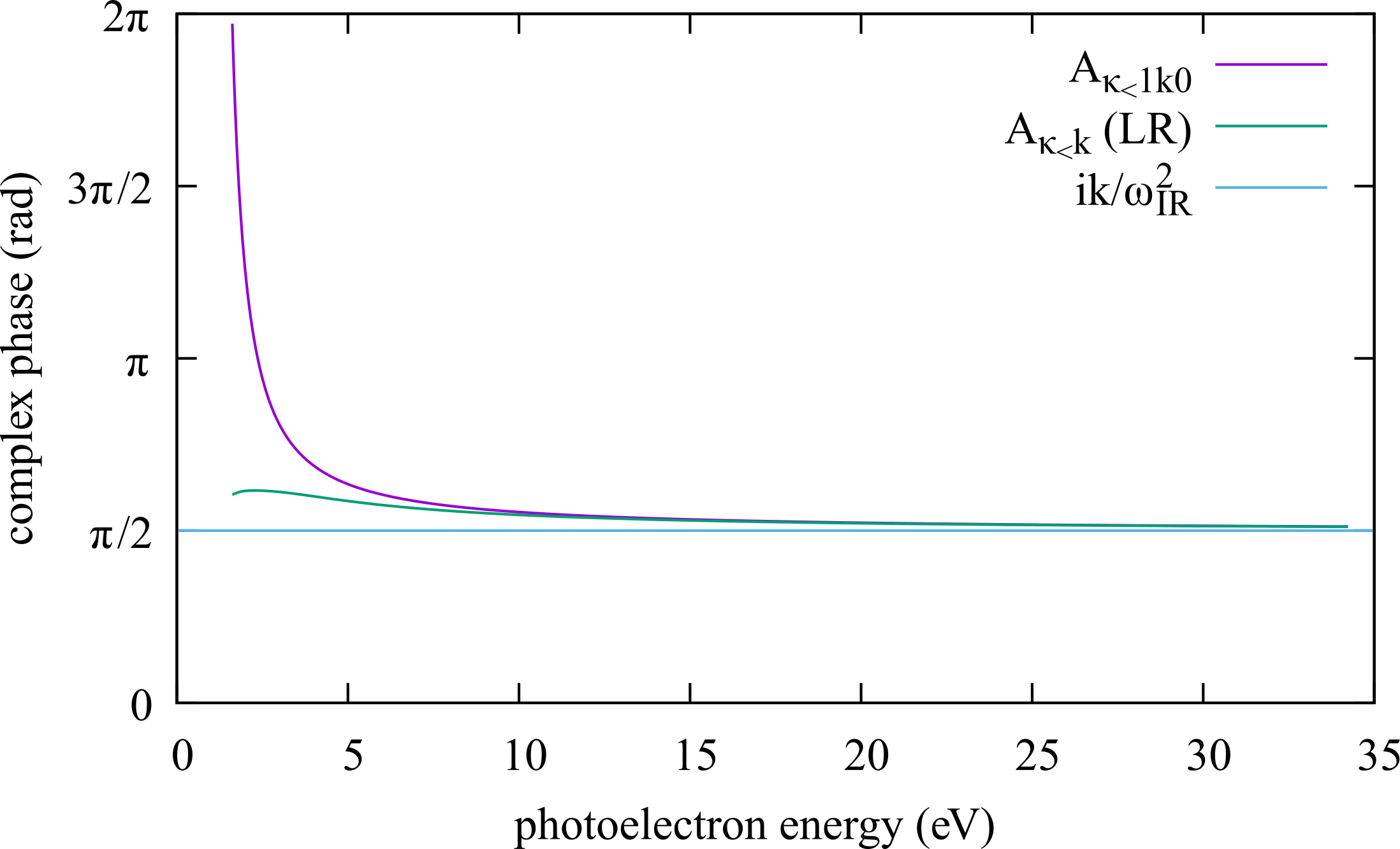}
    \caption{Complex phases of the continuum-continuum correction factors.}
    \label{fig:Akk}
\end{figure}

As discussed in~\cite{bendaAngularMomentumDependence2025}, a significantly more accurate form of the asymptotic approximation can be achieved by using
partial-wave-specific radial factors \(A_{\kappa\lambda kl}\) as
\begin{equation}
    T^{(2)}(\bm{k}, \bm{\varepsilon}) \approx \sum_{lm\lambda\mu} A_{\kappa \lambda k l} Y_{lm}(\hat{\bm{k}})
    \langle lm | \hat{\bm{n}} \cdot \bm{\varepsilon} | \lambda\mu \rangle
    [ \hat{\bm{\varepsilon}} \cdot \bm{d}_{\lambda\mu}^{(1)}(\kappa) ] \,.
    \label{eq:T2pwasy-pw}
\end{equation}
(see Eq.~(16) of~\cite{bendaAngularMomentumDependence2025} for definition of \(A_{\kappa\lambda kl}\)).
Below, we use all these approximations at various points. The different kinds of continuum-continuum phases are plotted
in Fig.~\ref{fig:Akk}.

A similar factorization can be done for two-photon amplitude where it is the residual ion that interacts with the IR field rather
than the photoelectron. Then
\begin{equation}
    T^{(2)}(\bm{\kappa}, \bm{\varepsilon}) \approx A_{\kappa k}^\text{ion} [\bm{D} \cdot \hat{\bm{\varepsilon}}]
    [ \hat{\bm{\varepsilon}} \cdot \tilde{\bm{d}}^{(1)}(\kappa\hat{\bm{k}}) ] \,.
\end{equation}
The one-photon ionization dipole is now understood for a different state of the residual ion (i.e., the intermediate state) than
which is being detected. The dipole transition element between the intermediate residual ion state and the final residual ion
state is denoted as \(\bm{D}\). The radial integral, labeled in the main text also as \(A_{ii}\), can be simplified with a good accuracy as
\begin{equation}
    A_{\kappa k}^\text{ion} \approx \frac{1}{\Delta_{BC} - \omega_\text{IR}} \,.
\end{equation}
Assuming that we are discussing only the XUV+IR process, \(\Delta_{BC}\) is the amount of energy that the IR field has to supply in order
to excite the intermediate state of the residual ion to the final state.

\subsubsection*{Competing pathways}

The significant jump in the RABBIT delay associated with the $C^2\Sigma^+_g$ state of CO\textsubscript{2}\textsuperscript{+}
is caused by destructive interference between the two contributing XUV\(+\)IR pathways:
one where the IR photon is absorbed by the residual ion (``ii''), and one where the IR photon is absorbed by the
photoelectron (``cc''). In the molecular frame we have for the corresponding RABBIT phase:
\begin{align}
    \phi_C^\text{mol}
    &= \arg T_{\text{XUV--IR},C}^{(2)} T_{\text{XUV+IR},C}^{(2)*} \\
    &= \arg T_{\text{XUV--IR},C}^{(2)} - \arg \left[
        T_{\text{XUV+IR},C}^{(2)\text{cc}} + T_{\text{XUV+IR},C}^{(2)\text{ii}}
    \right] \\
    &\approx \arg T_{\text{XUV--IR},C}^{(2)} - \arg \left[
        \frac{\mathrm{i}\bm{k}\cdot\hat{\bm{\varepsilon}}_\text{IR}}{\omega_\text{IR}^2}
            [\hat{\bm{\varepsilon}}_\text{XUV} \cdot \bm{d}_C^{(1)}] +
        \frac{\bm{D}_{BC} \cdot \hat{\bm{\varepsilon}}_\text{IR}}{\Delta_{BC} - \omega_\text{IR}}
            [\hat{\bm{\varepsilon}}_\text{XUV} \cdot \bm{d}_B^{(1)}]
    \right].
    \label{eq:phi-mol}
\end{align}
While the ``ii'' pathway dominates at very low energies, the ``cc'' pathway is dominant at high energies.
The switching is where the destructive interference occurs. Note that the last equation~\eqref{eq:phi-mol}
is a high-energy limit and its accuracy close to the threshold is very limited. It is more accurate to use the partial-wave
asymptotic approximation as developed in~\cite{bendaAngularMomentumDependence2025}.

\begin{figure}[htbp]
    \centering
    \includegraphics[width=0.4\textwidth]{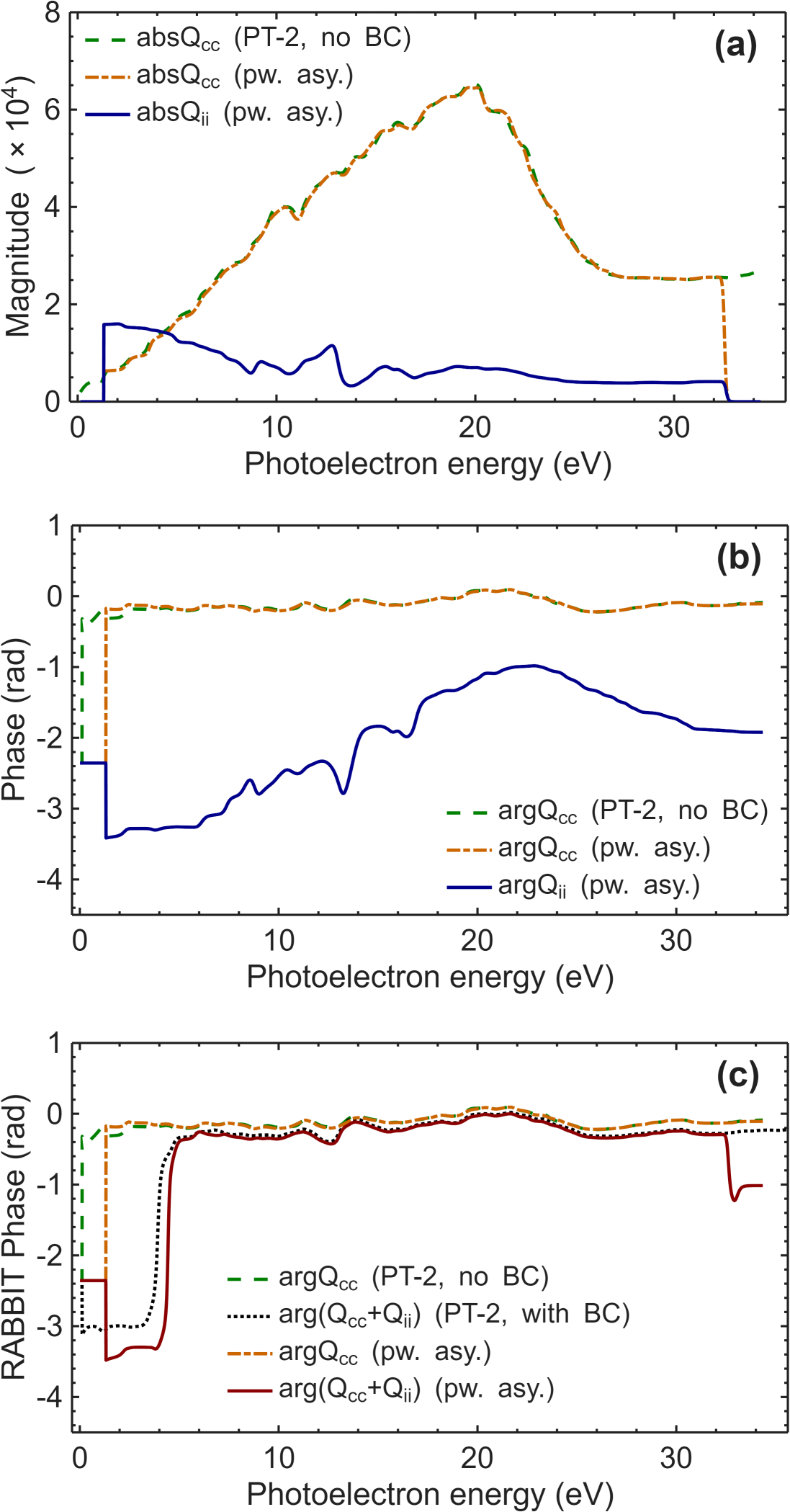}
    \caption{Magnitudes and phases of the orientation-averaged and emission-integrated free-free and ion-ion transition
    contributions to the complete RABBIT interference term, \(Q = Q_{cc} + Q_{ii}\), as introduced earlier. The labels
    ``pw. asy.'' refer to the approximation done in Eqs.~\eqref{eq:Qffpwasy}, \eqref{eq:Qiipwasy},
    while ``PT-2'' means the full two-photon
    R-matrix calculation in the second order of perturbation theory.}
    \label{fig:pwasy}
\end{figure}

Moreover, in the experiments we measure photoelectron angular distributions emitted from the
orientationally averaged sample of molecules. Denoting integration over emission directions and averaging over orientations
by angle brackets, we have:
\begin{align}
    \phi_C^\text{lab}
    &= \arg \left\langle T_{\text{XUV--IR},C}^{(2)} T_{\text{XUV+IR},C}^{(2)*} \right\rangle \\
    &= \arg \left[
        \left\langle T_{\text{XUV--IR},C}^{(2)} T_{\text{XUV+IR},C}^{(2)\text{cc}*} \right\rangle
        + \left\langle T_{\text{XUV--IR},C}^{(2)} T_{\text{XUV+IR},C}^{(2)\text{ii}*} \right\rangle
    \right] \\
    &= \arg [Q_{cc} + Q_{ii}] .
    \label{eq:phi-lab}
\end{align}
The two contributing terms are
\begin{align}
    Q_{cc} &= \left\langle T_{\text{XUV--IR},C}^{(2)} T_{\text{XUV+IR},C}^{(2)\text{cc}*} \right\rangle
    \approx \sum_{\substack{abcd\\ \lambda\mu l m}}
        I_{abcd} A_{\kappa\lambda kl}^{*} T_{\text{XUV--IR},C;lm,ab}^{(2)} d_{C;\lambda\mu,c}^{(1)*}
        \langle \lambda \mu | \hat{n}_d | lm \rangle \,, \label{eq:Qffpwasy} \\
    Q_{ii} &= \left\langle T_{\text{XUV--IR},C}^{(2)} T_{\text{XUV+IR},C}^{(2)\text{ii}*} \right\rangle
    \approx \frac{1}{\Delta_{BC} - \omega_\text{IR}} \sum_{\substack{abcd\\ l m}}
        I_{abcd} T_{\text{XUV--IR},C;lm,ab}^{(2)} d_{B;lm,c}^{(1)*}
        D_{BC;d} \,, \label{eq:Qiipwasy}
\end{align}
where
\(I_{abcd} = (4\pi)^{-1} \int \epsilon_{\text{XUV};a} \epsilon_{\text{XUV};b} \epsilon_{\text{IR};c} \epsilon_{\text{IR};d} \mathrm{d}\Omega =
(\delta_{ab}\delta_{cd} + \delta_{ac}\delta_{bd} + \delta_{ad}\delta_{bc})/15\) comes from averaging over
molecular orientations
and the angular integral \(\langle \lambda \mu | \hat{n}_d | lm \rangle\)  can be expressed using an integral
of three real spherical harmonics as \(\sqrt{4\pi/3} \langle X_{\lambda\mu} | X_{1d} | X_{lm} \rangle\).
The term $\Delta_{BC} = \mathcal{E}_C - \mathcal{E}_B$ indicates the energy difference between the two cationic states $B^2\Sigma^+_u$ and $C^2\Sigma^+_g$.
Clearly, the term \(Q_{ii}\) retains the constants of proportionality 
\(1/(\Delta_{BC} - \omega_\text{IR})\) and interferes with \(Q_{cc}\)
exactly as described above.
Only the partial waves are now mixed in an opaque way due to the angular integrations.
The corresponding magnitudes and phases of \(Q_{cc}\) and \(Q_{ii}\) are plotted in Fig.~\ref{fig:pwasy}.

\begin{figure}[htbp]
    \centering
    \includegraphics[width=0.4\textwidth]{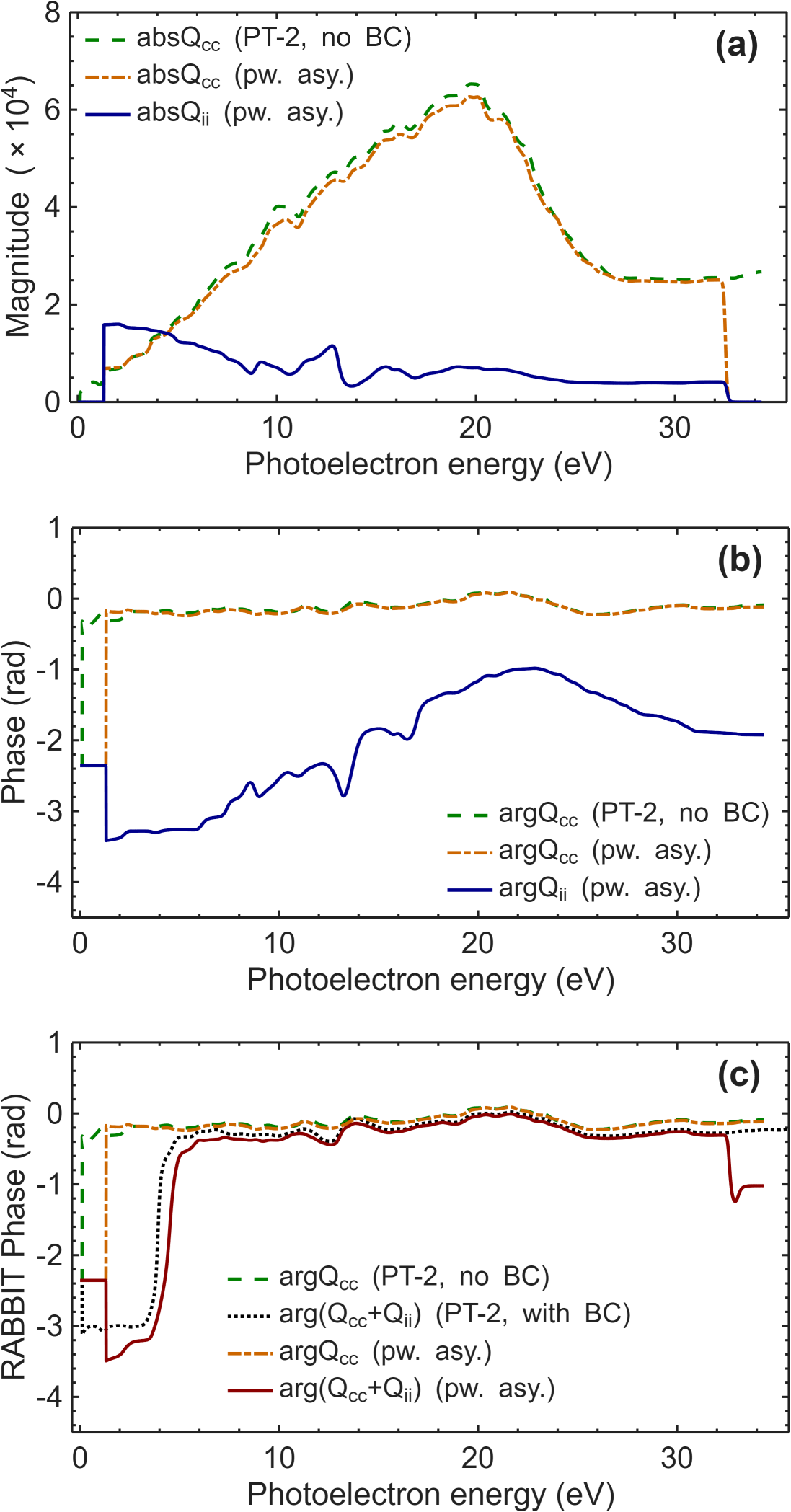}
    \caption{Magnitudes and phases of the orientation-averaged and emission-integrated free-free and ion-ion transition
    contributions to the complete RABBIT interference term, \(Q = Q_{cc} + Q_{ii}\), as introduced earlier. The labels
    ``pw. asy.'' refer to the approximation done in Eqs.~\eqref{eq:Qffasy}, \eqref{eq:Qiiasy},
    while ``PT-2'' means the full two-photon
    R-matrix calculation in the second order of perturbation theory.}
    \label{fig:asy}
\end{figure}

We can even go one step further with the approximation and still obtain qualitatively satisfactory result if we use the same
free-free correction factor \(A_{\kappa 1 k 0}\) universally for all partial waves. Then
\begin{align}
    Q_{cc} &\approx A_{\kappa1 k0}^{*} \sum_{\substack{abcd\\ \lambda\mu l m}}
        I_{abcd} T_{\text{XUV--IR},C;lm,ab}^{(2)} d_{C;\lambda\mu,c}^{(1)*}
        \langle \lambda \mu | \hat{n}_d | lm \rangle \,, \label{eq:Qffasy} \\
    Q_{ii} &\approx \frac{1}{\Delta_{BC} - \omega_\text{IR}} \sum_{\substack{abcd\\ l m}}
        I_{abcd} T_{\text{XUV--IR},C;lm,ab}^{(2)} d_{B;lm,c}^{(1)*}
        D_{BC;d} \,, \label{eq:Qiiasy}
\end{align}
The corresponding magnitudes and phases of \(Q_{cc}\) and \(Q_{ii}\) are plotted in Fig.~\ref{fig:asy}.
However, any further simplification (e.g., replacement of \(A_{\kappa 1 k 0}\) with the Dahlstr\"om factor \(A_{\kappa k}\)
or with the limit \(\mathrm{i}k/\omega^2\)) does not reproduce the correct size and position of the step.

The destructive interference depends on two aspects:
\begin{enumerate}
 \item \textit{The magnitudes of the two-photon ionization amplitudes for the ``ii'' and ``cc'' processes
    are comparable.}\\
    This is caused particularly by the opposite energy dependence of the XUV cross sections~\cite{masinElectronCorrelationsPrecollision2018}.
    While the cross section of the $B^2\Sigma^+_u$ state starts at large values and decreases at higher energies,
    the cross sections into the $C^2\Sigma^+_g$ state is small close to the threshold and gets progressively larger
    with the increasing energy. This might be related to the initial slow ramp-up of
    the contribution of the relatively high-\(\ell\) dominant partial wave.
 \item \textit{The phases of the two amplitudes are opposite.}\\
    The phase of the ``ii'' part differs from the phase of the ``cc'' approximately by \(\pi\). There are two contributions
    to this difference: Half of it comes from the radial continuum-continuum integral and is related to
    Fourier transform of the dipole operator. The second half arises in phase difference between dominant
    partial-wave contributions to the one-photon ionization amplitudes of states $B^2\Sigma^+_u$ and $C^2\Sigma^+_g$.
\end{enumerate}
Some selected points are discussed below in detail.

\subsection*{Phase analysis}

The magnitudes and phases of the interfering terms \(Q_{cc}\) and \(Q_{ii}\) are plotted in Figs.~\ref{fig:pwasy} and~\ref{fig:asy}
for IR wavelength
\(\lambda = 1030\)~nm. It is clear that the two terms have a comparable magnitude and opposite phase below the total absorbed
energy of 25~eV. This energy coincides with the steep change in the RABBIT phase. The bottom panel compares the RABBIT phase
in the full model (with and without the $B-C$ coupling) to the prediction of the asymptotic, high-energy approximation.
From the figure it is clear that the asymptotic approach works very well at energies above 25~eV, but keeps sufficient accuracy
even at energies below 25~eV.

The phase of the term \(Q_{cc}\) is approximately zero, or at least much closer to zero than the phase of the term \(Q_{ii}\).
This is expected because this term corresponds to the phase difference between the standard RABBIT XUV--IR and XUV+IR pathways
through the continuum. Both pathways acquire similar phases, albeit not identical, and the difference is small, typically corresponding
to tens of attoseconds.

The phase of the term \(Q_{ii}\) is very close to \(\pi\) (or, equivalently, \(-\pi\)), corresponding to an opposite sign. This
is understandable if we break down this term further by applying asymptotic factorization also to the XUV--IR amplitude. We may write in the partial wave and Cartesian component expansion
\begin{equation}
    Q_{ii} \approx \frac{1}{\Delta_{BC} - \omega_\text{IR}} \sum_{\substack{abcd\\ lm\lambda\mu}} I_{abcd} A_{\kappa' \lambda k l}
    d_{B;lm,c}^{(1)*}(\kappa') \langle lm | \hat{n}_b | \lambda \mu\rangle d_{C,\lambda\mu,a}^{(1)}(\kappa) D_{BC,d}
    \label{eq:Qiisimp}
\end{equation}
For \(\sim1000\)-nm IR field, the fraction has a complex phase of zero. (Interestingly, for a \(800\)-nm IR field it would switch
sign and remove the destructive interference and the jump in delays altogether.) The complex phase of the radial integral
\(A_{\kappa' \lambda k l}\) can be approximated by \(+\pi/2\), see Eq.~\eqref{eq:Akasy}.

\begin{figure}[htbp]
    \centering
    \includegraphics[width=0.8\textwidth]{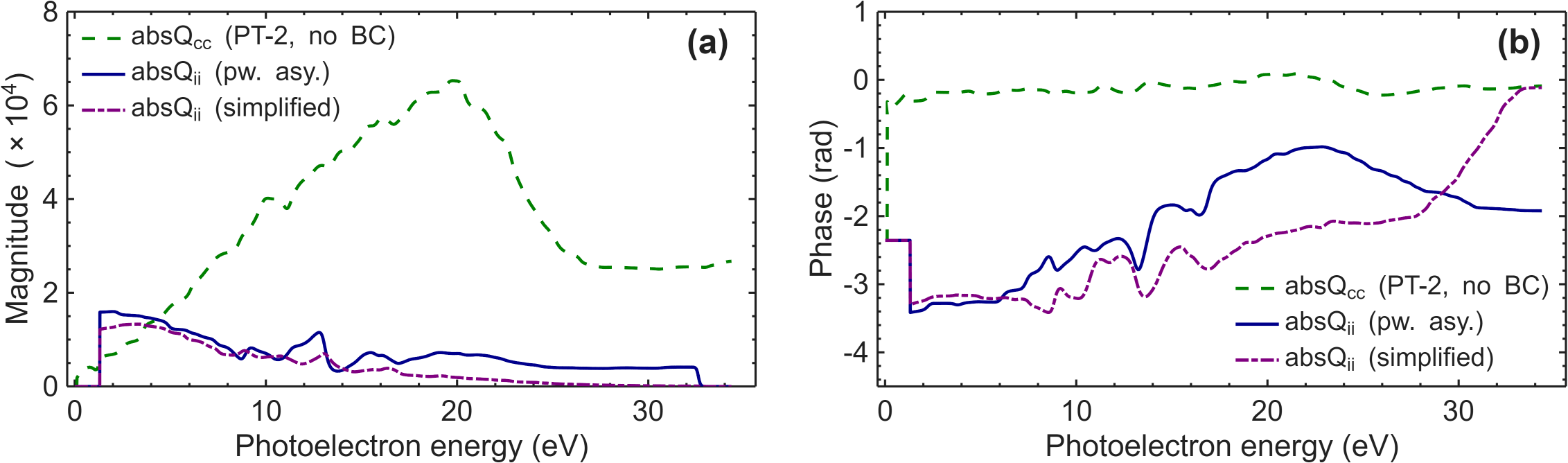}
    \caption{Magnitudes and phases of the orientation-averaged and emission-integrated free-free and ion-ion transition
    contributions to the complete RABBIT interference term, \(Q = Q_{cc} + Q_{ii}\), as introduced earlier. The labels
    ``pw. asy.'' refer to the approximation done in Eq.~\eqref{eq:Qffasy}, \eqref{eq:Qiiasy}, while ``PT-2'' means the full two-photon
    R-matrix calculation in the second order of perturbation theory. The label ``simplified'' refers to the restriction
    to a single partial wave process: \(d\)-wave in \(B\) channel to \(f\)-wave in \(C\) channel.}
    \label{fig:simplified}
\end{figure}

\begin{figure}[htbp]
    \centering
    \includegraphics[width=0.8\textwidth]{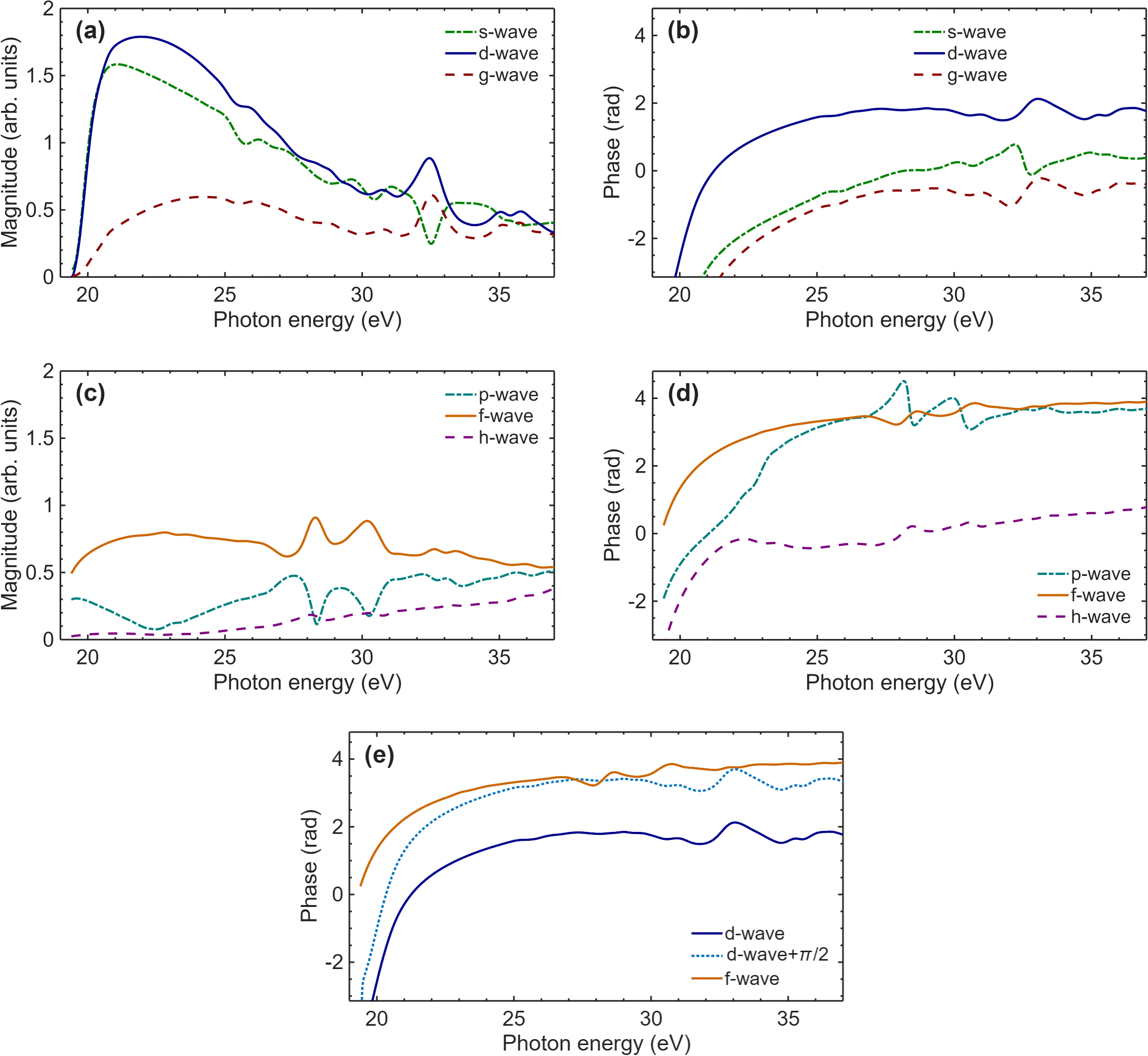}
    \caption{Magnitudes and phases of the partial-wave resolved one-photon ionization dipoles. Only partial waves for
    \(m = 0\) are shown. (Magnitudes for dipoles with \(m \ne 0\) are smaller than those for \(m = 0\) and are neglected in this
    analysis.) The shifted curve in the last
    panel illustrates that the phase of \(d\)-wave component in \(B\) channel is behind in phase roughly by \(\pi/2\) compared to
    the \(f\)-wave component in \(C\) channel.}
    \label{fig:pws}
\end{figure}

The remaining part can be significantly
simplified by restricting to the case \(l = 2\), \(\lambda = 3\), \(m = \mu = 0\). The angular momentum \(l = 2\) corresponds
to the dominant partial wave contribution in the \(B\) channel, while the angular momentum \(\lambda = 3\) corresponds to the
dominant partial wave contribution in the \(C\) channel. The dominant partial waves actually do correspond to \(m = \mu = 0\),
so this restriction is meaningful. The restriction to zero \(m\) and \(\mu\) is motivated by cylindrical
symmetry of the molecule. This approximation distorts the \(Q_{ii}\) energy dependence, but not so much at low energies,
see Fig.~\ref{fig:simplified}. This is the key observation: At low energies, the only relevant contribution to \(Q_{ii}\)
comes from the combination of a \(d\)-wave in the $B^2\Sigma^+_u$ channel and the \(f\)-wave in the $C^2\Sigma^+_g$ channel. These two
partial waves have almost constant phase difference: \(+\pi/2\), see Fig.~\ref{fig:pws}. It is apparent from Fig.~\ref{fig:pws}(a) that the ionization signal in the intermediate $B^2\Sigma^+_u$ channel is also strongly contributed by the \(s\)-wave. However, the latter does not contribute to the \(Q_{ii}\) term too much. In order to contribute, there would need to be a dipole-transition-allowed partial wave (i.e., the \(p\)-wave) in the $C^2\Sigma^+_g$ channel, see the bracket in Eq.~\eqref{eq:Qiisimp}, but the \(p\)-wave is
suppressed compared to the dominant \(f\)-wave.

Having only one combination of partial waves contribute to Eq.~\eqref{eq:Qiisimp} allows straightforward computation of the overall phase of the integral.
Given that the phases appear in Eq.~\eqref{eq:Qiisimp} with opposite signs, only the relative phase contributes to the formula.
Together with the continuum-continuum phase we are indeed getting the expected \(\arg Q_{ii} \approx \pm\pi\).

Overall, the phases can be expressed as:
\begin{align}
    \arg Q_{cc} &= \arg \underbrace{\langle T_{\text{XUV+IR}}^{(2)cc*} T_{\text{XUV--IR}}^{(2)cc} \rangle}_{\sim 0} \,, \\
    \arg Q_{ii} &\approx \arg \underbrace{\frac{1}{\Delta_{BC} - \omega_\text{IR}}}_{0}
        \underbrace{A_{\kappa 3 k 2}}_{\sim \pi/2}
        \underbrace{I_{zzzz} d_{B;20,z}^{(1)*}(\kappa') \langle 20 | \hat{n}_z | 30 \rangle d_{C;30,z}^{(1)}(\kappa) D_{BC,z}}_{\sim \pi/2} \,,
    \label{eq:argQii}
\end{align}
where \(I_{zzzz} = 3/15\) and \(\langle 20 | \hat{n}_z | 30 \rangle = 3/\sqrt{35}\). The dominance of the \(z\) components
can be interpreted as that the majority of the ``ii'' signal is coming from molecules oriented along the polarization, regardless
of the emission direction. This is plausible, as the quantity being averaged over orientations is weighted by
\(D_{BC;z} \sim D_{BC} \cos\beta\), which suppresses contributions of molecules perpendicular to the polarization.

\subsection*{Preferred orientations}

\begin{figure}[htbp]
    \centering
    \includegraphics[width=0.7\textwidth]{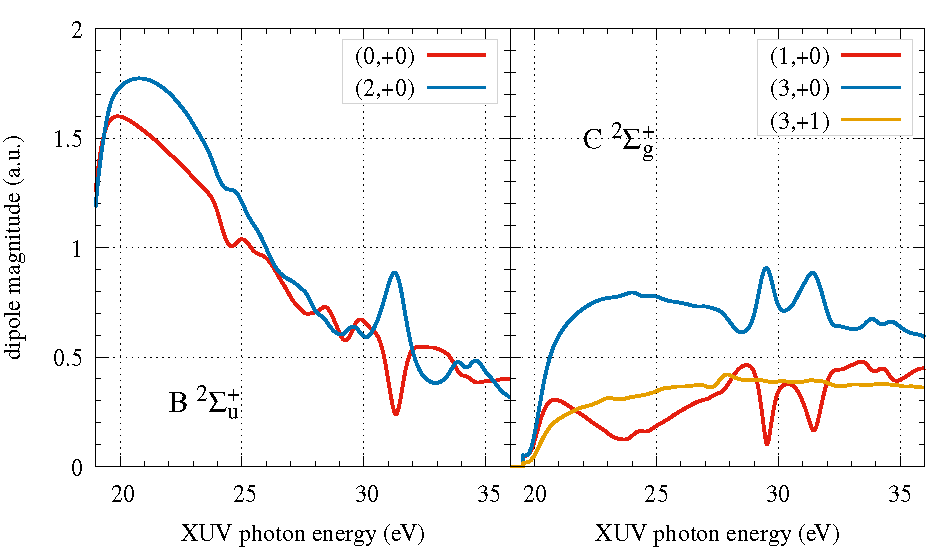}
    \caption{Magnitude of dominant partial-wave one-photon dipoles for ionization into \(B\) and \(C\) states at low energies.}
    \label{fig:dominant-dipoles}
\end{figure}

With the definition of the orientation averaging integral \(I_{abcd} = (\delta_{ab}\delta_{cd} + \delta_{ac}\delta_{bd} + \delta_{ad}\delta_{bc})/15\),
Eq.~\eqref{eq:Qiisimp} can be written also as
\begin{align}
    Q_{ii} \approx &\frac{1}{\Delta_{BC} - \omega_\text{IR}} \sum_{lm\lambda\mu} \frac{1}{15} A_{\kappa' \lambda k l}
        \left[ I_1 + I_2 + I_3 \right], \label{eq:Qiiavg} \\
    I_1 &= \left(\bm{d}_{C;\lambda\mu}^{(1)}(\kappa) \cdot \langle lm | \hat{\bm{n}} | \lambda \mu\rangle \right)
        \left(\bm{d}_{B,lm}^{(1)*}(\kappa') \cdot \bm{D}_{BC}\right), \\
    I_2 &= \left(\bm{d}_{B;lm}^{(1)*}(\kappa') \cdot \bm{d}_{C,\lambda\mu}^{(1)}(\kappa)\right)
        \left(\langle lm | \hat{\bm{n}} | \lambda \mu\rangle \cdot \bm{D}_{BC}\right), \\
    I_3 &= \left(\bm{d}_{C,\lambda\mu}^{(1)}(\kappa) \cdot \bm{D}_{BC}\right)
        \left(\bm{d}_{B;lm}^{(1)*}(\kappa') \cdot \langle lm | \hat{\bm{n}} | \lambda \mu\rangle \right).
\end{align}
We proceed in molecular frame, where \(z\)-axis is parallel with the molecular axis. Assuming real spherical harmonic basis,
\(\bm{n} = (n_x, n_y, n_z) \equiv (n_+, n_-, n_0)\), the dominant partial one-photon transition dipoles are
\begin{align}
    \bm{d}_{B;0,+0}^{(1)} &= (0, 0, d_{B;0,+0,z}^{(1)}) \,, \\
    \bm{d}_{B;2,+0}^{(1)} &= (0, 0, d_{B;2,+0,z}^{(1)}) \,, \\
    \bm{d}_{C;3,+0}^{(1)} &= (0, 0, d_{C;3,+0,z}^{(1)}) \,, \\
    \bm{d}_{C;3,+1}^{(1)} &= (d_{C;3,+1,x}^{(1)}, 0, 0) \,, \\
    \bm{d}_{C;3,-1}^{(1)} &= (0, d_{C;3,-1,y}^{(1)}, 0) \,.
\end{align}
see also Figs.~\ref{fig:pws} and~\ref{fig:dominant-dipoles}, while the residual ion \(B\)--\(C\) transition dipole is
\begin{equation}
    \bm{D}_{BC} = (0, 0, D_{BC}) \,.
\end{equation}
The transition dipole for $B^2\Sigma^+_u$-state and for the photoelectron partial wave \((0,+0)\) is irrelevant, as it is not connected by
a dipole-allowed transition to any partial wave relevant for the $C^2\Sigma^+_g$-state signal. That is, for the \(s\)-wave the
angular integral \(\langle 0,+0 | \hat{\bm{n}} | 3,\mu \rangle\) cannot be non-zero.
Clearly, of the three terms in Eq.~\eqref{eq:Qiiavg} the last two contribute only \(z\)-components in both factors.
Specifically,
\begin{align}
    I_2 &= d_{B;2,+0,z}^{(1)*} d_{C,3,+0,z}^{(1)} \langle 2,+0 | \hat{n}_z | 3,+0 \rangle D_{BC} \,, \\
    I_3 &= d_{C,3,+0,z}^{(1)} D_{BC} d_{B;2,+0,z}^{(1)*} \langle 2,+0 | \hat{n}_z | 3,+0 \rangle \,.
\end{align}
Only the first term of Eq.~\eqref{eq:Qiiavg} is nonzero also for \(x\) and \(y\) components, namely
\begin{gather}
    \sum_{\lambda\mu lm} A_{\kappa\lambda kl} \left(\bm{d}_{C;\lambda\mu}^{(1)}
    \cdot \langle lm | \hat{\bm{n}} | \lambda \mu\rangle \right)
    \left(\bm{d}_{B,lm}^{(1)*} \cdot \bm{D}_{BC}\right)
    \approx
    A_{\kappa3k2} \left[
        d_{C;3,+1,x}^{(1)} \langle 2,+0 | \hat{n}_x | 3,+1 \rangle \right. \nonumber \\
    + \left. d_{C;3,-1,y}^{(1)} \langle 2,+0 | \hat{n}_y | 3,-1 \rangle 
        + d_{C;3,+0,z}^{(1)} \langle 2,+0 | \hat{n}_z | 3,+0 \rangle
    \right] d_{B,2,+0,z}^{(1)*} D_{BC} \,.
\end{gather}
The same \(zzzz\) combination is thus present in \(I_1\), \(I_2\) and \(I_3\). Additionally, \(I_1\) also contains contributions of
some additional terms (polarization directions). However, the partial waves \((3,\pm1)\) in the $C^2\Sigma^+_g$ channel are not as strong as
the \((3,+0)\) partial wave, so their omission does not affect the result too much. Then only the \(zzzz\) term remains, again.
In conclusion, we can observe that:

At low energies, the dominant part of the \textit{ion-ion} component of the RABBIT signal into $C^2\Sigma^+_g$ state of
CO\textsubscript{2}\textsuperscript{+} comes from molecules oriented along the polarization vector. This is for two reasons:
\begin{itemize}
 \item The dominant photoionization signal associated with $B^2\Sigma^+_u$ and $C^2\Sigma^+_g$ states comes from molecules oriented parallel with
       the polarization vector.
 \item The \textit{ion-ion} pathway contributes the most when the transition dipole \(\bm{D}_{BC}\) is parallel with polarization.
\end{itemize}

\subsection*{Transition operators for the different RABBIT pathways}

The final $C^2\Sigma^+_g$-channel after XUV\(\pm\)IR interaction is accessible by the three well-known paths
\begin{enumerate}
 \item XUV\(+\)IR path from $C^2\Sigma^+_g$ through absorption in continuum,
 \item XUV\(-\)IR path from $C^2\Sigma^+_g$ through emission in continuum,
 \item XUV\(+\)IR path from $B^2\Sigma^+_u$ through absorption in residual ion.
\end{enumerate}
In all three cases, there is always some operator \(\hat{V}\) that mediates the transition from the XUV-ionized (intermediate)
state to the IR-affected (final) state and it can be trivially factorized to the two subspaces:
\begin{enumerate}
 \item \(\hat{V} = \hat{I}_\text{ion} \otimes \hat{D}_\text{pe}\);
    a product of an identity operator for the ion subspace and a dipole operator for the photoelectron subspace.
 \item \(\hat{V} = \hat{I}_\text{ion} \otimes \hat{D}_\text{pe}\);
    a product of an identity operator for the ion subspace and a dipole operator for the photoelectron subspace
    (same as above).
 \item \(\hat{V} = \hat{D}_\text{ion} \otimes \hat{I}_\text{pe}\);
    a product of a dipole operator for the ion subspace and an identity operator for the photoelectron subspace.
\end{enumerate}
The two-photon amplitude can be approximately written as a product of the XUV one-photon ionization
amplitude \(d^{(1)}\) and a IR-correction integral for the subsequent interaction with IR:
\begin{equation}
    T^{(2)} \sim A_\text{IR} d^{(1)} = A_\text{IR}^\text{ion} A_\text{IR}^\text{pe} d^{(1)} \,.
\end{equation}
The correction factor is a product of the ion-ion transition mediated either by \(\hat{I}_\text{ion}\) or by \(\hat{D}_\text{ion}\),
and of the continuum-continuum transition mediated either by \(\hat{I}_\text{pe}\) or by \(\hat{D}_\text{pe}\). In theory, the
above formula only holds (and even then only approximately) just in the partial wave resolution, but in practice it is often good
enough even for the complete amplitudes.

\subsection*{Contribution from a continuum transition}

The photoelectron IR-correction factor is, schematically, a matrix element of the continuum transition operator
\(\hat{O}_\text{pe}\) between the initial
and the final state of the photoelectron. For a given intermediate (\(\nu,\lambda,\mu\)) and final (\(n,l,m\))
combination of the state of the residual ion and of a photoelectron partial wave it reads
\begin{equation}
    A_{\text{IR}}^\text{pe} \sim
        -\frac{2}{\sqrt{k_\nu k_n}}
        \int_0^{+\infty} \sin [k_n r] \langle lm | \hat{O}_\text{pe} | \lambda\mu \rangle
        \exp [\mathrm{i}k_\nu r] \mathrm{d}r \,.
    \label{eq:A-IR-pe}
\end{equation}
Here we neglected the Coulomb field of the residual ion (compared to equation~(16) of~\cite{bendaAngularMomentumDependence2025}).
Additionally, we will neglect the positive-phase exponential in the sine,
and also take advantage of
\begin{equation}
    \langle lm | \hat{I}_\text{pe} | \lambda\mu \rangle \sim 1 \,, \qquad
    \langle lm | \hat{D}_\text{pe} | \lambda\mu \rangle \sim r \,,
\end{equation}
where the angular parts were omitted for brevity. The equation~\eqref{eq:A-IR-pe} then simplifies to a Fourier half-transform,
\begin{equation}
    A_{\text{IR}}^\text{pe} \sim \frac{1}{\mathrm{i}} \int_0^{+\infty} r^s \mathrm{e}^{\mathrm{i}(k_\nu - k_n)r} \mathrm{d}r
    = \mathrm{i}^{s} \frac{s!}{\sqrt{k_\nu k_n} (k_\nu - k_n)^{s+1}} \,.
\end{equation}
The conclusion is the following:

The IR probe essentially approximates the Fourier (half-)transform of the photoelectron IR coupling potential (which is identity
for the ``ii'' transition, radius for the ``cc'' transition). The phase is then directly related to the phase of this transform.

This is similar as in the first-order (Born) approximation to electron scattering.

Phase of the Fourier half-transform is the same as the phase of the full Fourier transform. So, for the two relevant processes
(absorption of IR in the ion and in the photoelectron, respectively) we have:
\begin{align}
    \text{ii}: 1 &\xrightarrow{\mathcal{F}} \delta(q) \,, \\
    \text{cc}: r &\xrightarrow{\mathcal{F}} \mathrm{i}\delta'(q) \,.
\end{align}
If, hypothetically, the coupling potential due to the IR field was quadrupolar, this would result in a full sign reversal:
\begin{align}
    \text{cc(q)}: r^2 &\xrightarrow{\mathcal{F}} \mathrm{i}^2\delta''(q) \,.
\end{align}
\bibliographystyle{apsrev4-2}

\bibliography{library.bib}